\documentclass[a4paper,twoside,prd,preprintnumbers,twocolumn]{revtex4}
\usepackage{amssymb,amsmath,bm,natbib}
\usepackage{color,xcolor}
\usepackage{slashed}
\usepackage{graphics}
\usepackage{graphicx}
\usepackage{xspace}
\usepackage[utf8]{inputenc}
\usepackage{tabularx}
\usepackage[caption=false]{subfig}
\usepackage{hyperref}
\usepackage{url}
\usepackage{dsfont}
\usepackage{float} 
\usepackage{cancel}
\usepackage[normalem]{ulem}
\usepackage[T1]{fontenc}

\newcommand{\GeV}{\,\text{GeV}}

\newcommand{\gev}{\,\text{GeV}}

\newcommand{\BKs}{$B\to K^*\mu^+\mu^-$\ }

\newcommand{\qsq}{\ensuremath{q^{2}}\xspace}

\newcommand{\knn}{\textit{knn}\xspace}
\newcommand{\Kp}{\ensuremath{K^{+}}\xspace}

\newcommand{\pip}{\ensuremath{\pi^{+}}\xspace}
\newcommand{\pim}{\ensuremath{\pi^{-}}\xspace}
\newcommand{\mkpi}{\ensuremath{m_{K\pi}}\xspace}
\newcommand{\Kstarz}{\ensuremath{K^{\ast{}0}}\xspace}

\newcommand{\soneac}{\ensuremath{\tilde{S}_{1a}^{c}}\xspace}
\newcommand{\stwoac}{\ensuremath{\tilde{S}_{2a}^{c}}\xspace}
\newcommand{\sonect}{\ensuremath{\tilde{S}_{1}^{c}}\xspace}
\newcommand{\stwoct}{\ensuremath{\tilde{S}_{2}^{c}}\xspace}
\newcommand{\sonest}{\ensuremath{\tilde{S}_{1}^{s}}\xspace}
\newcommand{\stwost}{\ensuremath{\tilde{S}_{2}^{s}}\xspace}

\newcommand{\CP}{\ensuremath{CP}\xspace}

\newcommand{\dd}{\ensuremath{{\rm d}}}

\begin{document}

\title{Model-independent unbinned analysis of $B \to K^*(\to K^+\pi^-)\mu^+\mu^-$: zeroes, bounds, Wilson coefficients and symmetries}

\author{Bernat Capdevila}
\affiliation{Universitat Aut\`onoma de Barcelona, 08193 Bellaterra, Barcelona,\\
DAMTP, University of Cambridge, Wilberforce Road, Cambridge, CB3 0WA, United Kingdom\\
Departament de Física Quàntica i Astrofísica, Universitat de Barcelona, Martí Franquès 1, E08028 Barcelona\\
Institut de Ciències del Cosmos (ICCUB), Universitat de Barcelona, Martí Franquès 1, E08028 Barcelona}
\author{Joaquim Matias}
\affiliation{Universitat Aut\`onoma de Barcelona, 08193 Bellaterra, Barcelona,\\
Institut de F\'{i}sica d'Altes Energies (IFAE), The Barcelona Institute of Science and Technology, Campus UAB, 08193 Bellaterra (Barcelona)}
\author{Mart\'in Novoa-Brunet}
\affiliation{Instituto de F\'isica Corpuscular,
Universitat de Val\'encia – Consejo Superior de Investigaciones Cient\'ificas,
Parc Cient\'ific, E-46980 Paterna, Valencia, Spain}
\author{Mitesh Patel}
\affiliation{Imperial College London, Prince Consort Road, London, United Kingdom}
\author{Mark Smith}
\affiliation{Imperial College London, Prince Consort Road, London, United Kingdom}

\begin{abstract}
We present a model-independent method to study the four-body decay $B\to K^*(\to K^+\pi^-)\mu^+\mu^-$, based on extracting continuous observables with a moments approach. 
The method allows the observables to be determined unbinned in both the dilepton and $K^+\pi^-$  invariant masses on which the decay dynamics depend. This will allow the method to shed new light on how the observables depend on the P- and S-wave contributions to the $K^+\pi^-$ system. This approach contrasts with the state-of-the-art analyses, which bin in dilepton and $K^+\pi^-$ mass, or use a model for the dependence of the underlying decay amplitudes on these masses. The method does not require making a statistical fit, and so avoids problems of biases and poor uncertainty estimation when dealing with small samples or a large number of fit parameters.
We provide the Standard Model predictions for the unbinned optimised observables, derive new geometrical bounds on their values
and study the robustness of these bounds in the presence of a scalar new physics contribution.
We explore the zero-crossing points of $P_2$ and $P_{4,5}^\prime$ observables as a function of a new physics contribution to the dominant vector Wilson coefficient, $C_9^{\rm NP}$. 
We also discuss the conditions that can be used to test the theoretical model of the amplitudes needed for an experimental amplitude analysis.
Finally, as an illustration, we show how the proposed method might be used to extract  the zero-crossing points, make a comparison with the bounds and test a non-trivial relation between the observable values.
\end{abstract}
\maketitle

\section{Introduction}
\label{intro}

Anomalies have been observed in the measurements of a range of $B$-meson decay modes. The most persistent effects are the deviations in the $P_5^\prime$~\cite{Descotes-Genon:2012isb} 
angular observable belonging to the $b\to s \mu^+\mu^-$ decay \mbox{$B \to K^*(\to K^+\pi^-)\mu^+\mu^-$}, 
the deficit  in 
the branching fractions of several other $b\to s \ell^+\ell^-$ processes (with $\ell=e,\mu$)~\cite{LHCb:2014cxe,LHCb:2021zwz} and the Lepton Flavour Universality~(LFU) ratios $R_{D,D^*}$ with charged current $b\to c\tau\nu$ decays~\cite{LHCb:2023zxo,LHCb:2023uiv}.

The data of the four-body angular distribution \mbox{$B \to K^*(\to K^+\pi^-)\mu^+\mu^-$} have been commonly presented in bins of the dilepton invariant mass squared (denoted $q^2$) by the LHCb  collaboration~\cite{LHCb:2020lmf}, but also by the Belle~{\cite{Belle:2016fev}, CMS~\cite{CMS:2024atz} and ATLAS~{\cite{ATLAS:2018gqc} collaborations. Analyses are typically made within a window of $K^+\pi^-$ invariant mass, $m_{K\pi}$. 
It is possible to construct a full set of angular observables, optimised to have reduced sensitivity to hadronic effects~\cite{Descotes-Genon:2013vna,Capdevila:2017bsm}.
The LHCb collaboration have recently presented two analyses that determine the complete set of such observables unbinned in $q^2$~\cite{LHCb:2023gpo, LHCb:2024onj}.
\bigskip

\noindent The purpose of these analyses was to extract Wilson coefficients together with hadronic information directly from the data using an Effective Field Theory (EFT)-based functional form for the transversity amplitudes. These analyses differ in the way that the hadronic contributions are parameterised. On the one hand, Ref.~\cite{LHCb:2023gpo}  relies on the use of a $z$-expansion to parameterise the non-local charm-loop contributions and includes constraints from $B(B\to K^*J/\psi)$ and $B(B\to K^*\psi(2s))$ decays. In this analysis, two hypotheses are considered for the data included in the fit: (i) $q^2<0$, where constraints coming from the light-cone Operator Product Expansion~(OPE) calculation in Ref.~\cite{Gubernari:2020eft,Gubernari:2022hxn} are included; and (ii) $q^2>0$, where these constraints are excluded. By contrast, the alternative analysis of Ref.~\cite{LHCb:2024onj} uses a dispersive approach based on the resonance model of Ref.~\cite{Cornella:2020aoq} and data from the decay $B\to K^*\mu^+\mu^-$ alone. An interesting outcome of both of these analyses is that the data indicate that the charm-loop is insufficient to explain the observed anomalies, and a new physics contribution is required.
\clearpage

Our goal in the present paper is to report a new model-independent method to parameterise directly the unbinned observables, as opposed to the amplitudes, without any theoretical input. The results of this method will thus be a facsimile of the experimental data, which can then be used for further analysis of the underlying physics. We also show that this method can be used to test bounds on the observables and to extract zero-crossing points  that give additional information on the Wilson coefficients. We use simplified pseudo-experiments to illustrate how the method works. This new method is in contrast with the above analyses, which attempt to parameterise the various contributions to the transversity amplitudes, i.e. local/non-local form factors, resonance lineshapes etc. It also contrasts with another proposed method to directly parameterise the amplitudes with \qsq-dependent polynomials~\cite{Egede:2015kha} and the alternative method of Ref.~\cite{Beck:2025qxx}. In all of the above cases, the theoretical models and hypotheses used to construct the amplitudes are then inherited by any subsequent calculation of the observables.

The structure of the paper is as follows. In section~\ref{section2}, we present our Standard Model (SM) predictions for the optimised unbinned observables, together with a method to test the validity of any theoretical model of the amplitudes. In section~\ref{section3}, we discuss a new method to extract unbinned observables from data, which is underpinned by a moments approach.  In section~\ref{section4}, we present some examples of the information that can be extracted from the knowledge of the unbinned shape of the observables. In particular, we present new geometrical bounds on the optimised observables that can be used to test the robustness of an experimental analysis. We explore the sensitivity of the bounds, as well as a relation between the observables in the presence of a scalar. We then discuss the information that can be extracted directly 
from~the zeroes of $P_{2}$ and $P_{4,5}^\prime$ and, finally, conclude in section~\ref{conclusions}.

\section{SM predictions for the unbinned observables}\label{section2}

Electroweak transitions of $B$-mesons occur at characteristic energies far below the electroweak scale. This separation of energy scales implies that electroweak degrees of freedom are never resolved in such processes. Consequently, the physics at high energies can be effectively described at low energies within an EFT framework. This effective theory is commonly referred to in the literature as the Weak Effective Theory (WET)~\cite{Grinstein:1987vj,Buchalla:1995vs} or the Low Energy Effective Theory~(LEFT)~\cite{Jenkins_2018}. The structure of the corresponding effective Hamiltonian follows an OPE
\begin{equation}\label{eq:WETHamiltonian}
\mathcal{H}_\text{eff} = -\frac{4G_F}{\sqrt{2}} V_{tb}V_{ts}^* \sum_{i} \mathcal{C}_i(\mu) \mathcal{O}_i(\mu),
\end{equation}
where $G_F$ is the Fermi constant, $V_{tb}$ and $V_{ts}$ are CKM matrix elements, $\mathcal{C}_i(\mu)$ are the Wilson coefficients encoding the short-distance dynamics, and $\mathcal{O}_i(\mu)$ are the corresponding effective operators. The scale $\mu$ represents the renormalisation scale at which the Hamiltonian is evaluated. In addition to the SM operators, the effective Hamiltonian may also include  operators that encapsulate structures not generated in the SM, such as right-handed currents or scalar interactions, which arise in various New Physics (NP) scenarios. The most relevant operators for the following discussion are
\begin{eqnarray}
\begin{aligned}
\mathcal{O}_7^{(\prime)} &= \dfrac{e}{16\pi^2}(\bar{s}\sigma_{\mu\nu}P_{R(L)}b)F^{\mu\nu},\\
{\mathcal{O}}_{9\ell}^{(\prime)} &= \dfrac{e^2}{16\pi^2}(\bar{s} \gamma_{\mu} P_{L(R)} b) (\bar{\ell} \gamma^\mu \ell),\\
{\mathcal{O}}_{10\ell}^{(\prime)} &= \dfrac{e^2}{16\pi^2}(\bar{s} \gamma_{\mu} P_{L(R)} b) (\bar{\ell} \gamma^\mu \gamma_5 \ell),\\
{\mathcal{O}}_{S}^{(\prime)} &= \dfrac{e^2}{16\pi^2} m_b (\bar{s}P_{R(L)} b)(\bar{\ell}\ell),
\label{bslloperators}
\end{aligned}
\end{eqnarray}
where colour indices have been omitted, $P_{L,R} = (1 \mp \gamma_5)/2$ are chirality projection operators, $F^{\mu\nu}\equiv \partial^\mu A^\nu-\partial^\nu A^\mu$ is the electromagnetic field strength tensor ($A^\mu(x)$ being the photon field), and $\sigma^{\mu\nu}=\frac{i}{2}[\gamma^\mu,\gamma^\nu]$, with $\gamma^\mu$ denoting the gamma matrices in four dimensions. 

The Wilson coefficients for the most relevant operators of the effective Hamiltonian in Eq.~\eqref{eq:WETHamiltonian} within the SM, evaluated at $\mu_b=4.8$GeV, are given in Table~\ref{table:WilsonCoefficientsSMandNP}~\cite{Gambino:2003zm,Bobeth:2003at,Misiak:2006ab,Huber:2005ig,Huber:2007vv}.

This paper will be focused on $B$-meson decays into a $K^*$ meson  and a lepton pair $\ell^+\ell^-$. The relevant amplitude for this process has the structure
\begin{align}
\mathcal{M}(B\to K^*\ell^+\ell^-) =&\, \frac{G_F \alpha}{\sqrt{2}\pi} V_{tb}V_{ts}^* \Big[(\mathcal{A}_V^\mu + \mathcal{H}^\mu) \bar{u}_\ell \gamma_\mu v_\ell\nonumber\\ 
&+ \mathcal{A}_A^\mu \bar{u}_\ell \gamma_\mu \gamma_5 v_\ell\Big],
\end{align}
with
\begin{align}
\mathcal{A}_V^\mu =& -2m_b\dfrac{q_\nu}{q^2}\,{\cal C}_7 \langle K^* |\bar{s}\sigma^{\mu\nu}P_R b|B\rangle +{\cal C}_9 \langle K^* |\bar{s}\gamma^\mu P_L b|B\rangle, ~
\nonumber\\
\mathcal{A}_A^\mu =& \,\, {\cal C}_{10}\langle K^* | \bar{s}\gamma^\mu P_L b|B\rangle,~\label{eq:LocalFFs2}
\end{align}
and
\begin{align}~\label{eq:NonLocalFFs}
\mathcal{H}^\mu &= \dfrac{-16 i \pi^2}{q^2}\times&\\
&\sum_{i=1,\ldots,6,8} \mathcal{C}_i \int dx^4 e^{i q\cdot x} \langle K^*|T\{j^\mu_{\rm em}(x),\mathcal{O}_i(0)\}|B\rangle\nonumber,
\end{align}
where we assume no right-handed currents. The electromagnetic quark current is denoted as $j^\mu_{\rm em}$, $q^\mu$ represents the dilepton momentum, and $\mathcal{O}_{i=1,\ldots,6,8}$ ($\mathcal{C}_{i=1,\ldots,6,8}$) correspond to four-quark effective operators (Wilson coefficients) in the same effective Hamiltonian. 

The matrix elements in Eq.~\eqref{eq:LocalFFs2} can be expressed in terms of scalar structure functions known as local form factors. Our framework treats the local form factors in the low- and high-$q^2$ regions separately. Consequently, it does not attempt or rely on combined fits of theoretical calculations at low (or negative) $q^2$ with lattice data at large $q^2$. 

In the low-$q^2$ region, we employ the most recent calculations of the $B\to K^*$ form factors within the light-cone OPE. The form factors are computed using light-cone sum rules (LCSRs) with $B$-meson light-cone distribution amplitudes (LCDAs), incorporating terms up to twist four for both two- and three-particle distributions. As described in Ref.~\cite{Gubernari:2018wyi}, these results are then fitted to a $z$-expansion within the so-called BSZ parameterisation~\cite{Bharucha:2015bzk}. We use the resulting $z$-expansion parameterisation of the form factors and decompose them into three pieces: i)~soft form factors $\xi_{\perp,\parallel}$~\cite{Charles:1998dr,Beneke:2000wa}, which are the relevant structure functions emerging in the formal Heavy Quark Effective Theory (HQET) and Soft Collinear Effective Theory (SCET) limits ($m_b\to \infty$ and $q^2\to 0$); ii)  $O(\alpha_s)$ corrections $\Delta F^{\alpha_s}$, which are calculable within QCD factorization (QCDf)~\cite{Beneke:2000wa}; and iii) factorisable power corrections $\Delta F^\Lambda$~\cite{Descotes-Genon:2014uoa,Capdevila:2017ert},
\begin{equation}\label{eq:FFnlo}
F(q^2) = F^\infty(\xi_\perp(q^2),\xi_\|(q^2))+\Delta F^{\alpha_s}(q^2)+\Delta F^\Lambda(q^2).
\end{equation}
The latter, representing part of the $O(\Lambda_{\rm QCD}/m_b)$ power corrections to QCDf, is parameterised as an expansion in $q^2/m_B^2$~\cite{Jager:2012uw},
\begin{equation}
\Delta F^{\Lambda}(q^2) = a_F+b_F\,\frac{q^2}{m_B^2}+c_F\,\frac{q^4}{m_B^4}+\ldots\;,
\end{equation}
where $F=V,A_{0,1,2},T_{1,2,3}$. For each form factor, the parameters $a_F$, $b_F$, and $c_F$ are determined from a fit to the form factor values, taking into account the corresponding structure in the HQET/SCET limit, as well as the associated $O(\alpha_s)$ corrections, following the schematic structure implied in Eq.~\eqref{eq:FFnlo}. Further details on the fit strategy and results for the factorisable power corrections, as extracted from KMPW~\cite{Khodjamirian:2010vf} and BSZ~\cite{Bharucha:2015bzk} form factors, can be found in Refs.~\cite{Descotes-Genon:2014uoa,Capdevila:2017ert}.

In contrast, in the high-$q^2$ region we use lattice QCD calculations of $B\to K^*$ form factors~\cite{Horgan:2013hoa}, which are subsequently fitted using a $z$-expansion within the BCL parameterisation~\cite{Bourrely:2008za}.

\begin{table}
\begin{tabular}{|c|c|c|c|}
  \hline
  Model & ${\cal C}_7^{\rm eff}(\mu_b)$ & ${\cal C}_9^{\rm eff}(\mu_b)-Y(q^2)$ & ${\cal C}_{10}(\mu_b)$  \\
  \hline
  SM & -0.29 & +4.08 & -4.31  \\
  \hline
\end{tabular}
\bigskip

\begin{tabular}{|c|c|c|c|c|c|c|}
  \hline
  Model & ${\cal C}_7^{NP}$ & ${\cal C}_9^{U}$ & ${\cal C}_{10}^{U}$ & ${\cal C}_{7^\prime}$ & ${\cal C}_{9^\prime}^{U}$ & ${\cal C}_{10^\prime}^U$ \\
\hline
  NP1 &  & -1.17 &  &  &  &  \\
\hline
  NP2 & +0.00 & -1.21 & +0.07 & +0.01 & -0.04 & -0.06 \\
  \hline
\end{tabular}
\caption{Main SM and NP Wilson coefficients at a scale $\mu_b=4.8$ GeV. The remaining SM Wilson coefficients can be found in Ref.~\cite{Descotes-Genon:2013vna}. NP1 hypothesis corresponds to the preferred LFU scenario discussed in Ref.~\cite{Alguero:2023jeh}, while NP2 corresponds to the 6D LFU hypothesis.}\label{table:WilsonCoefficientsSMandNP} 
\end{table}

While the contributions from local form factors are included in $\mathcal{A}_{V,A}^\mu$, non-local quark loop contributions appear in the term $\mathcal{H}^\mu$. In the low-$q^2$ region, these effects are taken into account by recasting the phenomenological parameterisation provided in Eq.~(7.14) of Ref.~\cite{Khodjamirian:2010vf}. This parameterisation is itself based on the light-cone OPE and LCSR calculations of the non-local form factor in the region $q^2\ll 4m_c^2$, and its corresponding matching to the dispersion relation proposed in the same paper. In practice, this is implemented as a shift in the factorisable charm-quark loop function $Y(q^2)$~\cite{Buras:1993xp}
\begin{equation}\label{eq:nonlocalffnlo}
 Y_i(q^2) = Y(q^2) + s_i Y_i^{LD}(q^2), \quad i=\perp,\parallel, 0,
\end{equation}
where the long-distance functions $Y_i^{LD}$, which depend on the $K^*$ polarization states $\perp$, $\parallel$, and $0$, are given by~\cite{Descotes-Genon:2015uva}
\begin{align}
Y_{LD}^\perp(q^2)&=\dfrac{a^\perp+b^\perp q^2(c^\perp -q^2)}{q^2(c^\perp -q^2)},\\
Y_{LD}^\parallel(q^2)&=\dfrac{a^{||}+b^{||} q^2(c^{||}-q^2)}{q^2(c^{||}-q^2)},
\end{align}
and similarly for the longitudinal amplitude, which contains no pole at $q^2=0$
\begin{equation}
Y_{LD}^0(q^2)=\dfrac{a^0+b^0 (q^2+s_0)(c^0 -q^2)}{(q^2+s_0)(c^0-q^2)},
\end{equation}
where $s_0=1$ GeV$^2$. The parameters $a^{\perp,\parallel,0}$, $b^{\perp,\parallel,0}$, and $c^{\perp,\parallel,0}$ are determined by ensuring that they reproduce the phenomenological model of Ref.~\cite{Khodjamirian:2010vf} in the $q^2$ range from $1\text{ GeV}^2$ to $9\text{ GeV}^2$. Specific values for these coefficients can be found in Ref.~\cite{Descotes-Genon:2015uva}. 
See also Ref.~\cite{Bordone:2024hui} for another treatment of long-distance amplitudes.

In order to be conservative, the $s_i$ parameters, which describe the interference between long and short distance in Eq.~\eqref{eq:nonlocalffnlo}, are varied independently for each amplitude within the range $[-1, 1]$. 

In the high-$q^2$ region several charmonium resonances appear. In order to avoid model-dependence biases we do not attempt to account for non-local contributions in this region. A common approach in the literature is to use quark-hadron duality \cite{Shifman:2000jv,Grinstein:2004vb,Beylich:2011aq,Lyon:2014hpa}, assuming that when integrated over a large bin the inclusion of these broad charmonium resonances averages to the short-distance contribution. 
However, this would not be the case in an unbinned analysis. Analyses, such as Ref.~\cite{LHCb:2024onj} use a model of Breit-Wigner amplitudes fitted to the data to try and explicitly include such resonances. Such an approach is inherently model dependent. In our proposed approach, model-independent observables are obtained, such that any estimation of the effect of the wide resonances can be evaluated a posteriori. In particular, should theoretical advances be made concerning the calculation of these resonances, the model-independent data may be reinterpreted.

\subsection{Definition of unbinned observables}

Using the previous framework we compute in this subsection the SM predictions for the unbinned optimised observables.

The definitions of the unbinned observables (counterparts of the binned ones defined in Ref.~\cite{Descotes-Genon:2013vna}) are given by
\begin{align}
P_1(q^2)&= \frac12 \frac{{} [J_3+\bar J_3]}{{}  [J_{2s}+\bar J_{2s}]}\ ,
& P_1^{\rm CP}(q^2)&= \frac12 \frac{{}  [J_3-\bar J_3]}{{}  [J_{2s}+\bar J_{2s}]}\ ,\nonumber
\\[1mm]
P_2(q^2) &= \frac18 \frac{[J_{6s}+\bar J_{6s}]}{  [J_{2s}+\bar J_{2s}]}\ ,
& P_2^{\rm CP}(q^2)&= \frac18 \frac{  [J_{6s}-\bar J_{6s}]}{ [J_{2s}+\bar J_{2s}]}\ , \nonumber \\[1mm]
{P_3}(q^2)&= -\frac14 \frac{ [J_9+\bar J_9]}{ [J_{2s}+\bar J_{2s}]}\ ,
& P_3^{\rm CP}(q^2) &= -\frac14 \frac{ [J_9-\bar J_9]}{ [J_{2s}+\bar J_{2s}]}\ ,\nonumber
\\[1mm]
P'_4(q^2) &= \frac{1}{{\cal N}^\prime}  [J_4+\bar J_4]\ ,
& {P'_4}^{\rm CP}(q^2) &= \frac{1}{{\cal N}^\prime}  [J_4-\bar J_4]\ ,\nonumber\\[1mm]
P'_5(q^2) &= \frac{1}{2{\cal N}^\prime}  [J_5+\bar J_5]\ ,
& {P'_5}^{\rm CP}(q^2) &= \frac{1}{2{\cal N}^\prime}  [J_5-\bar J_5]\ ,\nonumber\\[1mm]
{P'_6}(q^2) &= \frac{-1}{2{\cal N}^\prime}  [J_7+\bar J_7]\ ,
& {P'_6}^{\rm CP}(q^2) &= \frac{-1}{2{\cal N}^\prime}  [J_7-\bar J_7]\ ,\nonumber
\\[1mm]
{P'_8}(q^2) &= \frac{-1}{{\cal N}^\prime}  [J_8+\bar J_8]\ ,
& {P'_8}^{\rm CP}(q^2) &= \frac{-1}{{\cal N}^\prime}  [J_8-\bar J_8], \label{optimizedeqs}
\end{align}
together with \begin{equation}
F_L=-\frac{[J_{2c}+\bar{J}_{2c}]}{[d\Gamma/dq^2+d\bar{\Gamma}/dq^2]}
\end{equation}
and where the normalization ${\cal N}^\prime$ is defined as
\begin{equation}{\cal N}^\prime =\sqrt{- [J_{2s}+\bar J_{2s}]  [J_{2c}+\bar J_{2c}]}\ .\end{equation}

The uncertainties for the SM predictions are computed using a Monte Carlo sampling approach, which assumes a multivariate Gaussian distribution for the nuisance parameters described earlier, along with variations in the renormalisation scale, masses, and other relevant factors. The structure of the angular observables gives rise to non-linear behaviour, particularly stemming from the normalisation terms. A Gaussian scan of the parameters can therefore lead to non-Gaussian distributions of the observables. This is particularly the case when the normalisation is close to zero.

To extract meaningful uncertainties, we linearise the error propagation. This is effectively achieved by introducing a rescaling factor to the uncertainties associated with all input parameters. These rescaled uncertainties are then used as inputs for the corresponding probability distributions from which we generate the Monte Carlo replicas. Each replica is propagated through the mathematical structure of the observables to compute the full set of observables relevant to the analysis.

The resulting ensemble of observable samples is then used to produce predictions by computing the sample mean and covariance. When evaluating the covariance, the rescaling factor is reintroduced to avoid underestimating the overall uncertainty. We observe that a rescaling factor of $1/2$ is already sufficient to achieve this behaviour, and increasing it further does not lead to significant changes in the resulting distribution of observables. Nevertheless, our default choice for the rescaling factor is $1/3$.

For further details on this procedure, we refer the reader to Section~4.1 of Ref.~\cite{Descotes-Genon:2015uva}. The use of the full distribution of observables for the characterisation of relevant statistical metrics will be discussed in a forthcoming publication by some of the authors~\cite{Capdevila:2025aaa}.

\begin{figure}[!t]
\includegraphics[width=8cm,height=5.4cm]{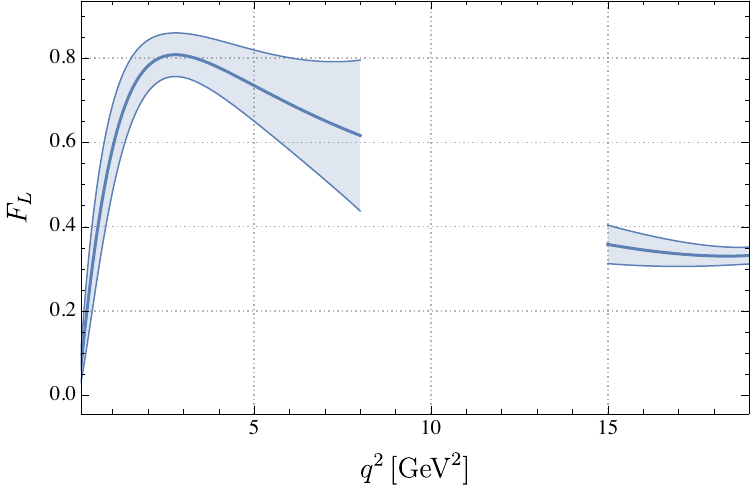}
\includegraphics[width=8cm,height=5.4cm]{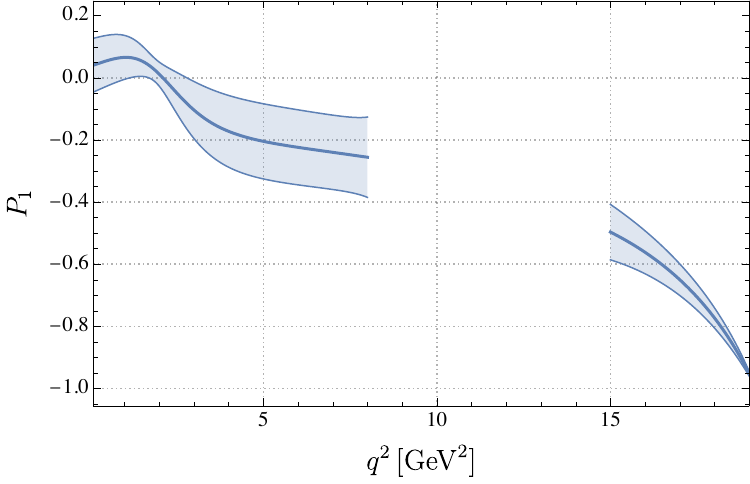}
\includegraphics[width=8cm,height=5.4cm]{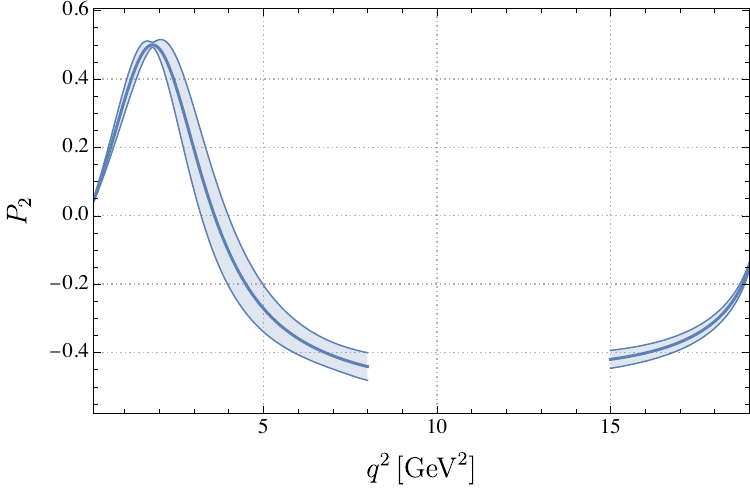}
\includegraphics[width=8cm,height=5.4cm]{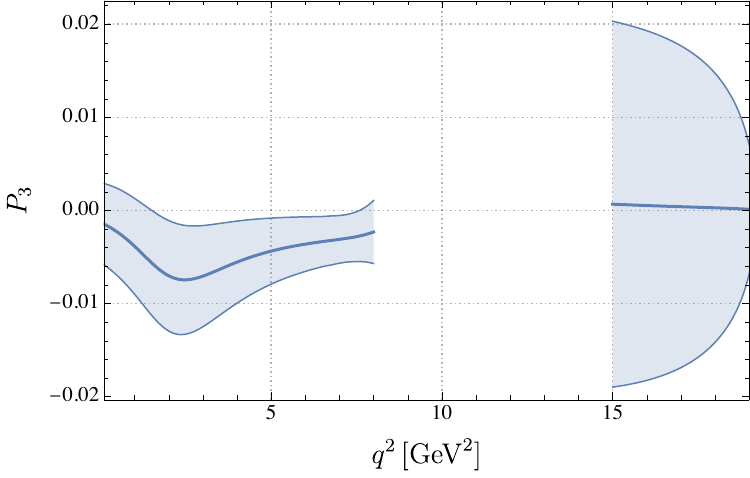}
\caption{SM prediction for the unbinned $F_L$ and $P_{1,2,3}$ optimised observables. }   \label{fig:opt123}    
\end{figure}
    \begin{figure} 
\includegraphics[width=8cm,height=5.4cm]{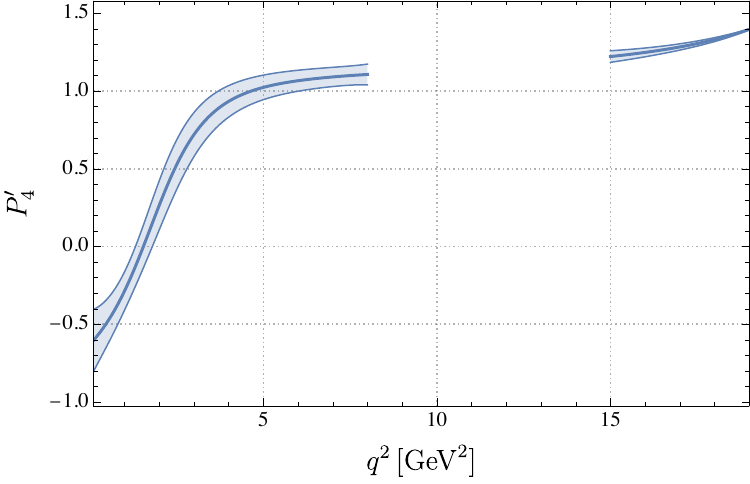}
\includegraphics[width=8cm,height=5.4cm]{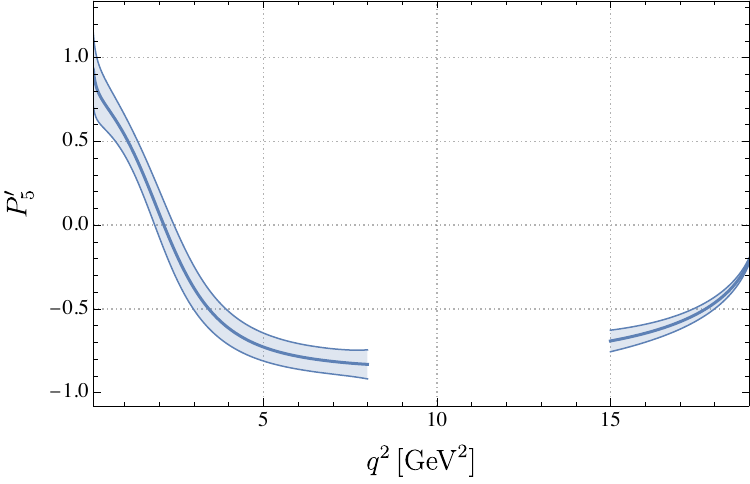}
\includegraphics[width=8cm,height=5.4cm]{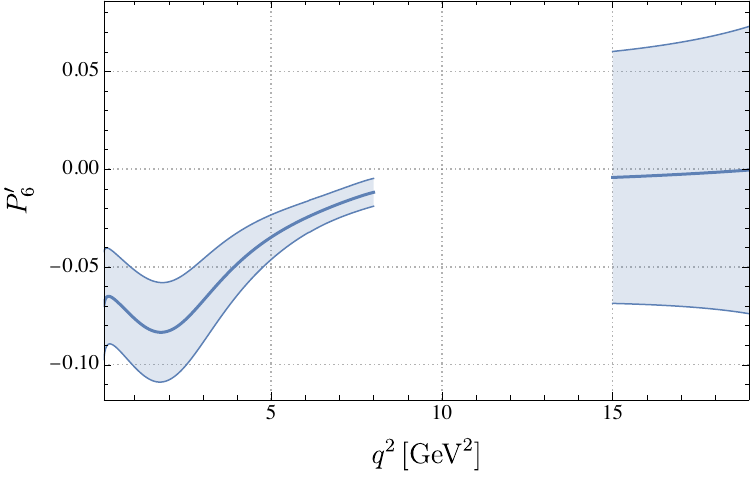}
\includegraphics[width=8cm,height=5.5cm]{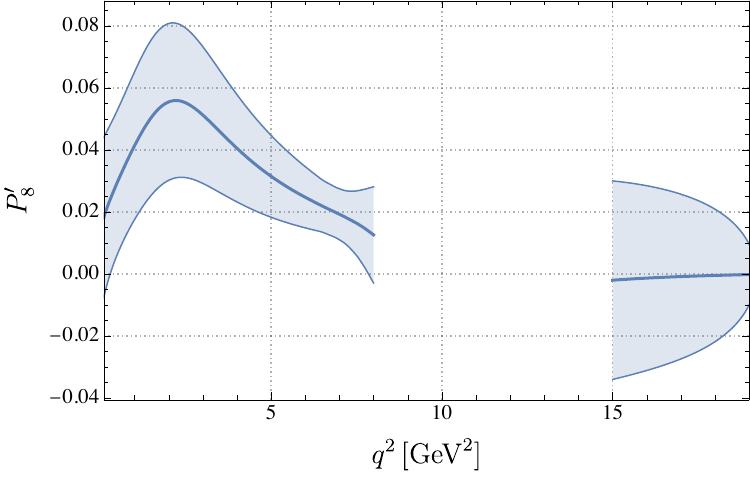}
\caption{SM prediction for the unbinned  $P_{4,5,6,8}^\prime$ optimised observables. } \label{fig:opt4568}    \end{figure}

The SM prediction for the optimised observables $P_i$ and $F_L$ are shown graphically in Figs.~\ref{fig:opt123} and \ref{fig:opt4568}. The $P_i^{\rm CP}$ are, to a high accuracy, very close to zero and we do not include them here.
For completeness we also provide the non-optimised observables $S_i$ and the branching fraction in Appendix~\ref{app:non-optimised}. The unbinned definitions of the $S_i$ observables can be easily derived following the same procedure as in Eq.~\eqref{optimizedeqs}.
Some remarks are in order here: 
\begin{enumerate}
 \item[i)] We have explicitly checked that binning the Monte Carlo sampling recovers the binned observable predictions presented in Ref.~\cite{Alguero:2023jeh}; 
 \item[ii)] Predictions are shown in the $[0.1, 8.0]\,\GeV^2$ (light quark resonances are neglected) and $[15.0,q^2_{\rm max}]\,\GeV^2$  regions in order to avoid the $c\bar{c}$ sharp resonance regions, which cannot be described by short distance physics. In the high-\qsq region, we assume quark hadron duality~\cite{Lyon:2014hpa}, due to the large number of broad resonances. However, particular caution needs to be taken in this region, as an unbinned analysis would capture the details of the resonances and thus quark hadron duality violations;  
 \item[iii)] The particular behaviour of the $P_2$ observable at low-$q^2$ is due to a convolution of the geometrical bound ($P_2\lesssim 1/2$) described in Eq.~\eqref{p2bound} below  and the asymmetric uncertainty; 
\item[iv)] The observables $P_3$ and $P_{6,8}^\prime$ are determined by the imaginary components of specific combinations of the helicity amplitudes, making them particularly sensitive to strong phases. The sudden increase in uncertainty seen in Figs.\,\ref{fig:opt123} and \ref{fig:opt4568} at high $q^2$ comes from the opening of the $c\bar{c}$ threshold, which introduces significant strong phases into $\mathcal{C}_9^{\text{eff}}$.
\end{enumerate}

The next issue to address is how to approach unbinned data.
Here there are two main options: i) either one parameterises the amplitudes with some well-motivated, but model-dependent, theoretical guidance and computes the observables from these amplitudes or ii) one determines the observables directly.
In the former approach, 
LHCb data are fitted using a model 
 for the amplitudes that incorporates both local and non-local contributions. From this fit, parameters associated with the non-local terms, as well as the Wilson coefficients that characterise the local contributions, are extracted. The extraction of these parameters depends on the underlying assumptions, such as the inclusion of potential contributions from NP in the form of scalars, tensors, or additional operators beyond the SM, the modelling of hadronic effects, and the treatment of the variables
 $m_{K\pi}$ and $q^2$. Once the amplitudes are determined, the observables can a posteriori be reconstructed.
There are different analyses in the literature parameterising the amplitudes~\cite{LHCb:2023gpo,LHCb:2024onj}.
In contrast, the latter approach, extracting observables directly from the distributions, avoids this model dependence.  This method allows us to obtain information without imposing a priori assumptions, enabling reinterpretation with new theoretical understanding and various tests of the internal consistency of the experimental analysis through the use of the observables themselves.
In the following subsections \ref{sec:conditions} and \ref{sec:interpretation}  we define the  tests that a consistent theoretical model of the amplitudes should fulfill. 
In section~\ref{section3}, we describe a model-independent method to determine unbinned observables using a moments approach.

\subsection{Testing  a theoretical model of the amplitudes: set of conditions} \label{sec:conditions}

A parameterisation of the amplitudes using an insufficient number of
degrees of freedom or with some of these degrees of freedom fixed to values inconsistent with the data will naturally lead to tension with the data.

One possible method to decide, in a systematic way, if a theoretical model for the amplitudes gives a good description of data is if it  fulfills a set of conditions.
To define these conditions, assuming a $K\pi$ system in a P-wave configuration, it is instructive to introduce a set of three complex 2-component vectors 
\begin{equation}
\label{eq:nvecs}
n_\|=\binom{A_\|^L }{A_\|^{R*} }\ ,\quad\!
n_\bot=\binom{A_\bot^L }{-A_\bot^{R*} }\ ,\quad\!
n_0=\binom{A_0^L }{A_0^{R*} }\ ,\quad\!
\end{equation}
that collect the L and R components of the amplitudes. If one includes the S-wave contribution, two more vectors are needed (see Ref.~\cite{Alguero:2021yus}) with corresponding additional conditions. 
 
 If a parameterisation of the amplitudes does not respect the conditions, for instance, by fixing some degrees of freedom with theory constraints in a way that is inconsistent with the data, 
this can manifest itself as the inability to reproduce one or more observables. 
In order to remedy this, more degrees of freedom must be included or some of the fixed degrees of freedom must be allowed to vary.

In the case where the mass of the leptons is neglected (referred to as the ``massless case''), with no scalar NP contribution, the number of degrees of freedom in a P-wave $B \to K^+\pi^-\mu^+\mu^-$ decay is eight (see Ref.~\cite{Egede:2010zc} and more detailed discussion in subsection~\ref{geonontrivial}). In the case where the lepton mass is not neglected (referred to as the ``massive case"), again with no scalar contribution,  there are ten degrees of freedom. 
Consequently, the number of conditions to test a theoretical model for the amplitudes in the massive case is ten. In the following, we provide the five conditions (three of them complex) that should be fulfilled  
\begin{eqnarray} \label{conditions}
{\mathrm{Cond.}}_1: \quad |n_\perp|^2&=& \frac{2}{\beta^2} J_{2s}+
\frac{1}{\beta^2} J_3,
\nonumber
\\[1mm]
{\mathrm{Cond.}}_2: \quad |n_\||^2&=& 
\frac{2}{\beta^2} J_{2s}-
\frac{1}{\beta^2} J_3,
\nonumber
 \\[1mm] 
{\mathrm{Cond.}}_{3} {\rm(Real\, and\, Imag)}: \quad n_\|^\dagger n_\perp&=&\frac{1}{2 \beta} J_{6s} {+} i \frac{1}{\beta^2} J_9,
\nonumber 
 \\[1mm]
{\mathrm{Cond.}}_{4} {\rm (Real\, and\, Imag)}: \quad n_\|^\dagger n_0&=&\frac{\sqrt{2}}{\beta^2} J_{4} { +} i \frac{1}{\beta \sqrt{2}} J_7,
\nonumber
 \\[1mm]
{\mathrm{Cond.}}_{5} {\rm (Real\, and\, Imag)}: \quad n_\perp^\dagger n_0&=&\frac{1}{\beta\sqrt{2}} J_{5} {+} i \frac{\sqrt{2}}{\beta^2} J_8, \nonumber\\[1mm]
\label{nJ}
\end{eqnarray}
where $\beta=\sqrt{1-4 m_\ell^2/q^2}$ and $m_\ell$ is the lepton mass. These conditions are common to the massless case, where $\beta=1$.

The two extra conditions in the massive case (conditions 6 and 7) that complete the system up to ten equations cannot be written in terms of $n_i$ and are given by
\begin{eqnarray} \label{eq3}
 4 \frac{m_\ell^2}{q^2} {\rm Re}(A_\perp^L A_\perp^{R*}+A_\|^L A_\|^{R*})=J_{1s}-\left(\frac{2+\beta_\ell^2}{\beta_\ell^2} \right) J_{2s},
 \cr
 4 \frac{m_\ell^2}{q^2} \left({|A_t|^2+2\rm Re}(A_0^L A_0^{R*})\right)=J_{1c}+\frac{1}{\beta_\ell^2}J_{2c}. \quad \quad \quad 
\end{eqnarray}
In the presence of scalar NP,  the last two conditions in Eq.~\eqref{nJ}  are affected  and also the last condition in Eq.~\eqref{eq3}. In both cases, the left hand side (LHS) of these conditions has to be modified. 
In a similar way, one can write a corresponding set of equations for the CP conjugate $J$ and $n_i$. 

Notice also that the condition 
\begin{equation}
|n_0|^2= -\frac{1}{\beta^2} J_{2c},
\end{equation}
is not included, because $n_0$ is not independent. Namely, the vector $n_0$ can be written in terms of $n_\|$ and $n_\perp$ \cite{Alguero:2021yus}
\begin{equation} \label{testc}
n_0=\alpha_0 n_\| + \beta_0 n_\perp,
\end{equation}
where 
\begin{eqnarray} \label{a0b0}
\alpha_0&=&\frac{|n_\perp|^2(n_\|^\dagger n_0)-(n_\|^\dagger n_\perp)(n_\perp^\dagger n_0)}{|n_\||^2|n_\perp|^2-|n_\perp^\dagger n_\||^2}, \nonumber \\[2mm]
\beta_0&=&\frac{|n_\||^2(n_\perp^\dagger n_0)-(n_\perp^\dagger n_\|)(n_\|^\dagger n_0)}{|n_\||^2|n_\perp|^2-|n_\perp^\dagger n_\||^2}. 
\end{eqnarray}

\subsection{Interpretation of conditions}
\label{sec:interpretation}

In order to test a given amplitude parameterisation using the conditions in 
Eq.~\eqref{conditions} and Eq.~\eqref{eq3}, 
on the LHS of these expressions one should insert 
the parameterisation of the amplitudes as a function of the relevant free parameters (phases and moduli); and on the RHS,  experimental values for the coefficients $J$. 
A fit for the free parameters can then be used to check if it is possible to  reproduce the RHS of each equation.

For illustrative purposes, an approximation to this test~\footnote{This is an approximation as lattice information is fitted together with the experimental data.} can be implemented by combining the two fits presented in Ref.~\cite{LHCb:2023gpo}:

\begin{itemize}

\item[Fit~I:] A fit to data using only $q^2>0$ information but adding lattice calculations. 
The fit for the observables in Ref.~\cite{LHCb:2023gpo} reproduces the binned data very well. We therefore take this fit for the observables as 
the ``data'', i.e.,  the RHS of the conditions in Eq.~\eqref{conditions} and Eq.~\eqref{eq3}.
 The results of this  fit are depicted in blue in Fig.~\ref{condition4} and in Fig.~\ref{condition5} in Appendix.~\ref{app:conditions}. 
\item[Fit~II:] A fit to data using, in addition, the $q^2<0$ information. 
This fit is highly impacted by theory points at $q^2 < 0$ and we reinterpret it as the LHS of the conditions.  The results of this fit are depicted in red in Fig.~\ref{condition4} and in Fig.~\ref{condition5}.

\end{itemize}

We observe in Fig.~\ref{condition5}  that conditions 1, 2, 3, 5, 6 and 7 exhibit a very good agreement between the LHS (red) and RHS (blue) of the conditions. This means that the experimental values of the $J_i$ coefficients can be reproduced within $\pm$1$\sigma$ for these conditions. However for condition 4 (see Fig.~\ref{condition4}), we observe a discrepancy in the imaginary part between the LHS and RHS of the condition, such that they are incompatible in the region from 0 to 4 GeV$^2$ at 95\% confidence level. 

This discrepancy indicates a problem in the parameterisation of ${\rm Im}(n_\|^\dagger n_0)$. Several reasons can be behind this problem: a) the strong phases fixed using theory at $q^2<0$ are in tension with the $B \to K^*(\to K^+\pi^-)\mu^+\mu^-$ data, or higher-order corrections to them are needed; there is a problem in b) the extrapolation to positive $q^2$ or c) the lattice calculation. However b) and c) are less plausible, given that the extrapolation and the lattice work well in the case of all the other conditions.

The tension observed in condition 4 is linked to the tension observed in the comparison between ${\rm Im}A_\|$ for $q^2>0$ and $q^2<0$ in Ref.~\cite{LHCb:2023gpo}.
To a much lesser extent there is also a tension in ${\rm Im}A_\perp$, as discussed below.

It is illustrative to observe that if one uses the explicit form of the amplitude, as given in Ref.~\cite{LHCb:2023gpo}:
\begin{eqnarray} \label{amplitude}
{ A}_\lambda^{L,R}&=&{\cal N}_\lambda \left( \left[C_9\mp C_{10} \right] {\cal F}_\lambda (q^2)+ \frac{2 m_b M_B}{q^2} [ C_7 {\cal F}_\lambda^T (q^2)\right.\nonumber \\[1mm] &&\left.-16 \pi^2 \frac{M_B}{m_b} {\cal H}_\lambda] \right)
\end{eqnarray}
where ${\cal N}_\lambda$ stands for the helicity dependent normalisation factor, ${\cal F}_\lambda$ and ${\cal F}^T_\lambda$ are the local form factors and ${\cal H}_\lambda$ the non-local correlator.
The expression above implies that ${\rm Im}A_i^L={\rm Im} A_i^R$ (with $i=\perp,\|,0$). 
  If there is a problem with ${\rm Im}A_\|$ it should be more clearly manifest in $P_6^\prime$. The reason is that
  this observable is proportional to $J_7$, which has the following structure
\begin{equation}
J_7=\sqrt{2}\beta ({\rm Im} A_\|^L ({\rm Re}A_0^R-{\rm Re}A_0^L)+{\rm Im}A_0^L ({\rm Re} A_\|^L-{\rm Re}A_\|^R)).
\end{equation}
Using Eq.\ref{amplitude} this implies
\begin{equation}
J_7=2 \sqrt{2}\beta C_{10} \left[
({\cal F}_0 {\cal N}_0) {\rm Im} A_\|^L -
({\cal F}_\| {\cal N}_\|){\rm Im}A_0^L \right].
\end{equation}
As this is a difference of two small quantities, it is particularly sensitive to their precision.
Another coefficient  of the angular distribution that is sensitive to ${\rm Im} A_\|^L$ is $J_9$. However in this case, 
\begin{equation}
J_9=\beta^2 \left[{\rm Im}A_\perp^L ({\rm Re}A_\|^L+{\rm Re} A_\|^R)- {\rm Im}A_\|^L ({\rm Re} A_\perp^L+{\rm Re} A_\perp^R)\right].
\end{equation}
Given that ${\rm Re} A_\perp$ and ${\rm Re}A_\|$ have  opposite signs, the $J_9$ coefficient is then not a difference between two small quantities as before, and is consequently less sensitive to the precision in the imaginary parts than $J_7$. A marginal tension is still observed in the imaginary component of condition 3 at  very low $q^2$ and in the imaginary component of condition 5 above 3 GeV$^2$, due to ${\rm Im}A_\perp$.   
  
On the contrary we should not expect any sensitivity to ${{\rm Im}A_\|}^L$ and ${{\rm Im}A_\|}^L$ in the $P_2$ and $P_5^\prime$ observables, as the dependence on the imaginary parts cancels exactly  for the two $J$'s involved in the numerator of these observables
\begin{eqnarray}
J_5&=&\sqrt{2}\beta ({\rm Re} A_0^L {\rm Re} A_\perp^L- {\rm Re} A_0^R {\rm Re} A_\perp^R), \nonumber \\[1mm]
J_{6s}&=&2\beta ({\rm Re} A_\|^L {\rm Re} A_\perp^L- {\rm Re} A_\|^R {\rm Re} A_\perp^R). 
\end{eqnarray}
 In the case of  $P_4^\prime$, the dependence  on
${\rm Im} A_\|^L$ via $J_4$
is hidden by the dominant real parts,
\begin{equation}
\!\!J_4=\frac{\beta^2}{\sqrt{2}} (2 {\rm Im} A_0^L {\rm Im} A_\|^L + {\rm Re} A_0^L {\rm Re} A_\|^L+{\rm Re} A_0^R {\rm Re} A_\|^R).
\end{equation}

\noindent In all other $J_i$, the real parts dominate over the imaginary ones. 

A possible way to fix the problem in condition 4 could be to leave one strong phase free in the amplitude $A_\|$ in Fit~II and let the observables (in particular $P_6^\prime$) fix the value of this phase. One should then understand how to obtain this value for the phase from a direct theoretical computation. The same thing could be done for $A_\perp$. 

\begin{figure} 
\includegraphics[width=0.95\linewidth]{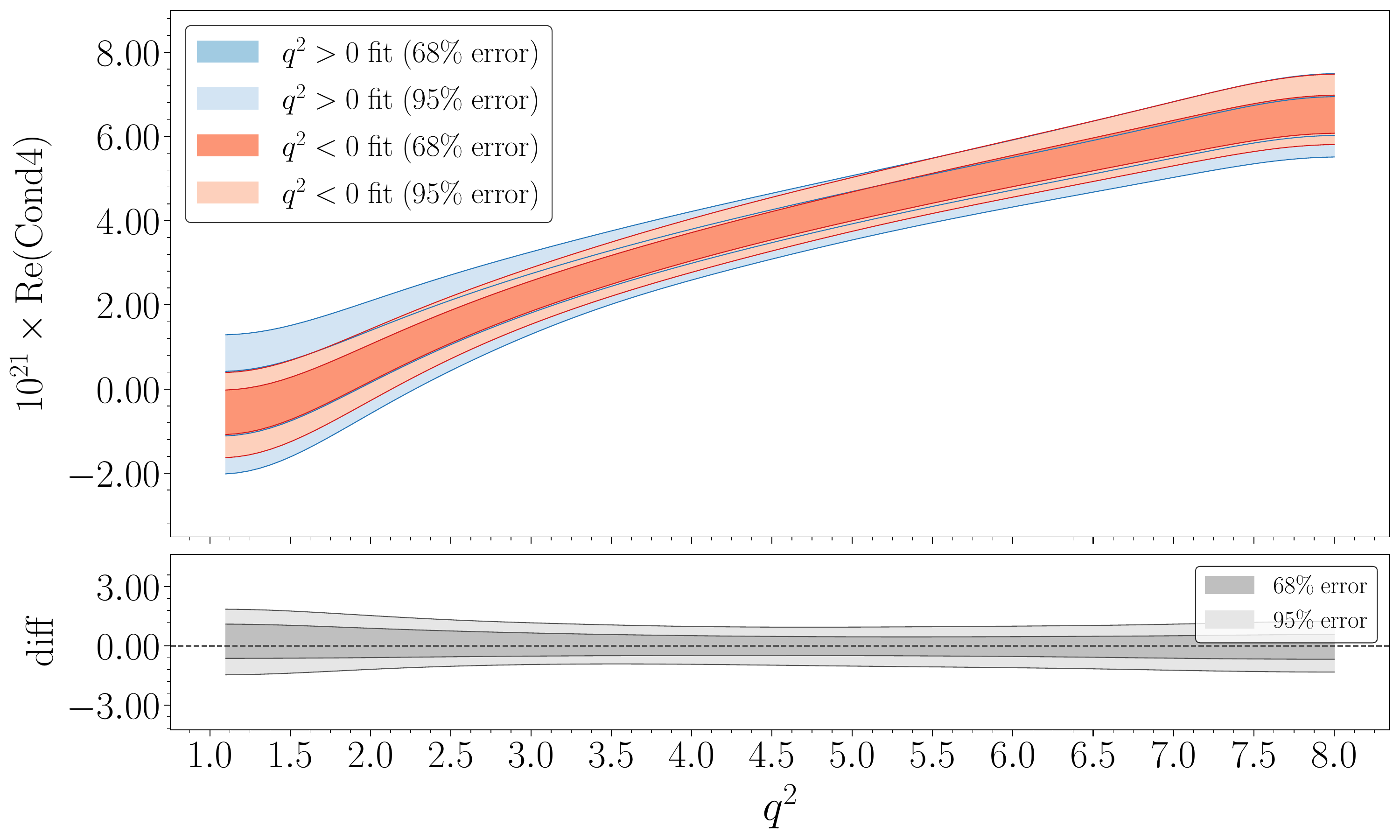}  
\includegraphics[width=0.95\linewidth]{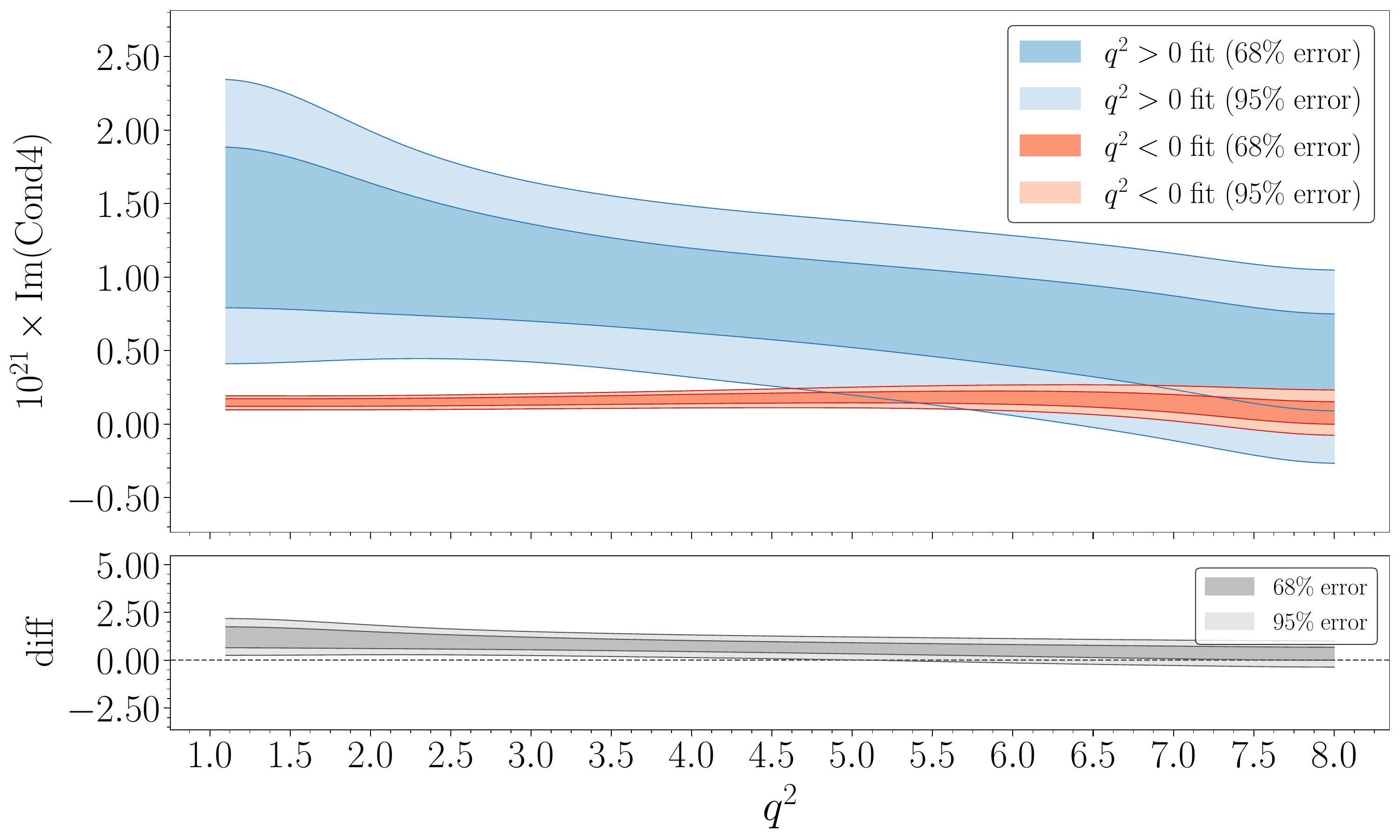}\caption{The upper panels display the real and imaginary parts of the ``condition'' $\mathrm{Cond.}_4$ in Eq.~\eqref{conditions}, computed from generated samples of the amplitudes following the distribution of fit results from the LHCb analysis of Ref.~\cite{LHCb:2023gpo}. We present results for the two variants of the analysis described in the text. The lower panel  shows the difference between the two curves in the upper panel.}
 \label{condition4}\end{figure}

\section{A new method to determine unbinned observables: A ``continuous'' moments method}\label{section3}

Previous measurements of the angular observables of the \BKs decay have binned the experimental data in \qsq and integrated the observables over the analysed window of $\Kp\pim$ invariant mass, \mkpi. These measurements include performing a fit of the three-dimensional angular PDF, with the observables as fit parameters~\cite{LHCb:2020lmf,CMS:2024atz}, or calculations of the corresponding angular moments in each bin~\cite{PhysRevD.91.114012,Aaij:2115087,Aaij:2216088}. By obtaining the complete basis of angular observables, the results are readily interpretable by third parties and may be averaged and re-interpreted in perpetuity. However, by binning the data in \qsq, some information about the underlying physics is invariably lost. Thus far, attempts to exploit the full \qsq dependence of the decay have involved trying to extract the theory parameters (Wilson coefficients, local and non-local form factor parameters etc.) directly~\cite{PhysRevD.109.052009,LHCb:2023gpo,LHCb:2024onj}. These measurements are model-dependent, which precludes easy re-interpretation with new theory insights or combination with other experimental data. Furthermore, the \mkpi dependence of the angular observables has not been studied and so experimental results have been interpreted with the approximation that the \Kstarz is a narrow resonance.

Here we introduce a novel method to extract the full set of model-independent angular observables continuously in \qsq and \mkpi. We do this by extending the previous binned method-of-moments technique~\cite{PhysRevD.91.114012} into an unbinned one. This is achieved with the use of a $k$-nearest neighbours (\knn) algorithm~\cite{knn}. The \knn with uniform weighting is exactly the definition of the normalised moment (the moment divided by the total weight), $M_{i}$, but instead of being binned, it is really a reflection of the density of the data in the locale of a given phase-space point.
The \textit{knn} takes the \textit{k} nearest events in a region of \qsq, \mkpi or both for the given point at which the data are being evaluated. Each of the $k$ events has a value of $\cos\theta_{K}$, $\cos\theta_{\ell}$  and $\phi$, and so for each the moment is calculated, i.e. each event has a value of $m_{5}=\sin2\theta_{K}\sin\theta_{\ell}\cos\phi$. The \knn algorithm sums the relevant moment ($m_{i}$) and divides by $k$ to calculate the average moment in that region.
Thus one can pick an arbitrary point in \qsq, \mkpi, or $(\qsq,\mkpi)$, and estimate the value of the angular moment there.

The result of the $k$-nearest neighbours algorithm may be augmented with further processing, for example, Gaussian process regression~\cite{scikit-learn} or a Savitzky-Golay filter~\cite{doi:10.1021/ac60214a047} (other filtering algorithms are available), to produce smoothly varying, continuous, numerical functions. For the studies presented here, a Savitzky-Golay filter is used with polynomial order four.
The variances of the observables are calculated with the second-order moments, for which the \knn algorithm is again used with subsequent filtering.
Covariances (and by implication correlations) may similarly be obtained with the second-order moments. The validity of the second-order moments as an estimate of the statistical uncertainty of the observables is checked with pseudo-experiments by comparing the range of values obtained when sampling with replacement, with the expected range from the calculated variance. The agreement between the two is good for all observables. An illustration of the determination of the observable $S_{5}$ using this method is shown in Fig.~4, resulting from a pseudo-experiment described below.

With this unbinned method, different points in the phase space are statistically correlated with each other. The degree of correlation depends on the number of neighbours in the \knn, the smoothing and how far apart the points are. When using the results of this method in further analysis for the underlying physics parameters, knowing these correlations is vital. Consequently, if the correlations are well understood, the precise choice of the number of neighbours or smoothing strength is to some extent a matter of aesthetics. It is recommended that experimental analyses publish unsmoothed distributions, such that theorists are free to make their own choices in their subsequent analyses. Few neighbours leads to a large apparent uncertainty at a single point but nearby points are relatively independent and so give equal statistical power. Conversely, a large number of neighbours reduces the apparent uncertainty but leads to large statistical correlations with neighbouring regions. What really matters is choosing the analysis parameters that maximise the sensitivity to what is being measured, and in particular allows for the resolution of the phase-space dependence of the observables. Care must be taken, as using too many neighbours can lead to a poor description of the true variables in regions where they vary quickly. However, for some analyses, such as finding the zero-crossing points in \qsq (discussed in the following sections), one may choose a larger number of neighbours, as the zeros lie in a \qsq region where the observables of interest are expected to vary smoothly.

\begin{figure}[!t]
    \centering
    \includegraphics[width=0.95\linewidth]{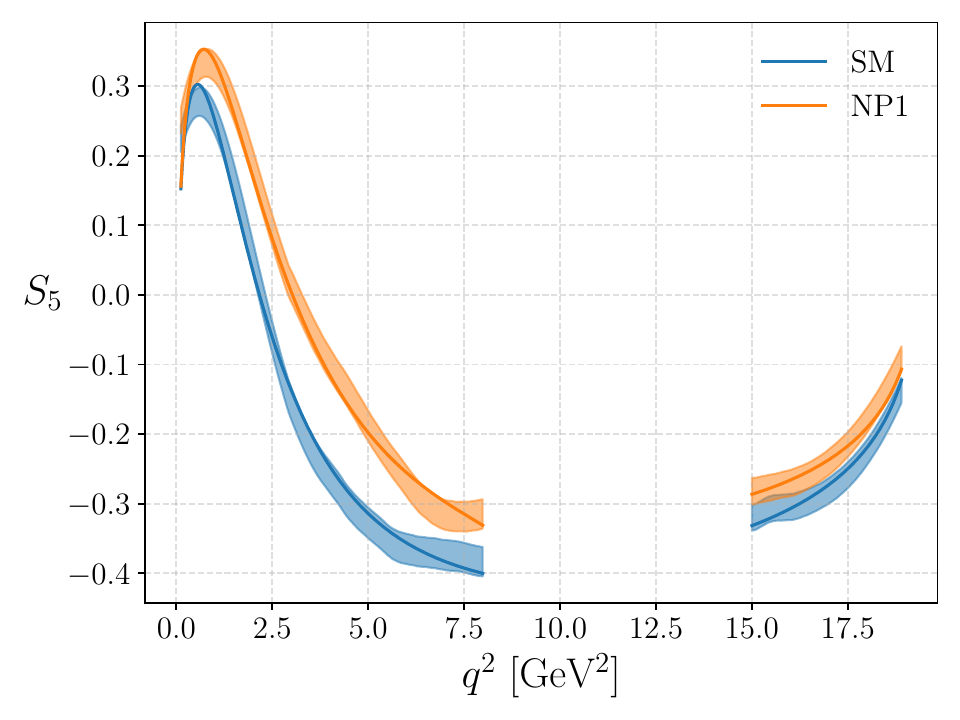}
    \includegraphics[width=0.95\linewidth]{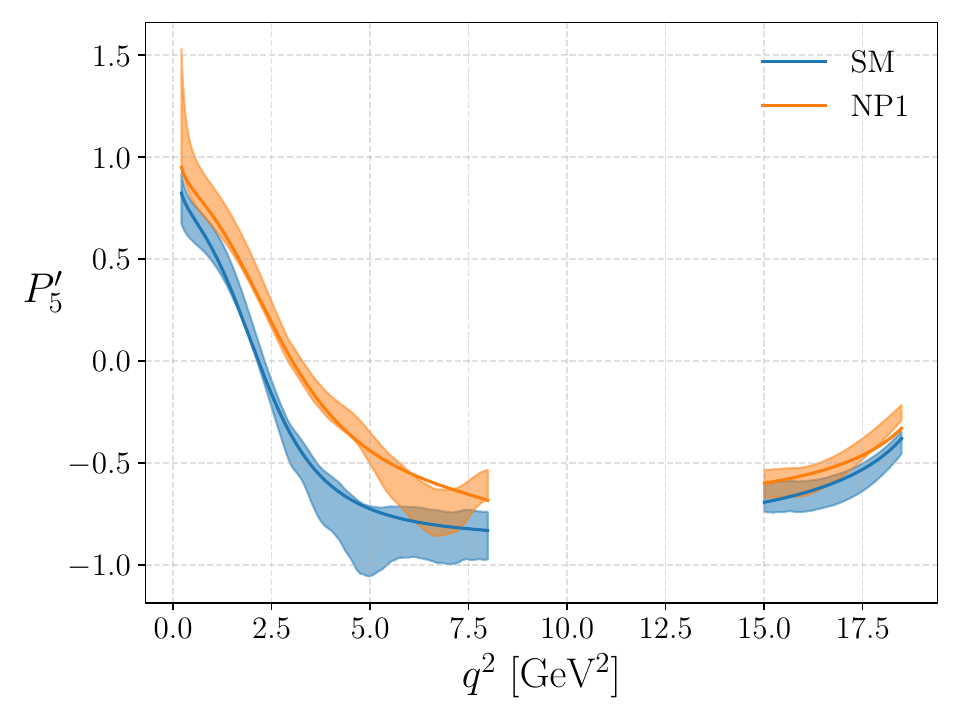}
    \caption{Extraction of (top) $S_{5}$ and (bottom) $P_{5}^{\prime}$ as a function of \qsq from two pseudo-experiments. The bands show the obtained $\pm 1\,\sigma$ uncertainty bounds for each pseudo-experiment. The solid lines are the predictions from which the respective data sets were generated: (blue) the SM and (orange) NP1, both from Table~\ref{table:WilsonCoefficientsSMandNP}.
    }
    \label{sec:experiment:fig:p5pexample}
\end{figure}

The output of the method is a large grid of the values of the angular observables at each of the sampled points in the variables of interest (\qsq, \mkpi or both), with the size of the grid depending on the number of sampled points, $N$. The choice of $N$ should be large enough such that all of the experimental data are exploited (i.e. there are no regions in the variable of interest that are not covered by the $knn$ output) but not so large that the output becomes unwieldy. By evaluating the correlations between observables and points in e.g. $\qsq$, one has a $MN\times MN$ covariance matrix, where $M$ is the number of angular observables (11 in the P-wave with massive leptons). This matrix is positive definite and sparse as regions of the variables that are separated by any distance will have negligible correlations. Inverting this matrix for subsequent fits of the physics parameters is necessarily computationally expensive but need only be done once. The sparse nature of the matrix implies that numerical methods may be used to approximate the inverse, such as conjugate gradient descent~\cite{Hestenes:1952mok}.

A key benefit of an approach using moments is that the observables are obtained with \textit{calculations}, rather than via a statistical likelihood fit. With a moments calculation one need not be concerned about fits that do not converge, poor statistical behaviour and biased likelihood estimators. Dealing with these has been a major factor in the previous generations of experimental analyses. With this unbinned method one can improve the \qsq resolution of the angular observables, or investigate the \mkpi dependence of the observables, without having to resort to fits in narrow bins with limited sample sizes, which would involve significant statistical complexity.


Previous experimental analyses have assumed that the leptons are massless, reducing the number of angular terms to be measured and so simplifying the task. However, it is found that the effects of lepton mass in the muon mode are non-negligible even at relatively large values of \qsq ($\sim 4\,\GeV^{2}$), as shown in Fig.~\ref{sec:experiment:fig:masses} (see also Appendix~\ref{app:lepmass} for a definition of a simpler lepton mass modified distribution that avoids the conflicting terms $J_{1}^{s}$ and $J_{1}^{c}$). In the absence of S-wave contributions, the full set of angular observables may be extracted with the unbinned moments, fully accounting for the effects of the lepton masses. The observables 
$S_{1}^{s}$, $S_{1}^{c}$, $S_{2}^{s}$, $S_{2}^{c}$ do not have orthogonal angular functions and must be extracted from a linear combination of the corresponding moments
\begin{align}
\label{eq:experiment:masslessmoments}
    S_{1}^{s} &= \frac{1}{16}\left( 21 M_{1}^{s} - 14 M_{1}^{c} + 15 M_{2}^{s} - 10 M_{2}^{c}\right), \nonumber \\
    S_{1}^{c} &= \frac{1}{8}\left( -7 M_{1}^{s} + 28 M_{1}^{c} - 5 M_{2}^{s} + 20 M_{2}^{c} \right) ,\nonumber\\
    S_{2}^{s} &= \frac{1}{16} \left( 15 M_{1}^{s} - 10 M_{1}^{c} + 45 M_{2}^{s} - 30 M_{2}^{c} \right), \nonumber \\
    S_{2}^{c} &= \frac{1}{8}\left( -5 M_{1}^{s} + 20 M_{1}^{c} - 15 M_{2}^{s} + 60 M_{2}^{c}\right) .
\end{align}

\begin{figure}
    \centering
    \includegraphics[width=0.98\linewidth]{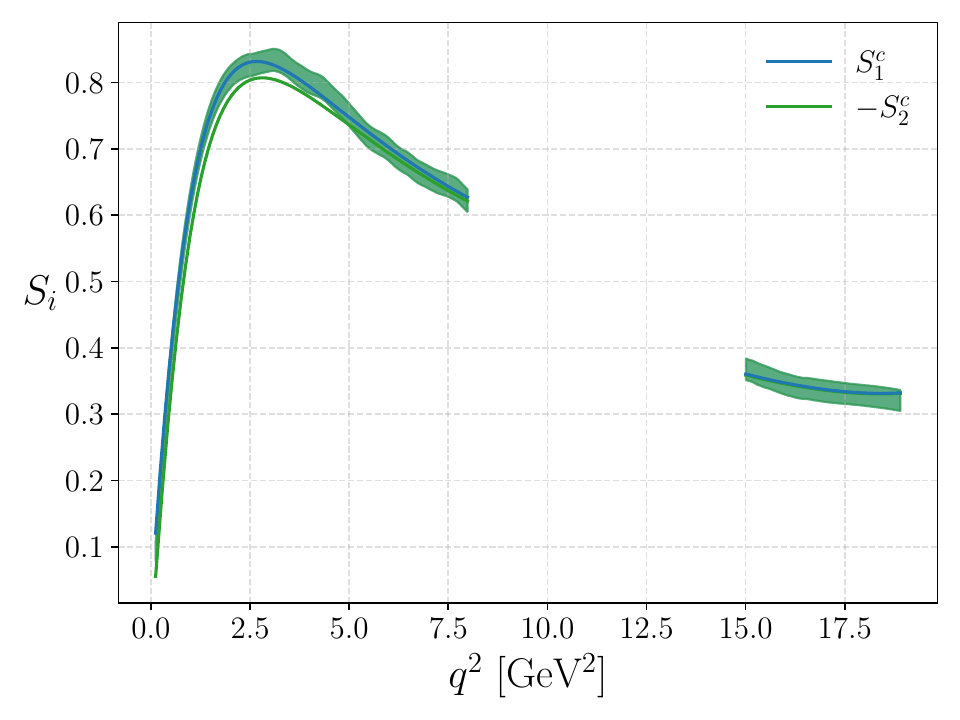}
    \caption{Comparison of $S_{1}^{c}$ and $-S_{2}^{c}$ for the SM prediction from Table~\ref{table:WilsonCoefficientsSMandNP}. The solid lines show the predictions and should align if the effects of the lepton mass are negligible. The band shows the extraction of these observables with the unbinned moments when assuming massless leptons.}
    \label{sec:experiment:fig:masses}
\end{figure}

As always, contamination from S-wave decays to $\Kp\pim\mu^{+}\mu^{-}$ introduces further complications. The first is that the normalised moments provided by the $knn$ algorithm have a factor of $\Gamma_{P} + \Gamma_{S}$ in the denominator, implying that the resulting observables are normalised by $\Gamma_{P}+\Gamma_{S}$, denoted by $\tilde{S}_{i}$ for the P-wave, rather than the usual normalisation by the P-wave only rate, $\Gamma_{P}$, that defines the $S_{i}$ observables. This may be readily rectified by multiplying the P-wave $\tilde{S_{i}}$ observables by a factor of $1/(1 - 2\tilde{S}_{1a}^{c} + \frac{2}{3}\tilde{S}_{2a}^{c})$ to recover the appropriate normalisation.

The second complication is that there is no single unique linear combination of moments that can be used to extract the complete set of angular observables if lepton masses are considered. The problem arises from terms such as $S_{1}^{s}\sin^{2}\theta_{K} + S_{1}^{c}\cos^{2}\theta_{K} + \tilde{S}_{1a}^{c}$; here one can scale $S_{1}^{s}$ and $S_{1}^{c}$ to accommodate changes in $\tilde{S}_{1a}^{c}$, giving rise to a degeneracy in the determination of these observables. The same problem applies for $S_{2}^{s}$, $S_{2}^{c}$ and $\tilde{S}_{2a}^{c}$.

There are two possible ways to handle this degeneracy. The first is to make approximations and introduce relations between the angular observables, as is done for the current generation of binned analyses. For $\qsq>4\gev^{2}$ such an approach introduces negligible bias. For lower \qsq values, one need not utilise all the relations pertaining to negligible lepton mass, instead choosing only those that result in negligible bias in the extracted observables. The second option is to use the information from the variable \mkpi to estimate the relative S- and P-wave contributions. This can be done as part of the moments calculation, assuming some analytic forms for the lineshapes of the P- and S-wave contributions. A full discussion of dealing with S-wave contamination is contained in Appendix~\ref{app:swaveandmasses} along with a simple study. Unless explicitly stated, the studies presented here only involve the P-wave component of the decay. 

Pseudo-data are generated and analysed to prove the principle of the new method. The signal yield from the LHCb experiment with $50\,{\rm fb^{-1}}$ of $pp$ collision data (expected by the end of LHCb upgrade I~\cite{LHCb:2023hlw}), across the full \qsq range, is estimated to be 130,000 candidates and so each pseudo-data set is generated with this quantity of signal. The angular distributions generated follow the SM and NP predictions in Table~\ref{table:WilsonCoefficientsSMandNP}. Here, backgrounds and the warping of the distributions of the kinematic variables due to detector acceptance and selection effects are neglected. In an analysis of collision data the treatment of such effects could follow that of previous moments analyses, such as in Ref.~\cite{Aaij:2115087}. Although, the yields mentioned here pertain to LHCb, the method could readily be used to analyse data from Belle II or the general purpose LHC detectors.

The optimised angular observables, $P_{i}$ and $P_{i}^{\prime}$, are not directly related to the moments but may be calculated from the extracted continuous functions of the non-optimised observables. As the optimised observables are ratios of the $S_{i}$, the normalising rate terms cancel and the $\tilde{S}_{i}$ observables may be used for such a calculation. Propagating the covariance of the moments to the optimised basis of observables is also straightforward. 
An example of the result of this method is shown in Fig.~\ref{sec:experiment:fig:p5pexample} for $S_{5}$ and the corresponding optimised observable $P_{5}^{\prime}$.
The method may be readily extended to assess the angular \CP-asymmetries by flipping the sign of the moment that contributes to the sum in the $knn$ for $B^{0}$ flavour decays, relative to $\bar{B}^{0}$.

The method has the ability to extract the dependence of the observables on \mkpi, or to consider \qsq and \mkpi together. Such a result is demonstrated in  Fig.~\ref{sec:experiment:fig:twodexample}.
By considering \mkpi, one may readily discern if the finite width of the \Kstarz leads to a variation of the angular distribution across the width of the \Kstarz. Additionally, by extracting the S-wave moments one is able for the first time to model-independently ascertain the lineshape of the S-wave component of the data. This is particularly important as it is not obvious that the various S-wave resonances (and the non-resonant S-wave contribution) should all have the same angular distributions.

\begin{figure}
    \centering
    \includegraphics[width=0.95\linewidth]{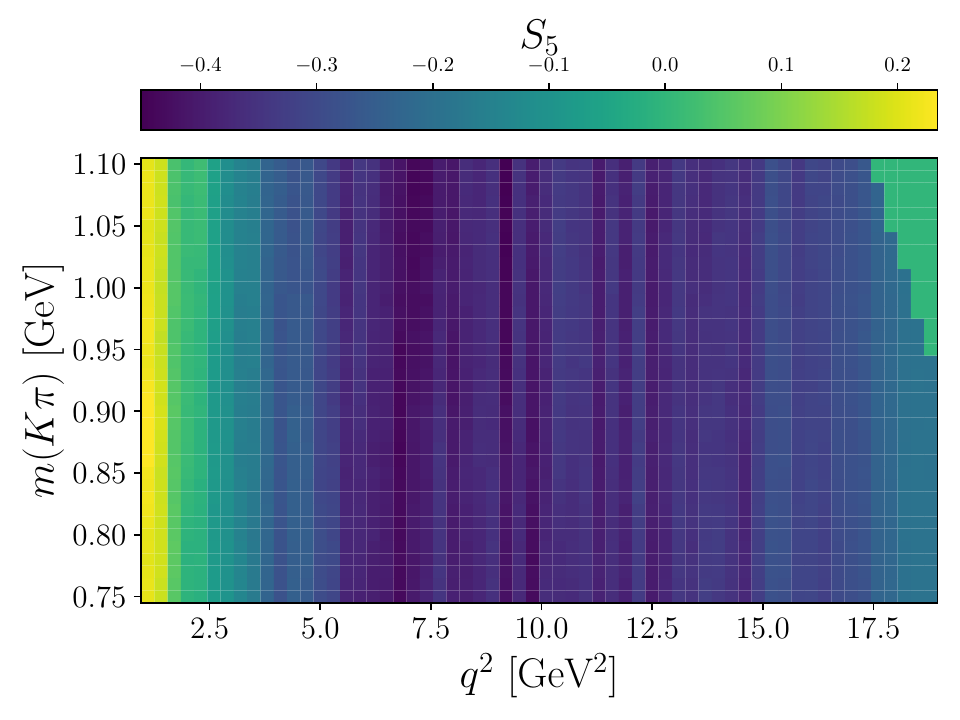}
    \caption{An example of the two-dimensional unbinned extraction of $S_{5}$ in the kinematic variables $q^{2}$ and \mkpi. The pseudo-experiment data set has been generated with a relativistic Breit-Wigner in \mkpi, which for this demonstration factorises with the angular distribution and \qsq (except for phase-space effects).
    }
    \label{sec:experiment:fig:twodexample}
\end{figure}

The zero-crossing points of the optimised observables that will be discussed from a theoretical point of view in the next section, may be readily ascertained with this method. An example of the extraction of the crossing points for $P_{2}$ and $P_{5}^{\prime}$ is shown in Fig.~\ref{sec:experiment:fig:zero}. As only the value of \qsq where these observables are zero is of interest, the crossing points are most easily discerned from the $\tilde{S}_{i}$ observables that come straight out of the $knn$ algorithm. These are shown for a single example pseudo-experiment; clearly there is discrimination between the SM and NP1 scenarios. Nevertheless, for a given single measurement the estimated uncertainty of the observable at a particular point in \qsq depends on the number neighbours and smoothing. Therefore, the uncertainty on the value of the crossing point must be be assessed with an ensemble of pseudo-experiments. The distribution of measured crossing points in $P_{2}$ and $P_{5}^{\prime}$, resulting from $\sim 10^{3}$ such experiments, is shown in Fig.~\ref{sec:experiment:fig:zerotoys}. The expected uncertainty on these values is the width of these distributions. For $P_{2}$ and $P_{5}^{\prime}$, separation between SM and NP is clear; although with these sample sizes it is not possible to distinguish between the two NP scenarios. Importantly, the method is unbiased i.e. the centres of the distributions align with the true values used to generate the pseudo-experiments.

\begin{figure}
    \centering
    \includegraphics[width=0.95\linewidth, page = 1]{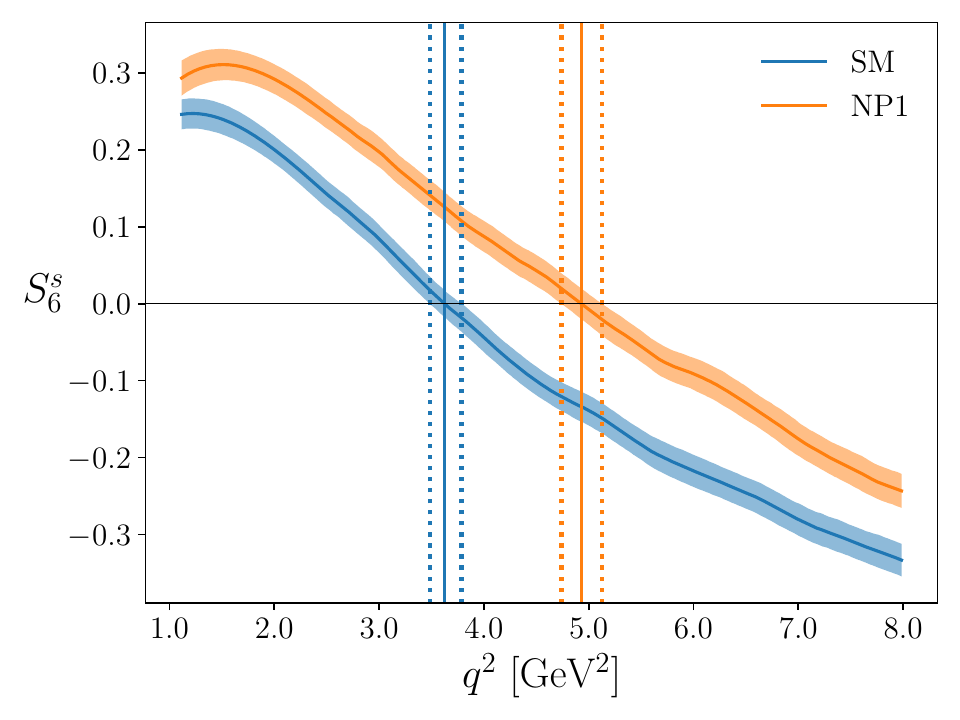}
    \includegraphics[width=0.95\linewidth, page = 2]{Plots/experiment/zero_crossing_trim.pdf}    
    \caption{Extraction of the zero-crossing points of the observables (top) $S_{6}^{s}$ and (bottom) $S_{5}$, corresponding to the zeros of $P_{2}$ and $P_{5}^{\prime}$ respectively, in a single pseudo-experiment. For each, an example of the (blue) SM and (orange) a NP model is shown. The vertical lines denote the obtained central values and intervals for where the observables cross zero.}
    \label{sec:experiment:fig:zero}
\end{figure}

\begin{figure}
    \centering
    \includegraphics[width=0.95\linewidth, page = 1]{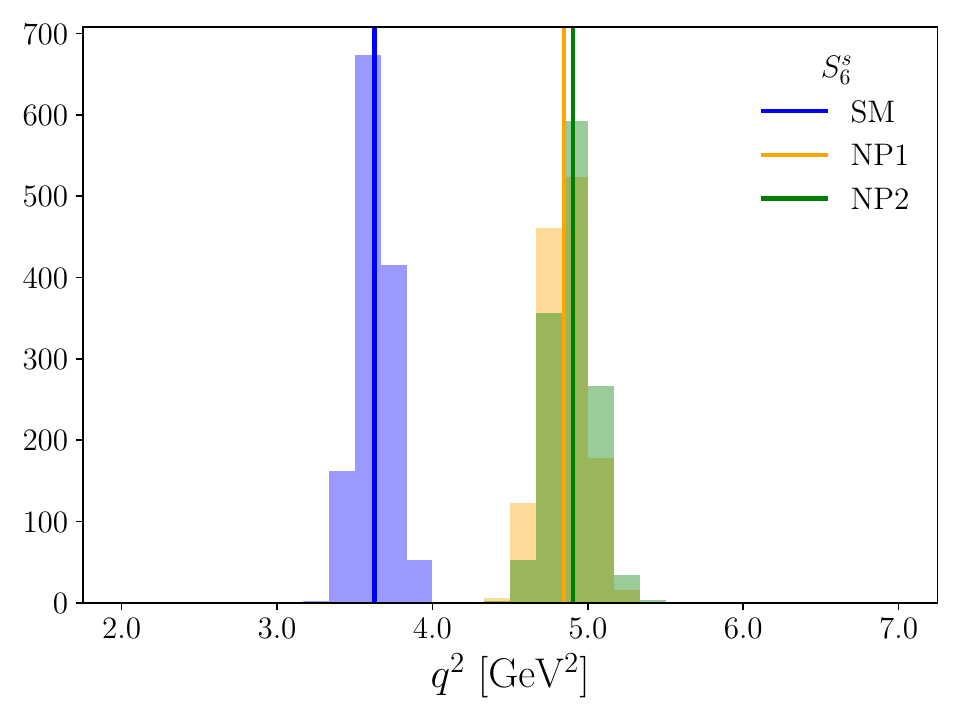}
    \includegraphics[width=0.95\linewidth, page = 2]{Plots/experiment/crossings_trim.pdf}    
    \caption{Expected precision for the extraction of the zero-crossing points. Shown are the variables (top) $S_{6}^{s}$  and (bottom) $S_{5}$, corresponding to $P_{2}$ and $P_{5}^{\prime}$, respectively.  Comparisons are made with (blue) the SM predictions, (orange) NP scenario 1 and (green) NP scenario 2. The distributions of the pseudo-experiment results are the translucent histograms, with the predicted values given by the solid lines.}
    \label{sec:experiment:fig:zerotoys}
\end{figure}

The geometrical bounds discussed in the next section may also be extracted in a slightly altered form. Taking the square of the bounds, to avoid potentially troublesome square roots, leads to the distributions in Fig.~\ref{sec:experiment:fig:bounds}. It is seen that, in these single pseudo-experiments, the uncertainty bands of the measurements may violate the geometric bounds due to the limited statistics of the data. This is not unexpected. The true values of the relations get very close to the bounds and the experimental measurements treat each angular observable as a statistically independent quantity, that has no prior for the physical limits.

\begin{figure}
    \centering
    \includegraphics[width=0.98\linewidth, page = 1]{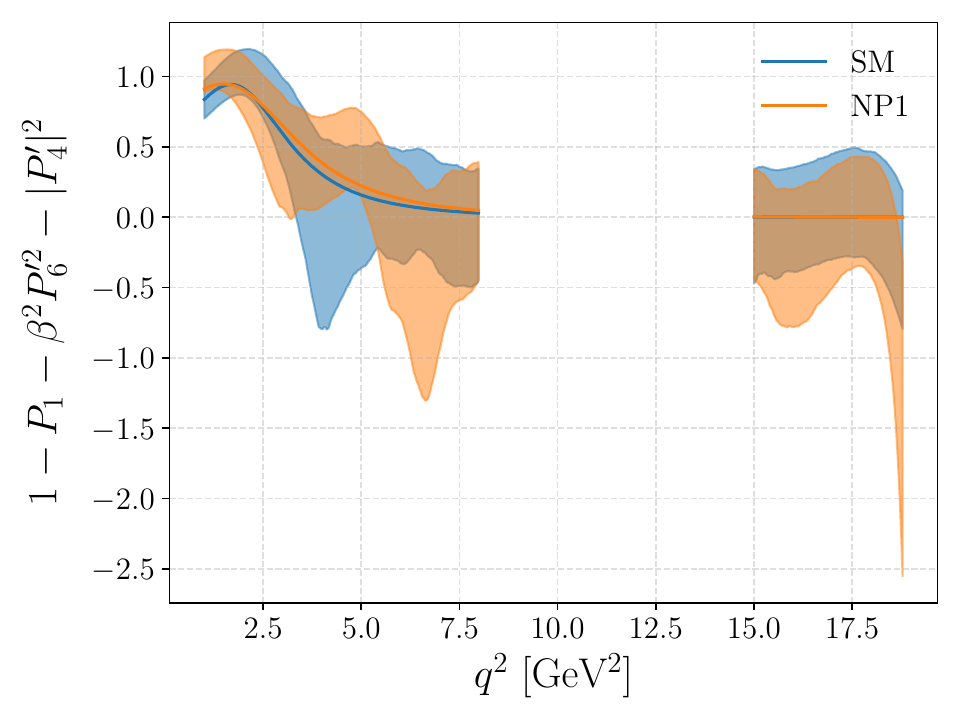}
    \includegraphics[width=0.98\linewidth, page = 2]{Plots/experiment/obs_bounds_trim.pdf}
    \includegraphics[width=0.98\linewidth, page = 3]{Plots/experiment/obs_bounds_trim.pdf}
    \caption{The square of the geometric bounds, as extracted from a single pseudo-experiment.}
    \label{sec:experiment:fig:bounds}
\end{figure}

\bigskip

\section{New geometrical bounds, analysis of zeroes and extracting Wilson coefficients}\label{section4}

For the first time, we have model-independent access to the shape in the relevant $q^2$-regions of all the observables. This opens a new window of sensitivity to the Wilson coefficients and the possibility to implement robustness tests on data. In this section, we explore from the theoretical side, possible ways to extract information from the knowledge of the shape of the observables. In particular, we discuss new geometrical bounds, a well-known relation among the observables and we obtain the zeroes of three of the optimised observables. We also explore the robustness of the bounds and the symmetry relation in the presence of a NP scalar contribution. 

\subsection{Description of new geometrical bounds and non-trivial relation with and without  a scalar}\label{geonontrivial}

In the absence of scalar contributions, as was assumed in the LHCb analysis of Ref.~\cite{LHCb:2023gpo}, and neglecting lepton masses, there are 11 coefficients for the P-wave part of the angular distribution. Two of them are trivially related (see Ref.~\cite{Egede:2010zc}) but there is also one non-trivial relation. This implies that there is a maximum of eight independent observables that correspond to the eight degrees of freedom mentioned in subsection~\ref{sec:conditions}. When taking into account the lepton mass terms, there are 11 coefficients and one relation is maintained (the non-trivial one, up to factors of $\beta$) and hence two extra degrees of freedom appear. Finally, in the presence of a scalar contribution and including lepton mass terms,  there are eight complex amplitudes, $A_{\perp,\|,0}^{L,R}$ and $A_{t,S}$ ($n_A=8$), twelve coefficients ($n_c=12)$ and in this case there is no non-trivial relation ($n_{rel}=0$) and hence twelve degrees of freedom. Notice that we are focusing in this paper on the P-wave, for the S-wave we refer the reader to Ref.~\cite{Alguero:2021yus}. 

The relations between the observables are a consequence of the symmetries of the angular distribution. For instance, in the massless case the relation
\begin{equation}
2 n_A - n_{sym}=n_c-n_{rel},
\end{equation}
implies that there must be four symmetries of the distribution ($n_{sym}=4$) because there are six complex amplitudes $A_{\perp,\|,0}^{L,R}$ ($n_A=6$),  eleven coefficients ($n_c=11)$ and the three relations ($n_{rel}=3$) explained above. The symmetries can be expressed in a compact form
\begin{equation}
n^\prime=U n,
\end{equation}
where $U$ is a unitary transformation.

At this point we will present two different type of tests that can be used in the context of unbinned analyses: i) geometrical tests associated with the structure of the observables; and ii) tests based on the non-trivial relation between the observables. 

The bounds and the non-trivial relation are strictly defined for unbinned observables, even if they have previously been applied only on binned observables~\cite{Matias:2014jua}. Since the non-trivial relation has been already discussed in several different papers~\cite{Egede:2010zc,Matias:2012xw,Matias:2014jua} we  refer the reader to these references for further details. Our interest here is to make the first tests of the relation directly with the unbinned amplitudes and observables, and to explore the size of the breaking of the relation and these bounds in the presence of a scalar contribution, where in principle some of them may no longer hold.

\onecolumngrid\
\begin{center}\
\begin{figure}[h]\ \includegraphics[width=0.3\textwidth]{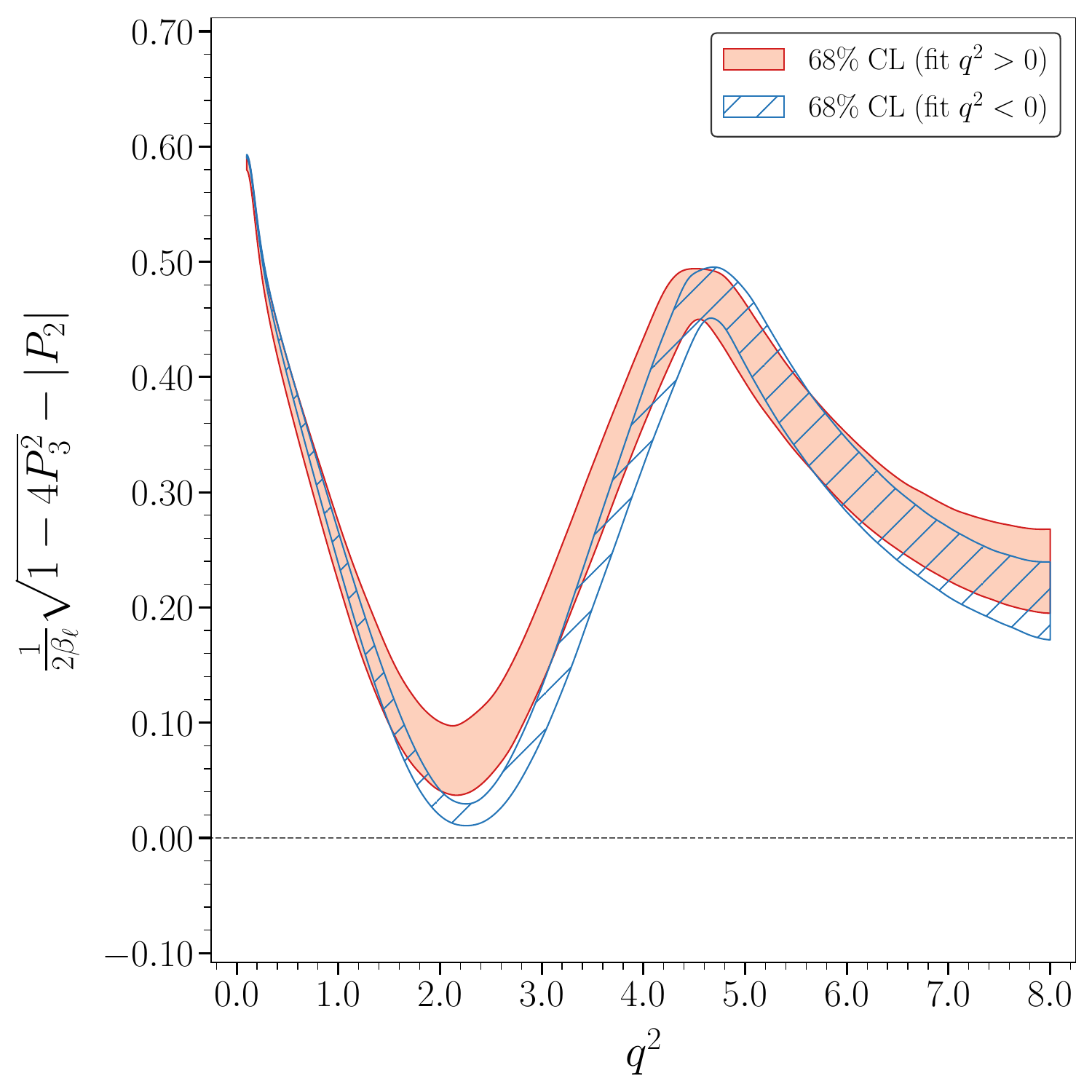}
\includegraphics[width=0.3\textwidth]{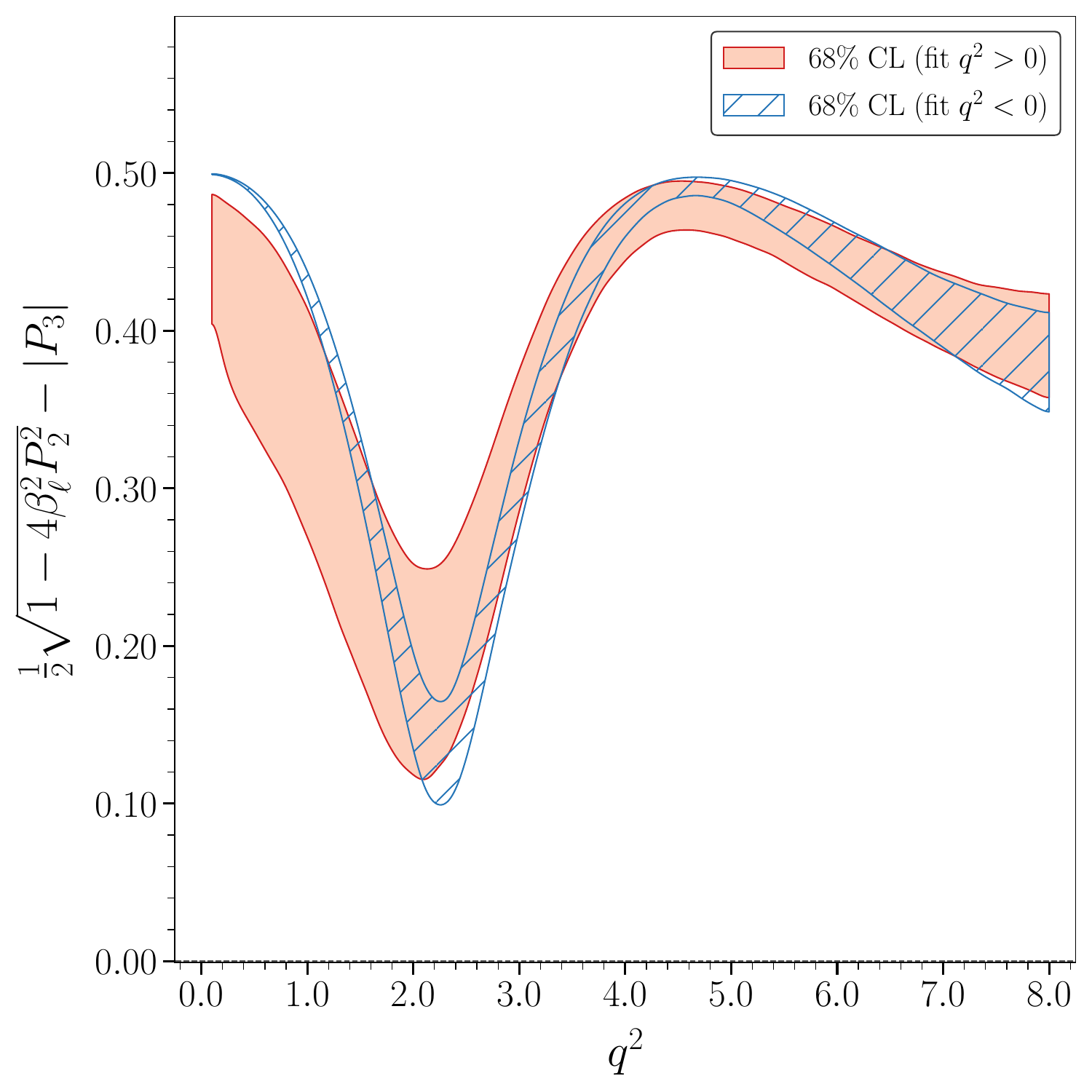}
\includegraphics[width=0.3\textwidth]{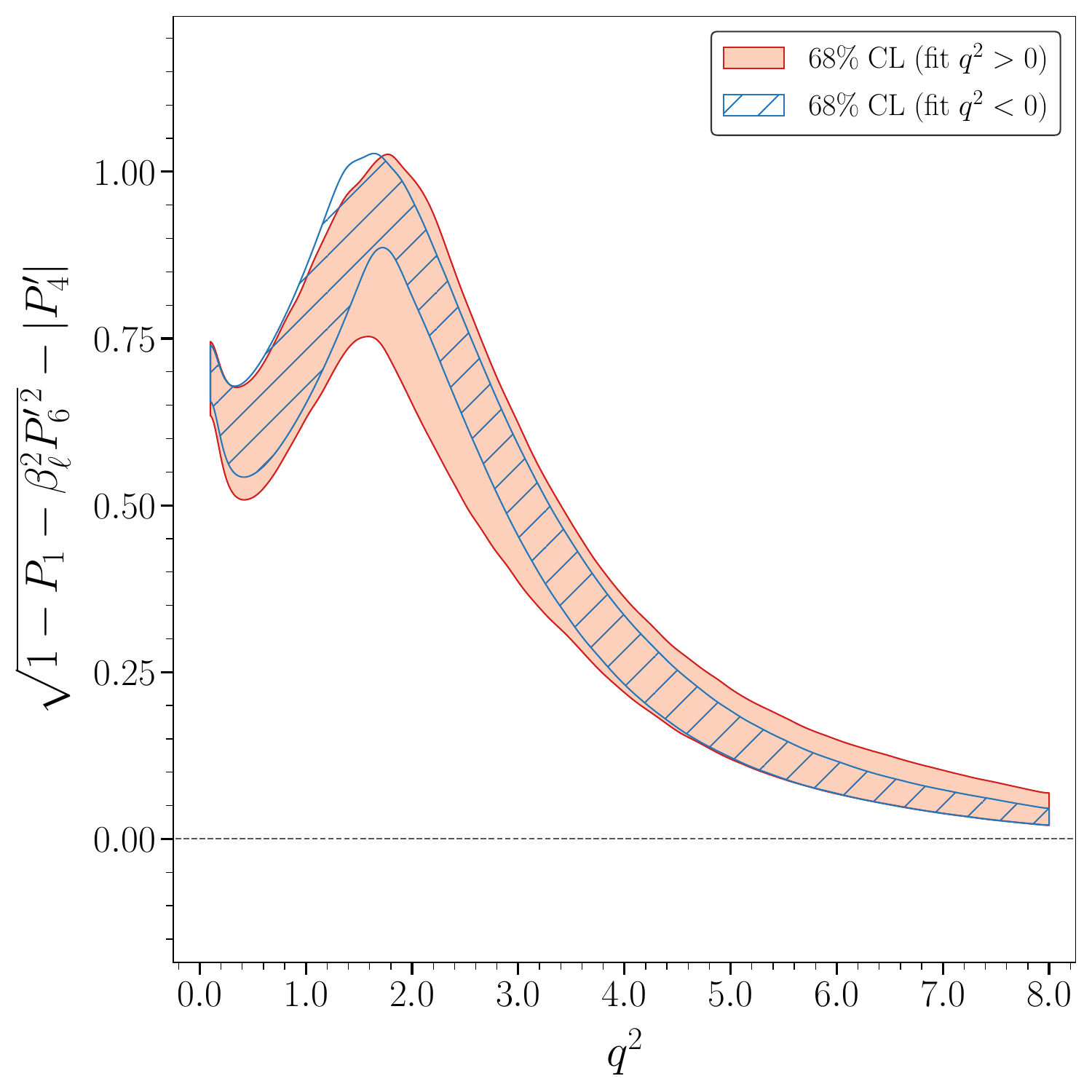}
\includegraphics[width=0.3\textwidth,]{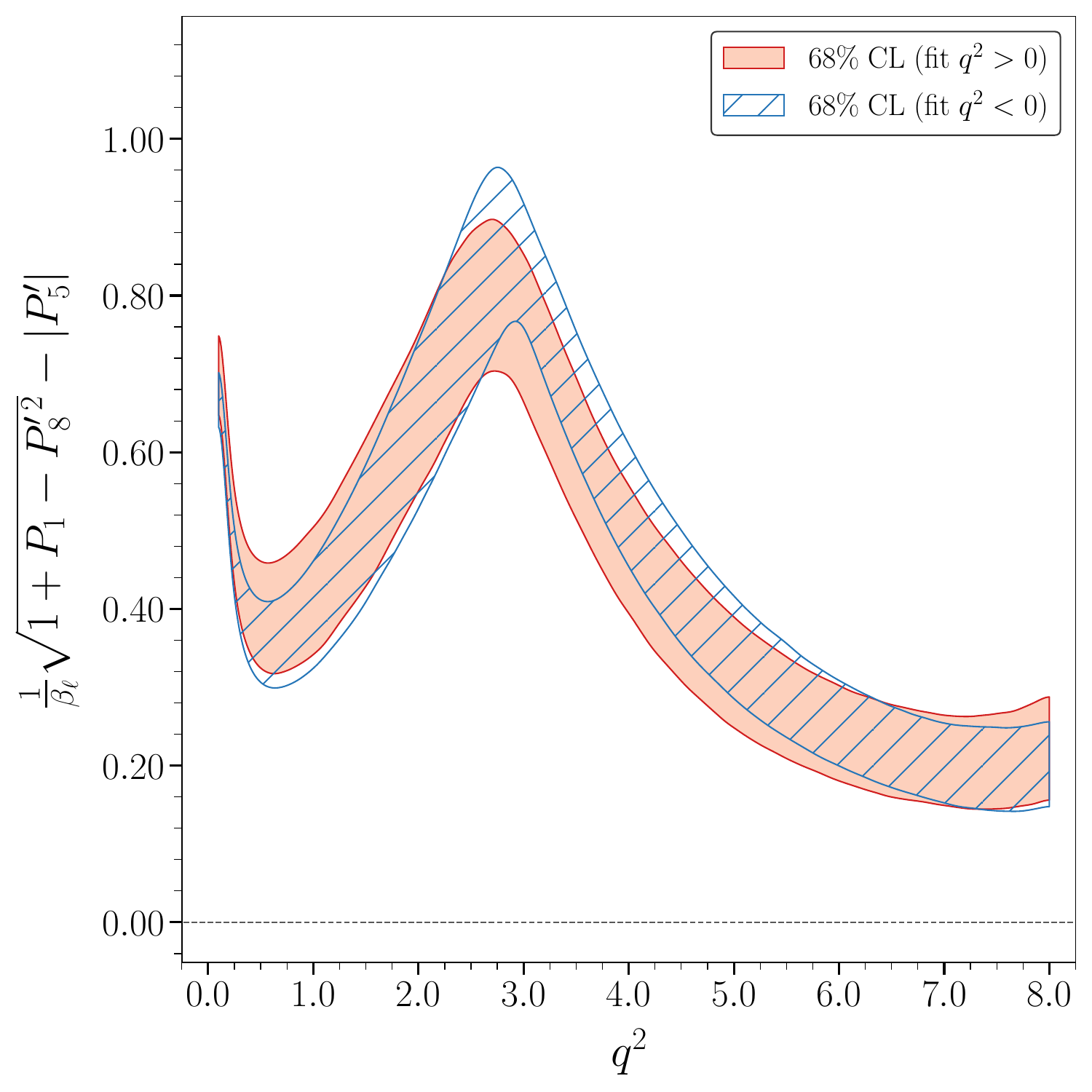}
\includegraphics[width=0.3\textwidth]{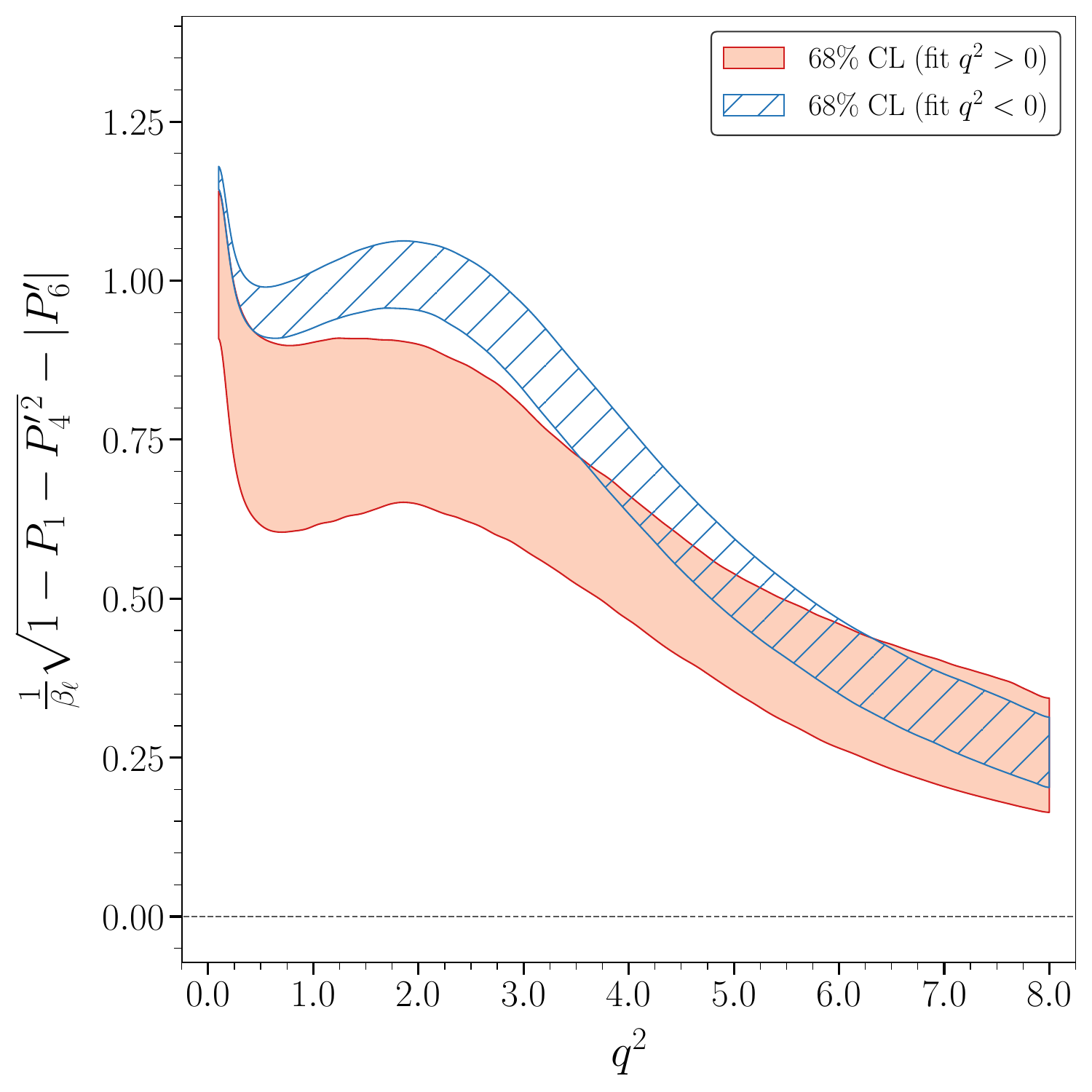}
\includegraphics[width=0.3\textwidth]{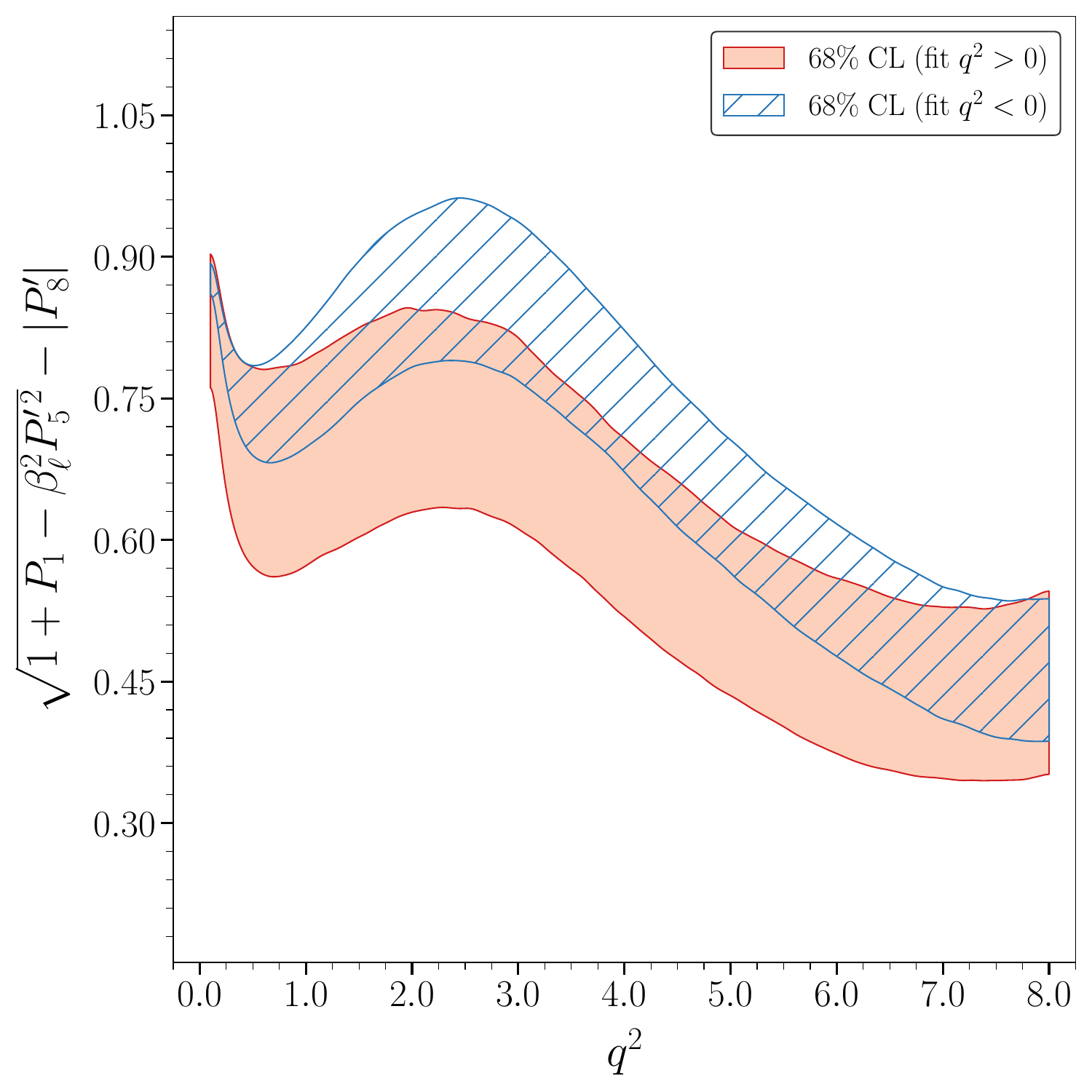}
\caption{Bounds on optimised observables using the amplitude parameterisation of Ref.~\cite{LHCb:2023gpo} in the two cases $q^2>0$ and $q^2<0$ discussed in subsection~\ref{sec:interpretation}. }\label{fig:boundsPP}\
\end{figure}\
\end{center}\
\twocolumngrid

\subsubsection{Geometrical tests associated with the structure of the observables}

The geometrical tests rely on the description of the observables in terms of the $n_i$ vectors and lead to bounds on the maximum size of the observables.  There are three such tests:

\begin{itemize}

\item[i)] Geometrical test I: 
One can combine 
the normalised observable~\cite{Matias:2012xw} \begin{equation}
    P_4=\frac{{\rm Re}(n_0^\dagger n_\|+ {\rm CP})}{\sqrt{(|n_\||^2+{\rm CP})(|n_0|^2+{\rm CP})}},
    \end{equation}
    where CP stands for the CP conjugate
    and the relation with the $P_4^\prime$ observable given by $P_4=P_4^\prime/\sqrt{1-P_1}$  
together with 
\begin{equation}
P_6=\frac{{\rm Im}(n_0^\dagger n_\|+{\rm CP})}{\sqrt{(|n_\||^2+{\rm CP})(|n_0|^2}+{\rm CP})},
\end{equation}
and its corresponding relation with the $P_6^\prime$ observable given by $P_6=\beta P_6^\prime/\sqrt{1-P_1}$ to obtain
\begin{equation}
\quad \quad |P_4+i P_6|^2\leq 1 \quad {\rm or}\quad |P_4^\prime+ i\beta  P_6^\prime|^2 \leq 1-P_1.
\end{equation}
This turns into
\begin{equation}
    |P_4^\prime| \leq \sqrt{1-P_1-\beta^2 P_6^{\prime 2}},
\end{equation}
and equivalently
\begin{equation} \label{p6pbound}
    |P_6^\prime| \leq \frac{1}{\beta}\sqrt{1-P_1- P_4^{\prime 2}}.
\end{equation}

\item[ii)] Geometrical test II: Also in the case of $P_5^\prime$ and $P_8^\prime$ following the same strategy and taking as a starting point
\begin{equation}
    P_5=\frac{{\rm Re}(n_0^\dagger n_\perp+ {\rm CP})}{\sqrt{(|n_\perp|^2+{\rm CP})(|n_0|^2+{\rm CP})}},
    \end{equation}
together with 
$P_5=\beta P_5^\prime/\sqrt{1+P_1}$ and also
\begin{equation}
P_8=\frac{{\rm Im}(n_0^\dagger n_\perp+{\rm CP})}{\sqrt{(|n_\perp|^2+{\rm CP})(|n_0|^2}+{\rm CP})},
\end{equation}
together with $P_8=P_8^\prime/\sqrt{1+P_1}$ one obtains
\begin{equation}
\quad \quad |P_5+i P_8|^2\leq 1 \quad {\rm or}\quad |\beta P_5^\prime+ i P_8^\prime|^2 \leq 1+P_1,
\end{equation}
\noindent which turns into
\begin{equation} \label{p5pbound}
    |P_5^\prime| \leq \frac{1}{\beta}\sqrt{1+P_1- P_8^{\prime 2}},
\end{equation}
\noindent and equivalently
\begin{equation}
    |P_8^\prime| \leq \sqrt{1+P_1- \beta^2 P_5^{\prime 2}}.
\end{equation}

\item[iii)] Geometrical test III: Finally, a bound on $P_2$ and $P_3$ can be derived following a different strategy. The starting point here is
\begin{equation}
    P_2=\frac{1}{\beta}\frac{{\rm Re}(n_\perp^\dagger n_\|+ {\rm CP})}
    { (|n_\perp|^2+|n_\||^2+{\rm CP})},
    \end{equation}
\noindent and
\begin{equation}
    P_3=\frac{{\rm Im}(n_\perp^\dagger n_\|+ {\rm CP})}
    { (|n_\perp|^2+|n_\||^2+{\rm CP})}.
    \end{equation}
\end{itemize}
Combining these expressions and using $(|n_\perp|-|n_\||)^2\geq 0$
one finds
\begin{equation}
    | P_2+i \frac{1}{\beta}P_3|\leq \frac{1}{2\beta},
\end{equation}
which implies
\begin{equation} \label{p2bound}
|P_2| \leq \frac{1}{2\beta}\sqrt{1-4 P_3^2},
\end{equation}
or alternatively,
\begin{equation}
|P_3| 
\leq \frac{1}{2}\sqrt{1-4\beta^2 P_2^2}.
\end{equation}

These bounds must be respected both in the SM and in presence of NP, and therefore they are a model-independent consistency test of the validity of an experimental analysis.

It is important to notice when comparing to data that the LHCb collaboration uses a different convention for the observables than the original definition in Ref.~\cite{Descotes-Genon:2013vna}. These differing conventions can be interchanged using, 
\begin{eqnarray}
&& P_1^{\rm LHCb}=P_1, P_{2,3}^{\rm LHCb}=-P_{2,3}, 
P_4^{\prime \rm LHCb}=-1/2 P_4^{\prime},\nonumber \\[2mm] 
&&  P_{5,6}^{\prime \rm LHCb}=P_{5,6}^{\prime}, P_8^{\prime \rm LHCb}=-1/2 P_8^{\prime}.
\end{eqnarray}
The previous bounds and the relations discussed below are expressed in the original basis.

In Fig.~\ref{fig:boundsPP} the bounds are applied in the case of the amplitude analysis of Ref.~\cite{LHCb:2023gpo} discussed above. In an amplitude analysis  the bounds  must be fulfilled by construction and they therefore serve as a cross-check of the experimental analysis procedure. 
As the bound  will continue to be respected in any amplitude analysis, even in the presence  of a NP scalar, the bound cannot be used to detect any scalar NP contribution. 

Fig.~\ref{fig:boundsPP} illustrates that the bounds all become systematically stronger for large values of $q^2$.
This is  particularly noticeable in the case of $P_4^\prime$, where the bound is quasi-saturated
at $q^2=8$~GeV$^2$.  The $P_2$ observable also saturates the bound around 2~GeV$^2$, where the observable hits its maximum value of $P_2\sim 1/2$.

In contrast to the situation with amplitudes, measurements of observables - such as those proposed in section~\ref{section3} - can violate some bounds in the presence of a scalar contribution. Our next point is to explore the extent of this violation.

In the presence of a scalar contribution, the bounds on $P_5^\prime$ and $P_6^\prime$ are modified and become  
\begin{equation}  \label{p5pboundscalar}    |P_5^\prime+\omega_r| \leq \frac{1}{\beta}\sqrt{1+P_1- P_8^{\prime 2}},
\end{equation}
and
\begin{equation}  \label{p6pboundscalar}    |P_6^\prime+\omega_i| \leq \frac{1}{\beta}\sqrt{1-P_1- P_4^{\prime 2}},
\end{equation}
where 
\begin{equation} \label{omegar}
\omega_r=
+\frac{1}{ N}\sqrt{2}  \frac{m_\ell}{\beta \sqrt{q^2}} [ {\rm Re}(A_\|^L A_S^*+A_\|^{R*} A_S) +{\rm CP}],
\end{equation}
and 
\begin{equation} \label{omegai}
\omega_i=
+\frac{1}{ N}\sqrt{2}  \frac{m_\ell}{\beta \sqrt{q^2}} [ {\rm Im}(A_\perp^L A_S^*-A_\perp^{R*} A_S) +{\rm CP}],
\end{equation}
with 
\begin{equation} \label{defN} N=\sqrt{(|n_\perp|^2+|n_\||^2+{\rm CP})(|n_0|^2+{\rm CP})}.
\end{equation}

The terms $\omega_{r,i}$ contain the scalar contributions. If the signs of $P_5^\prime$ and $\omega_r$ are the same the bound in Eq.~\eqref{p5pbound}  may be violated in an analysis measuring observables, 
 indicating the presence of a scalar NP contribution according to Eq.~\eqref{p5pboundscalar}.  Similarly, if the signs of $P_6^\prime$ and $\omega_i$ are the same, the bound in Eq.~\eqref{p6pbound} may also be violated according to Eq.~\eqref{p6pboundscalar}. 
All other bounds are unaffected by the presence of a scalar.

In the following, we quantify how large the breaking of the bounds and of the non-trivial relation can be 
under the assumption of a scalar NP contribution with all other amplitudes SM-like. 
The goal is to find out if such a NP contribution is detectable or, on the contrary, if the bounds are very robust and cannot be broken in a detectable way.

\subsubsection{Tests based on the non-trivial relation between the observables}

Another consistency test of an analysis  can be made using the non-trivial relation between the coefficients of the angular distribution. Under the assumption of massive leptons, this relation is given~\cite{Egede:2010zc}
\begin{align}\label{relationJ}
&\!\!J_{2c}=\, \frac{\beta_\ell^2 J_{6s} (J_4 J_5 + J_7 J_8) + J_9 (\beta_\ell^2 J_5 J_7 - 4 J_4 J_8)}{4 J_{2s}^{2} -  \left( J_3^2+ \beta_\ell^2 J_{6s}^{2}/4 +  J_9^2 \right)}\!\!\,\, \\
&- \,\frac{ (2 J_{2s}+ J_3) \left(4 J_4^2+\beta_\ell^2 J_7^2\right) + ( 2 J_{2s} - J_3) \left(\beta_\ell^2 J_5^2+4 J_8^2 \right)}{8 J_{2s}^{2} -  \left(2 J_3^2+ \beta_\ell^2 J_{6s}^{2}/2 + 2 J_9^2 \right)}. \quad \quad \nonumber    
\end{align} 
This relation has been already discussed and presented in~Refs.~\cite{Matias:2012xw,Egede:2010zc,Matias:2014jua} in the case of binned observables, where it is necessary to add a systematic uncertainty to handle the variation of the observables within each bin. However, this is the first time it can be tested directly on unbinned observables. 

A non-trivial relation can also be obtained for the CP-conjugate coefficients, which leads to a formally identical combination.

At this point there are two options: 1) to test directly the relation in Eq.~\eqref{relationJ}, which is given in terms of $J_i$; or 2) 
to re-express it in terms of the $P_i$ and $P_i^{\rm CP}$ observables, as shown below. The latter is implemented by inverting the definitions in Eq.~\eqref{optimizedeqs}.
There is a minor difference between the two options, because in the first case a breaking of Eq.~\eqref{relationJ}  can only be due to an internal problem of the experimental analysis, or to a large new scalar contribution. In the second case,  if we use the expression in terms of $P_i$ and $P_i^{\rm CP}$ and then assume that all $P_i^{\rm CP}$ are negligible, we introduce a third possible source of breaking if there exists large NP weak phases in Nature. In any case, this relation is always a very efficient cross-check of an experimental analysis.

We have explicitly checked that, up to machine precision, the relation of Eq.~\eqref{relationJ} using the $J$'s obtained from the fit of Ref.~\cite{LHCb:2023gpo} is fulfilled, considering either the $q^2<0$ or $q^2>0$ cases. Since the $J_i$ are obtained from an amplitude analysis where no scalar contribution is introduced, the relation in Eq.~\eqref{relationJ} must, by construction, be fulfilled. However, this implies nothing about the presence or absence of a scalar amplitude. 

In order to express Eq.~\eqref{relationJ} in terms of optimised observables we use the definitions of these observables in Eq.~\eqref{optimizedeqs}. 
One can then write down one relation for $\bar{P_i}=P_i+P_i^{\rm CP}$ and one for $\hat{P_i}=P_i-P_i^{\rm 
 CP}$. Taking the example of  $P_2$, we can write an equation for $\bar{P}_2$ and one for $\hat{P}_2$ as a function of all the other observables. The corresponding equations are~\cite{Matias:2014jua}
\begin{widetext}
\begin{equation}  \label{relation}  \bar{P}_{2}=P_2+P_2^{\rm CP}=+\frac{1}{2 {\bar k_1}} \bigg[ ( {\bar P_4^\prime} {\bar P_5^\prime} + \delta_{1}) + \frac{1}{\beta }\sqrt{(-1+{\bar P_1} + {\bar P_4^{\prime 2}})(-1-{\bar P_1} + \beta^2 {\bar P_5^{\prime 2}}) +\delta_{2} + \delta_3 {\bar P_1} +\delta_4 {\bar P_1}^2  }\bigg],  \end{equation}
\end{widetext}
and
\begin{widetext}
\begin{equation}  \label{relation2}  \hat{P}_{2}=P_2-P_2^{\rm CP}=+\frac{1}{2 {\hat k_1}} \bigg[ ( {\hat P_4^\prime} {\hat P_5^\prime} + \hat{\delta}_{1}) + \frac{1}{\beta }\sqrt{(-1+{\hat P_1} + {\hat P_4^{\prime 2}})(-1-{\hat P_1} + \beta^2 {\hat P_5^{\prime 2}}) +\hat{\delta}_{2} + \hat{\delta}_3 {\bar P_1} +\hat{\delta}_4 {\hat P_1}^2  }\bigg],  \end{equation}
\end{widetext}
where in these expressions $P_{2}$ stands for the value of the observable obtained from the relation, which should be exactly the same as the measured observable $P_2$ in the absence of a scalar contribution. 
The definitions of all $\bar{k}_1$ and ${\delta}_i$ can be found in Ref.~\cite{Matias:2014jua}.
The corresponding $\hat{\delta}_i$ coefficients are simply obtained from the $\delta$'s in Ref.~\cite{Matias:2014jua}  by changing ``bars'' to ``hats''.
It is important to emphasise that this relation mixes $P_i$ and $P_i^{\rm CP}$ observables. In the SM, however, the $P_i^{\rm CP}$ are negligible and this implies that, to an excellent approximation, the relation can be simplified by taking all $P_i^{ \rm CP}\to 0$. 
The simplified relation with $P_i^{\rm CP}\to 0$ can be broken in the presence of a large weak new physics phase that induces a non negligible $P_i^{\rm CP}$, or in the presence of a new scalar contribution. However, in this paper we will not consider further the possibility of a new weak phase, and instead  will focus on analysing the breaking of this relation due to the presence of a NP scalar.

A simplified version of Eqs.~\ref{relation} and~\ref{relation2}, fulfilled to a high accuracy, is obtained in the approximation $P_i^{\rm CP} \to 0$, $P_{3}$ and $P_{6,8}^\prime\to 0$, which is equivalent to assuming no  large NP weak and strong phases. This simplified version is given by 
\begin{equation}
P_2=\frac{1}{2} [P_4^\prime P_5^\prime+\Omega], \label{approximate}
\end{equation}
where $\Omega=\frac{1}{\beta}\sqrt{(-1+P_1+{P_4^\prime}^2)(-1-P_1+\beta^2 {P_5^\prime}^2})$.
For illustration, we show in Fig.~\ref{Wfunction} the expression in  Eq.~\eqref{approximate} rewritten in quadratic form using the results from one pseudo-experiment.

As discussed at the beginning of this subsection~\ref{geonontrivial}, in the presence of a NP scalar the non-trivial relation between the observables is broken. Here we estimate, for the first time, the amount of breaking of the relation due to the presence of a scalar by combining theory inputs with data on the branching fraction of ${B_s\to \mu^+\mu^-}$.
If we allow for the presence of a scalar,
Eq.~\eqref{relation} and Eq.~\eqref{relation2} can be generalised by introducing the following substitutions in Eq.~\eqref{relation} 
\begin{equation}
\bar{P_5^\prime} \to \bar{P_5^\prime}+\bar{\omega}_r,\quad \quad
\bar{P_6^\prime} \to \bar{P_6^\prime}+\bar{\omega}_i,
\end{equation}
and in Eq.~\eqref{relation2}
\begin{equation}
\hat{P_5^\prime} \to \hat{P_5^\prime}+{\hat{\omega}_r}, \quad \quad
\hat{P_6^\prime} \to \hat{P_6^\prime}+\hat{\omega}_i, \end{equation}
where $\hat{\omega}_{r,i}=\omega_{r,i}-\omega^{\rm CP}_{r,i}$ and 
$\bar{\omega}_{r,i}=\omega_{r,i}+\omega^{\rm CP}_{r,i}$ with 
\begin{equation}
\omega_r^{\rm CP}=
+\frac{1}{ N}\sqrt{2}  \frac{m_\ell}{\beta \sqrt{q^2}} [ {\rm Re}(A_\|^L A_S^*+A_\|^{R*} A_S) -{\rm CP}],
\end{equation}
and 
\begin{equation}
\omega_i^{\rm CP}=
+\frac{1}{ N}\sqrt{2}  \frac{m_\ell}{\beta \sqrt{q^2}} [ {\rm Im}(A_\perp^L A_S^*-A_\perp^{R*} A_S) -{\rm CP}],
\end{equation}
are the CP counterparts of the $\omega_{r,i}$ defined in Eq.~\eqref{omegar} and Eq.~\eqref{omegai}, with $N$ given by Eq.~\eqref{defN}. By expanding the terms in $\omega$, one can understand the maximal amount of breaking expected in the presence of a NP scalar.

\begin{figure}[b]
\includegraphics[width=8cm,height=5cm, page = 4]{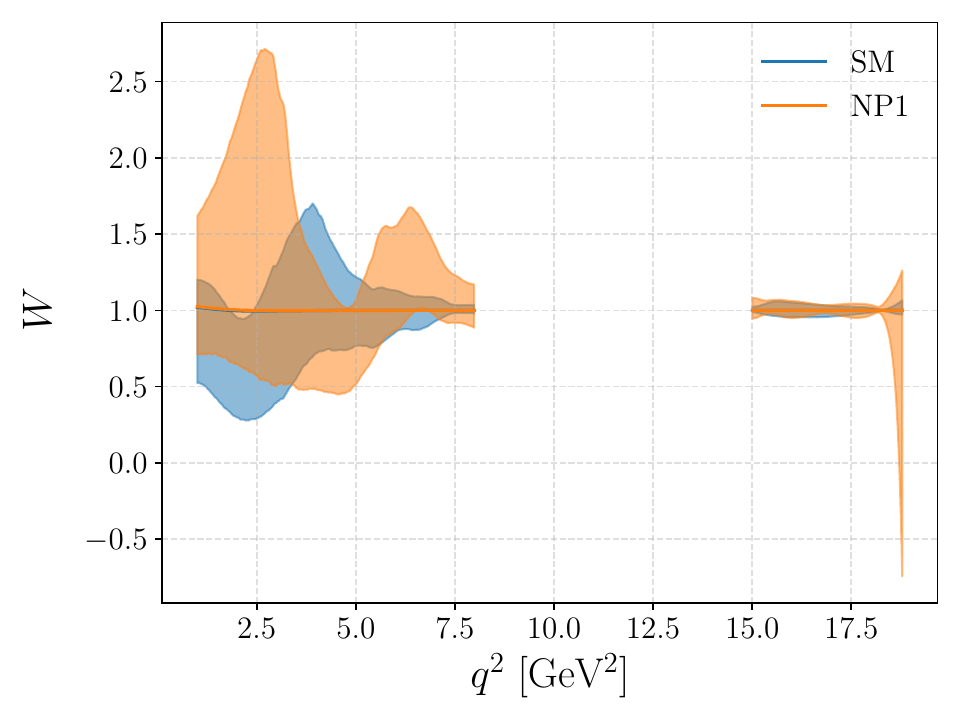} 
\caption{Plot of the non-trivial relation rewritten as $W=P_1^2+P_4^{\prime 2}+\beta^2 (4 P_2^2-4 P_2 P_4^\prime P_5^\prime+P_5^{\prime 2})+P_1 (P_4^{\prime 2}-\beta^2 P_5^{\prime 2})
$, where $W=1$ for all $q^2$. Only one pseudo-experiment is used for illustrative purposes. } \label{Wfunction}\end{figure}

\begin{figure}[!t]
\includegraphics[width=\columnwidth]{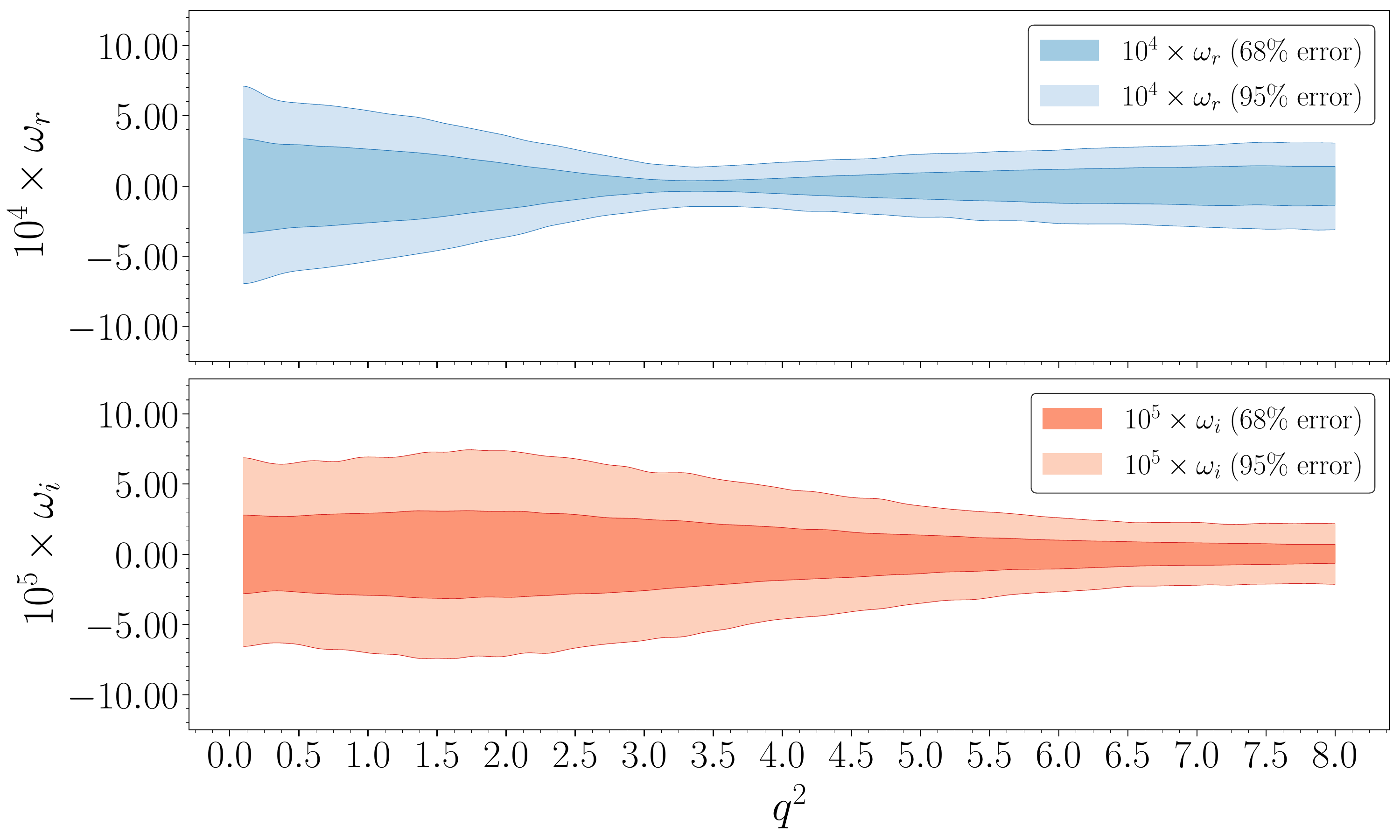}
\caption{
1$\sigma$ and 2$\sigma$ confidence intervals for $\omega_r$ (top panel) and $\omega_i$ (bottom panel) as functions of $q^2$ based on $B_s \to\mu^+\mu^-$ constraints.} \label{fig:omegas}
\end{figure}

\begin{figure}
\includegraphics[width=\columnwidth]{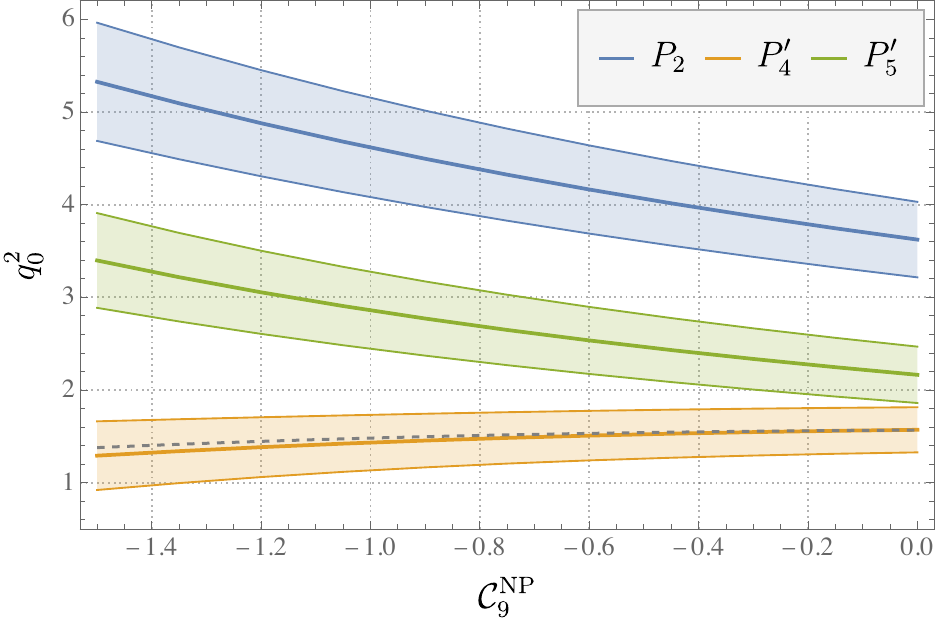}
\caption{Plot of the position of the zero of $P_{2}$ and $P_{4,5}^\prime$ versus $C_9^{\rm NP}$. The dashed line corresponds to the approximate expression of Eq.~\eqref{zerop4paprox}. }
\label{figPzeroes}
\end{figure}

As an illustration, assuming neither large NP weak phases nor large strong phases, one can take  
 the simplest form of the relation  and expand in $\omega_r$.  Assuming a small amount of breaking in $\omega_r$, one finds
\begin{equation} \label{p2basic}
P_2=\frac{1}{2}\left[P_4^\prime P_5^\prime+\Omega\right]+\Delta ,
\end{equation}
 and
\begin{equation}\label{eq:delta}
\Delta=\frac{\omega_r}{2}\left(P_4^\prime-\beta P_5^\prime \sqrt{\frac{-1+P_1+{P_4^\prime}^2}{-1-P_1+\beta^2 {P_5^\prime}^2}}\right).
\end{equation}
Next, we evaluate the size of the breaking parameters $\omega_{r,i}$. In Fig.~\ref{fig:omegas}, we present the expected sizes of $\omega_r$ and $\omega_i$ after including all sources of uncertainty and varying $\mathcal{C}_S$ according to a Gaussian distribution centred at the best-fit point of Table~\ref{table:ScalarBsmumufit}, with its standard deviation corresponding to the 1$\sigma$ confidence interval quoted in the same table. 
 
The small size of $\omega_{r,i}$, and therefore of $\Delta$ in Eq.~\eqref{eq:delta}, means that one cannot discern if there is a scalar contribution to $B \to K^*\mu^+\mu^-$ from the breaking of the relations and the bounds. 
This is consistent with previous determinations of these contributions within WET analyses of $b\to s\ell^+\ell^-$ data~\cite{Hurth:2023jwr,Beaujean:2015gba,Bobeth:2012vn}.

In summary, we have found that the bounds and the non-trivial relation assuming the absence of large new weak phases  should be fulfilled to a high-accuracy, making them a robust internal cross-check of any experimental analysis.

\subsection{Zero-crossing points of the optimised observables}\label{subsec:zeroes}

Knowledge of the shape of the observables $P_{1,2}$ and $P_{4,5}^\prime$  provides  information on zero-crossing points, which can be used to determine the Wilson coefficients.

The best known zero-crossing point is that of the forward-backward asymmetry, which is equivalent to the zero-crossing point of the $P_2$ observable. 
In the limit of the initial hadron being heavy and the final meson having a large energy, i.e., at small $q^2$, the form factors are related by
\begin{equation}
\frac{T_1(q^2)}{V(q^2)}=\frac{m_B^2}{(m_B+m_{K^*})^2}\frac{T_2(q^2)}{A_1(q^2)}
    \end{equation}
and using that $T_1(q^2)/V(q^2)=m_B/(m_B+m_{K^*})$ (at leading order in $\alpha_s$) in this same limit, all form factor dependence drops out and one gets,
\begin{equation}
{q_0^2}^{LO}\Big\rvert_
{P_2}=-2 m_b m_B \frac{{\cal C}_7^{\rm eff}}{{\rm Re} {\cal C}_9^{\rm eff}(q_0^2)}. 
\end{equation}
This is the well known expression of the position of the zero-crossing point at leading order~\cite{Ali:1994bf}. However, in this case, the next-to-leading order correction is substantial and moves the crossing-point in the SM from $\simeq 3$ GeV$^2$ to  $q_0^2=3.60\pm 0.40$ GeV$^2$. Unfortunately, the explicit expression for the next-to-leading order zero-crossing point is a complicated function of form factors, hadronic nuisance parameters, as well as the Wilson coefficients.  
The global analysis of Ref.~\cite{Alguero:2023jeh} points to NP only in the Wilson coefficient ${\cal C}_9$, with marginal contributions to the other Wilson coefficients. We therefore predict, at next-to-leading order, the  theoretical position of the zero of $P_2$ as a function of ${\cal C}_9^{\rm NP}$, as shown in Fig.~\ref{figPzeroes}. A strong dependence of the zero-crossing points of $P_2$ and $P_5^\prime$ on ${\cal C}_9^{\rm NP}$ is observed but there is a marginal dependence on the crossing point of $P_4^\prime$. We have not added the corresponding zero-crossing point of $P_1$, because it is relatively flat and therefore has significant uncertainties. Nonetheless, we explain below a further interesting feature of the $P_1$ zero-crossing point. 

Using the non-trivial symmetry relation we can find an alternative way of cross-checking the position of the zero of $P_2$ in the SM and in the presence of NP. 
It consists of finding where the following function is zero
\begin{equation}
X_{P_2}=P_4^\prime P_5^\prime+\Omega .
\end{equation}
This is true up to corrections of ${\cal O}(P_i^{\rm CP}, P_3, P_{6,8}^\prime)$, i.e., the same approximation as in Eq.~\eqref{approximate}.

The zero-crossing point of $P_5^\prime$ in the SM  is $q_0^2=2.16\pm 0.29$ GeV$^2$. Again here  the exact expression is rather cumbersome, so instead of writing it explicitly we present the exact zero-crossing point as a function of ${\cal C}_9^{\rm NP}$ in Fig.~\ref{figPzeroes}.
As in the  case of $P_2$, there also exists for $P_5^\prime$ an alternative way to determine its zero using the expression,
\begin{equation}
X_{P_5^\prime}=-P_2+\frac{1}{2\beta} \sqrt{(-1+P_1+P_4^{\prime 2})(-1-P_1)}.
\end{equation}

An interesting property is found when comparing the zeroes of $P_5^\prime$ and $P_1$. To a high accuracy both observables have the same zero-crossing point in the SM, i.e.
\begin{equation}
 {q_0^2}\Big\rvert_{P_5^\prime}^{\rm SM}\simeq q_0^2\Big\rvert_{P_1}^{\rm SM}.
 \end{equation}

\begin{table}
\begin{tabular}{|c|c|c|c|}
  \hline
  Scenario & best-fit-point & 1$\sigma$ & 2$\sigma$\\
  \hline
  $\mathcal{C}_S$~($\mathrm{GeV}^{-1}$) & 0.000  & [-0.0084, 0.0084] & [-0.0129, 0.0129]\\
  \hline
\end{tabular}
\caption{Best-fit point, 1$\sigma$, and 2$\sigma$ confidence intervals for the determination of the scalar Wilson coefficient $\mathcal{C}_S$ from a fit to the experimental measurement of the branching fraction of the $B_s\to\mu^+\mu^-$ decay mode. }\label{table:ScalarBsmumufit} 
\end{table}

On the contrary, in the presence of NP in ${\cal C}_9^{\rm NP}$
the zeroes of $P_1$ and $P_5^\prime$ are affected differently. Specifically, when adding a ${\cal C}_9^{\rm NP}=-1.17$, the central value of the zero-crossing point of $P_1$ moves marginally by $+0.10$ GeV$^2$, while the crossing point of $P_5^\prime$ moves by $+0.85$ GeV$^2$.
 This means that, independently of the exact position of the zeroes of $P_1$ and $P_5^\prime$, if these crossing points do not coincide it indicates
 a NP contribution to ${\cal C}_9$. Unfortunately when adding uncertainties to the zero-crossing point of $P_1$, the mild dependence on $C_9$ makes this difficult to exploit in experiments.

A byproduct of this is that at this common zero in the SM, using the simplified form of the relation between the observables, one finds

\begin{equation}
{P_4^{\prime}}^2+4\beta^2 {P_2^2}\Big\rvert_{P_5^\prime,P_1}^{\rm SM}\simeq 1.
\end{equation}
 
Finally, for the case of $P_4^\prime$, given that the position of its zero is even closer to the limit of large energy for the $K^*$ (small $q^2$) the expressions in the limit of large $E_{K^*}$ work accurately. At next-to-leading order, the zero-crossing point of $P_4^\prime$ in the SM is  found at $q_0^2=1.57\pm 0.25$ GeV$^2$ and in the presence of a NP contribution $C_9^{\rm NP}\sim -1$ it becomes even closer to $q^2 \simeq 0$ GeV$^2$.

In this limit, the position of the zero-crossing point for $P_4^\prime$ in the SM is found to be
\begin{equation} \label{zerop4paprox}
{q_0^2}\Big\rvert_{P_4^\prime}=\frac{-2 m_b {m_B}\,{\cal C}_7^{\rm eff}(2 m_b\,{\cal C}_7^{\rm eff}+ m_B\,{\rm Re} {\cal C}_9^{\rm eff}(q_0^2))}{m_B ({|{\cal C}_{9}^{\rm eff}(q_0^2)|}^2+{\cal C}_{10}^2){+\,2m_b\,{\cal C}_7^{\rm eff}{\rm Re}{\cal C}_9^{\rm eff}(q_0^2)}}\end{equation}
in excellent agreement with the next-to-leading order SM prediction, as can be seen in Fig.~\ref{figPzeroes}. (See, for instance, Ref.~\cite{ Kumar:2014bna} for explicit leading order expressions of the zero-crossing points of other optimised observables.)

The $C_{9}$ dependence of the zero of $P_{4}^{\prime}$ is small, such that the ability of experimental data to distinguish SM and NP is limited. However, as shown in Figs.~\ref{sec:experiment:fig:zero} and~\ref{sec:experiment:fig:zerotoys}, the new experimental method presented in this paper will have excellent ability to resolve expected NP effects in the observables $P_{2}$ and $P_{5}^{\prime}$.

\bigskip
\section{Conclusions and Outlook}
\label{conclusions}
In this paper we present a model-independent unbinned method, based on a moments approach, to extract angular observables in the four-body decay mode $B\to K^*(\to K^+\pi^-)\mu^+\mu^-$. This method is an alternative to the present approach of putting the data into fixed bins of \qsq and maintains the model independence of the observables. It will also allow for the first studies of the \mkpi dependence of the observables, shedding new light on how the P- and S-wave observables depend on the \mkpi kinematic variable. The method is in contrast to different amplitude analyses that have been presented, which are model-dependent. In the case of experimental amplitude analyses, we define 
a procedure to test the validity of a theoretical model for the amplitudes. 
As an illustration, we apply this test to the fit of Ref.~\cite{Gubernari:2020eft}, fixing the theory parameters at $q^2<0$. We show that one of the conditions is not fulfilled and suggest several possible reasons for this.

We provide the SM predictions for all the unbinned optimised observables and derive a set of new geometrical bounds on these observables. We explore the impact of a NP scalar on these bounds and on a non-trivial relation between the unbinned observables, considering the constraint from the branching fraction of the decay $B_s \to \mu^+\mu^-$. We find that the impact of scalar NP is negligible and therefore the bounds and the relation are robust, and can be used as a cross-check of an experimental analysis procedure. 

We also provide next-to-leading order predictions for the zero crossing-points of the optimised observables $P_2$ and $P_{4,5}^\prime$ as a function of $C_9^{\rm NP}$, and show that they can be extracted with the proposed moments approach. 

In a forthcoming paper~\cite{CMNPS}, we will investigate the precision with which the Wilson coefficients can be extracted with the unbinned moments approach and compare to a traditional binned fit.
\bigskip

\acknowledgments{}

The authors acknowledge Konstantinos Petridis and Michael McCann for enlightening discussions. B.C. would like to thank the Phenomenology Group at DAMTP for useful discussions at the early stages of this work. J.M. gratefully acknowledges the financial support from ICREA under the ICREA Academia programme 2018 and to AGAUR under the Icrea Academia programme 2024 and from Departament de Recerca i Universitats de la Generalitat de Catalunya. J.M. also received financial support from the Spanish Ministry of Science, Innovation and Universities (project PID2020-112965GB-I00) and from the IPPP Diva Award 2024.
The work of B.C. is supported by the Margarita Salas postdoctoral program funded by the European Union-NextGenerationEU and the La Caixa Junior Leader fellowship from the ``la Caixa” Foundation (ID 100010434, fellowship code LCF/BQ/PI24/12040024).
MN acknowledges the financial support by the Spanish Government (Agencia Estatal de Investigaci\'on MCIN/AEI/10.13039/501100011033)  and the European Union NextGenerationEU/PRTR through the “Juan de la Cierva” program (Grant No. JDC2022-048787-I), Grants No. PID2020-114473GB-I00 and  No. PID2023-146220NB-I00. M.P. and M.S. acknowledge financial support from the UK Science and Technology Facilities Council (STFC).

\newpage
\clearpage

\appendix
\begin{widetext}
\section{SM UNBINNED PREDICTIONS OF NON-OPTIMISED OBSERVABLES} \label{app:non-optimised}
For completeness, we provide in Fig.~\ref{fig:non-optimised} also the SM predictions of the unbinned non-optimised observables.
\begin{figure}[!h]
\includegraphics[width=8cm,height=5.2cm]{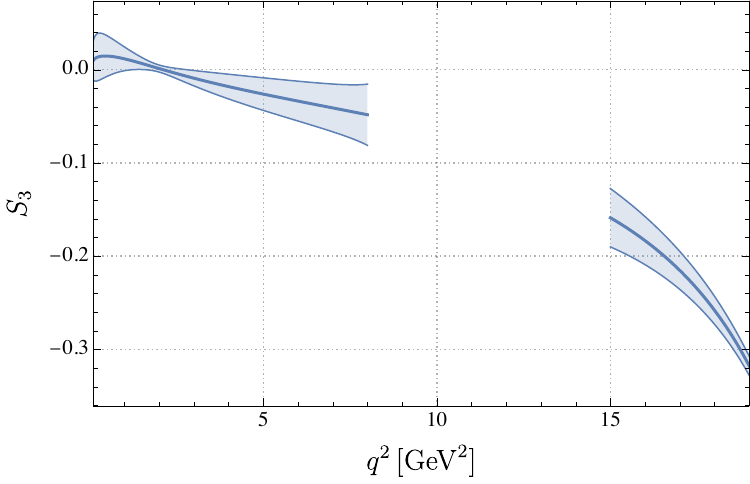}
\includegraphics[width=8cm,height=5.2cm]{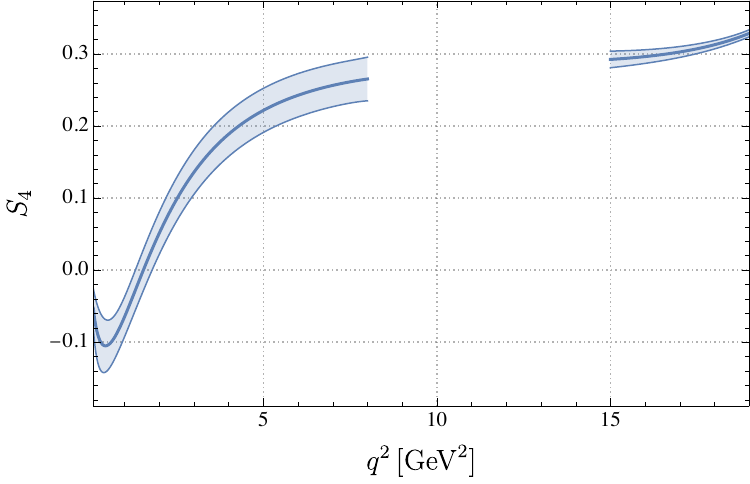}
\includegraphics[width=8cm,height=5.2cm]{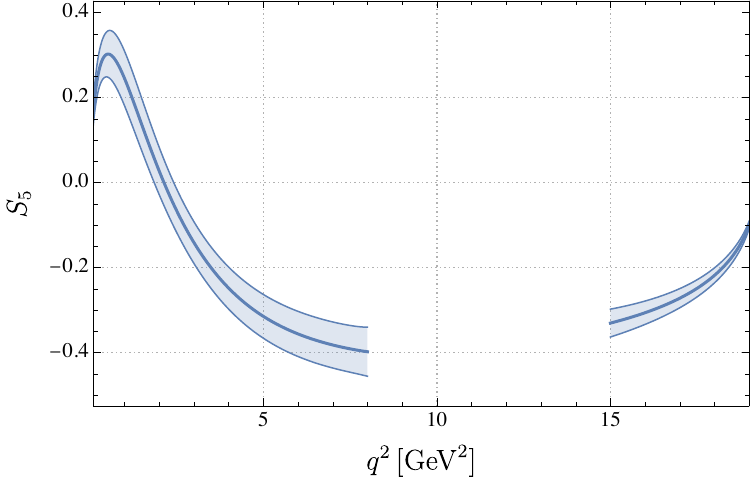}
\includegraphics[width=8cm,height=5.2cm]{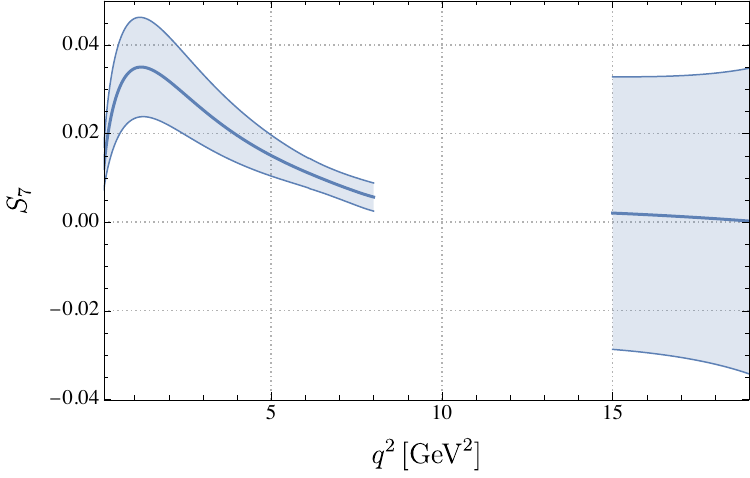}
\includegraphics[width=8cm,height=5.2cm]{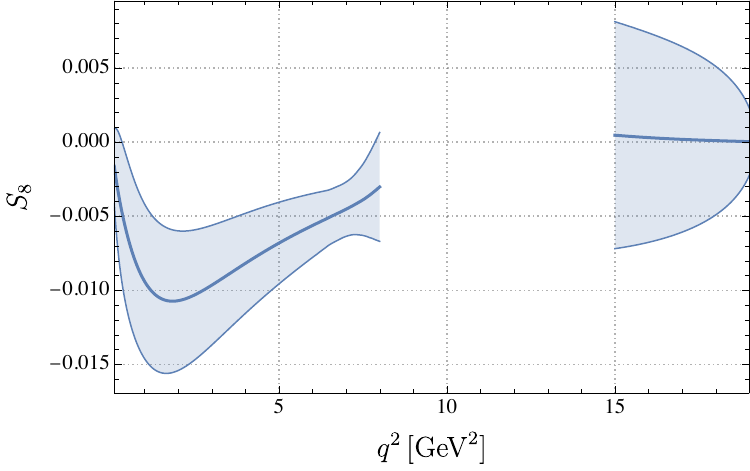}
\includegraphics[width=8cm,height=5.2cm]{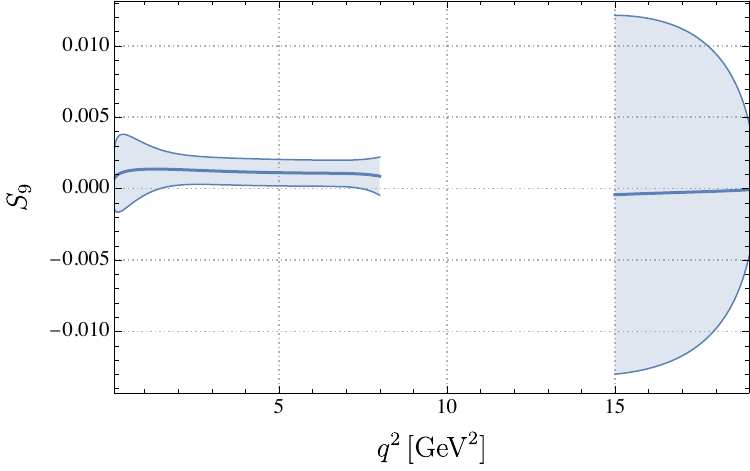}
\,\, \hspace*{-0.0cm}\includegraphics[width=7.8cm,height=5.2cm]{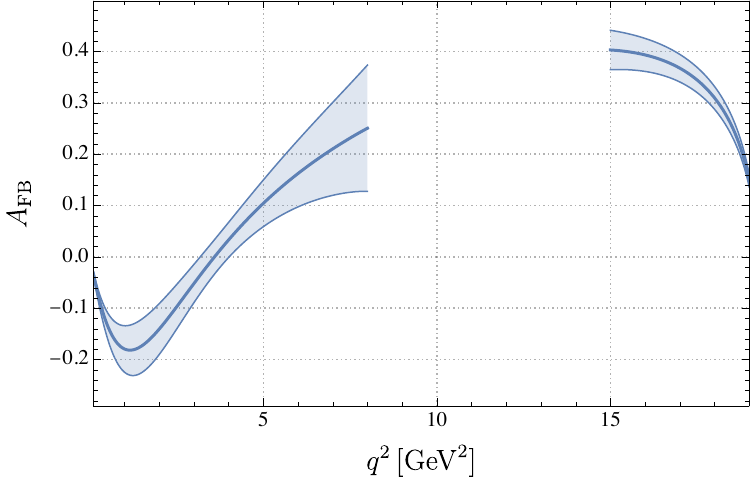}\, \, \, \, \,
\includegraphics[width=7.4cm,height=5.2cm]{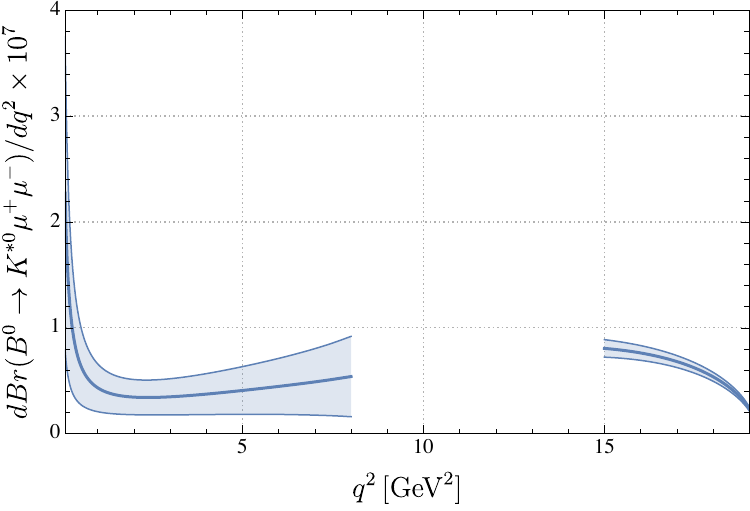}
\caption{SM prediction for the unbinned non-optimised observables $S_{3,4,5,7,8,9}$, $A_{\rm FB}$ and the branching fraction.} \label{fig:non-optimised}    
\end{figure}
\end{widetext}

\clearpage

\begin{widetext}
\section{CONDITIONS}
\label{app:conditions}
The rest of the conditions that exhibit a perfect agreement are presented in Fig.~\ref{condition5}.
\vspace*{-0.2cm}
\begin{figure}[!h]
    \includegraphics[height=150pt]{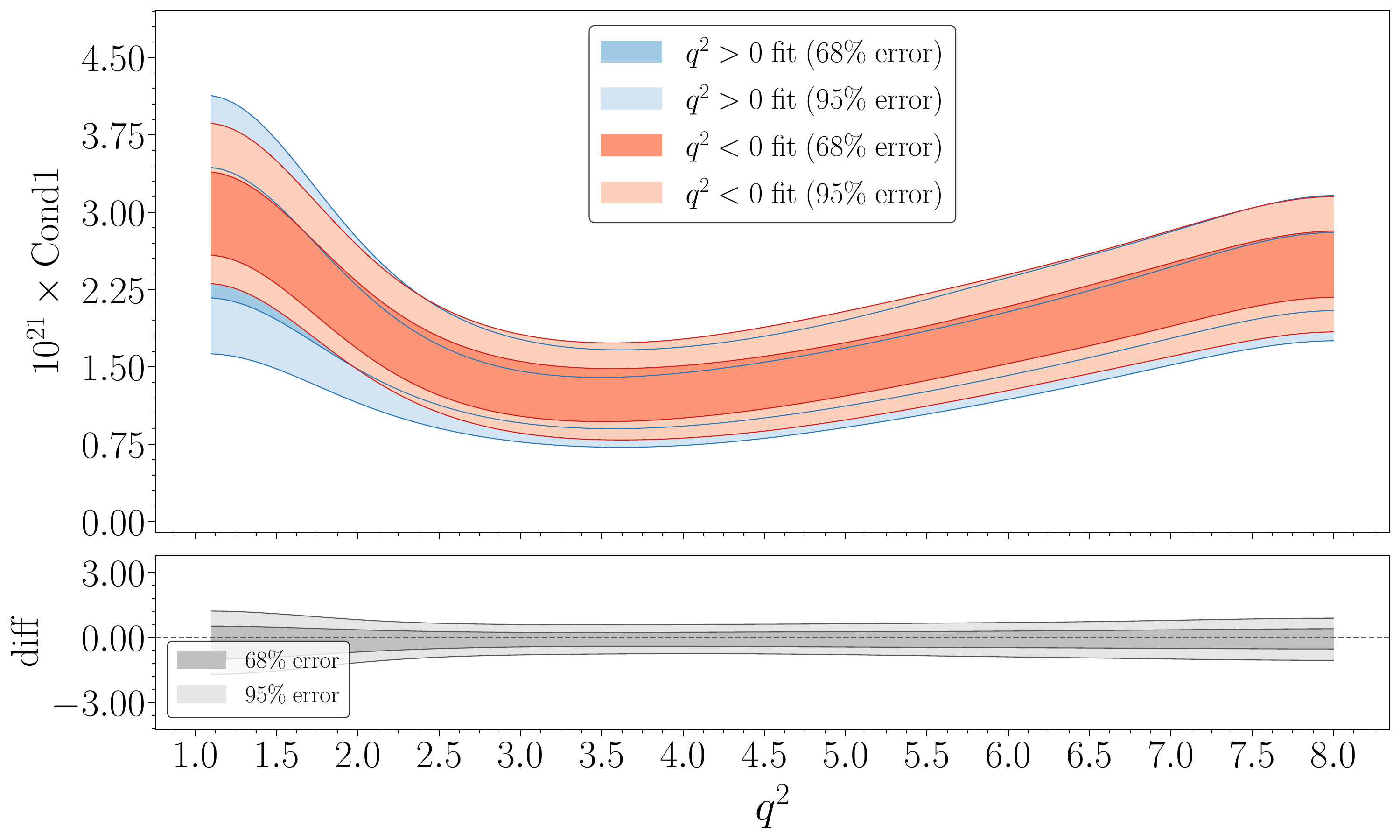}
    \includegraphics[height=150pt]{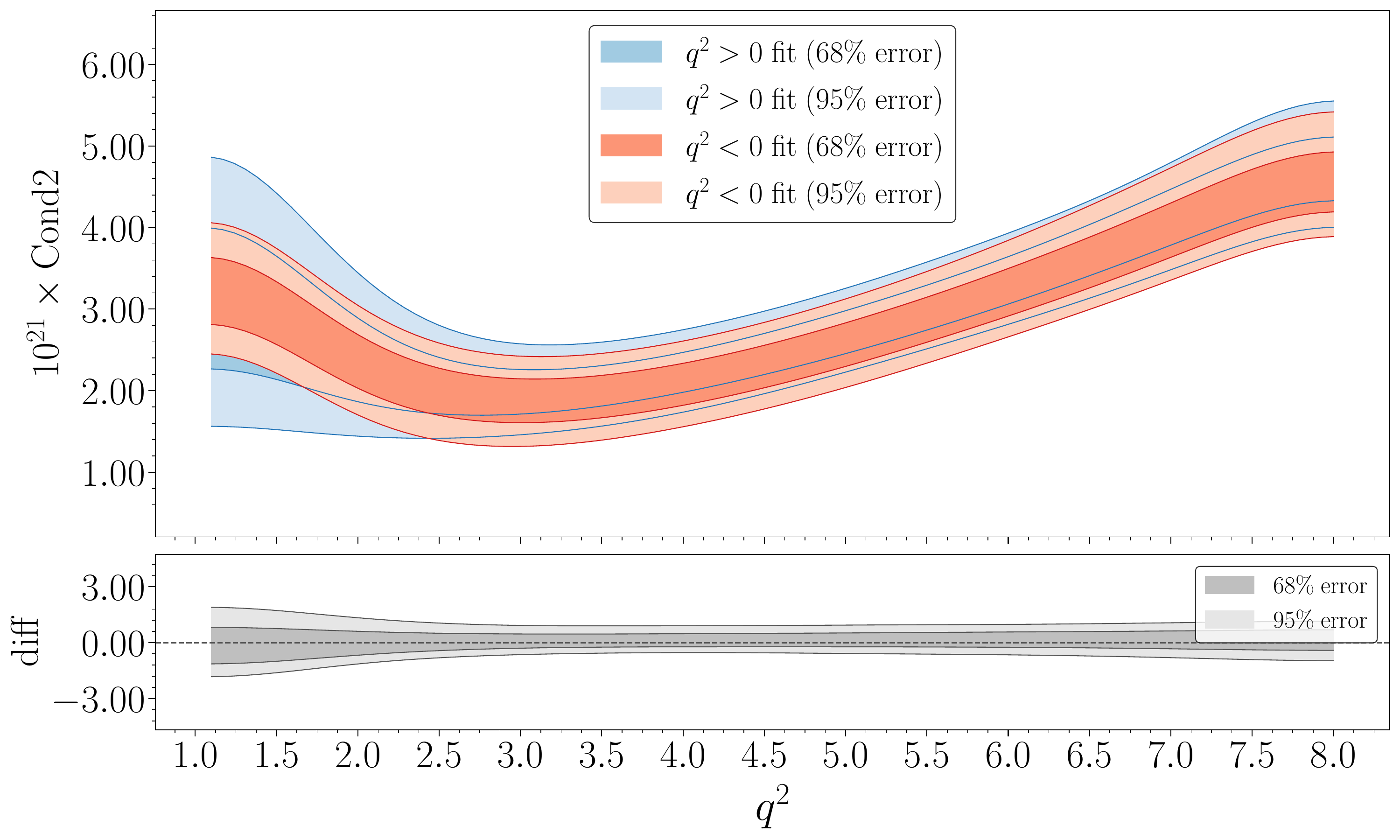}  
%
\includegraphics[height=150pt]{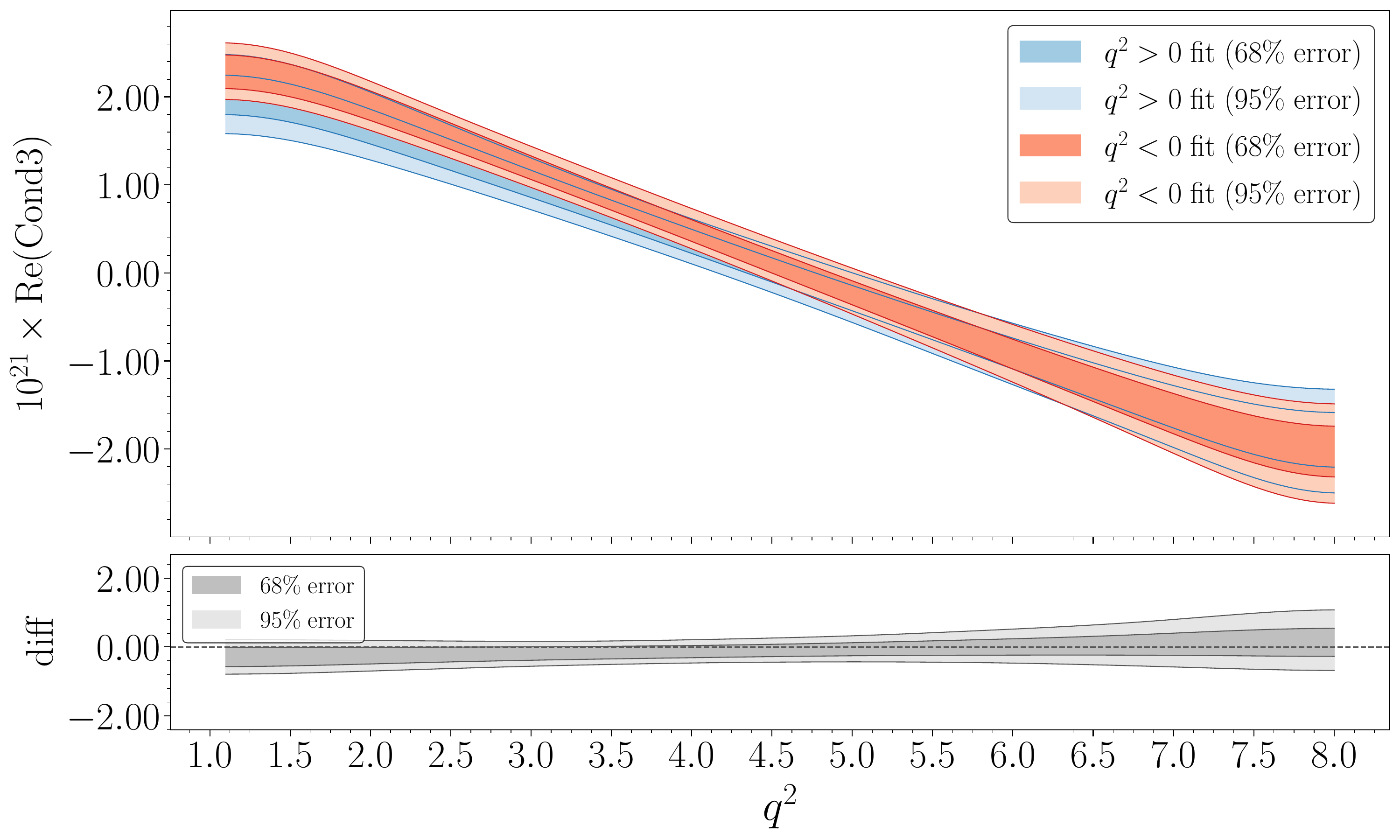}  
\includegraphics[height=150pt]{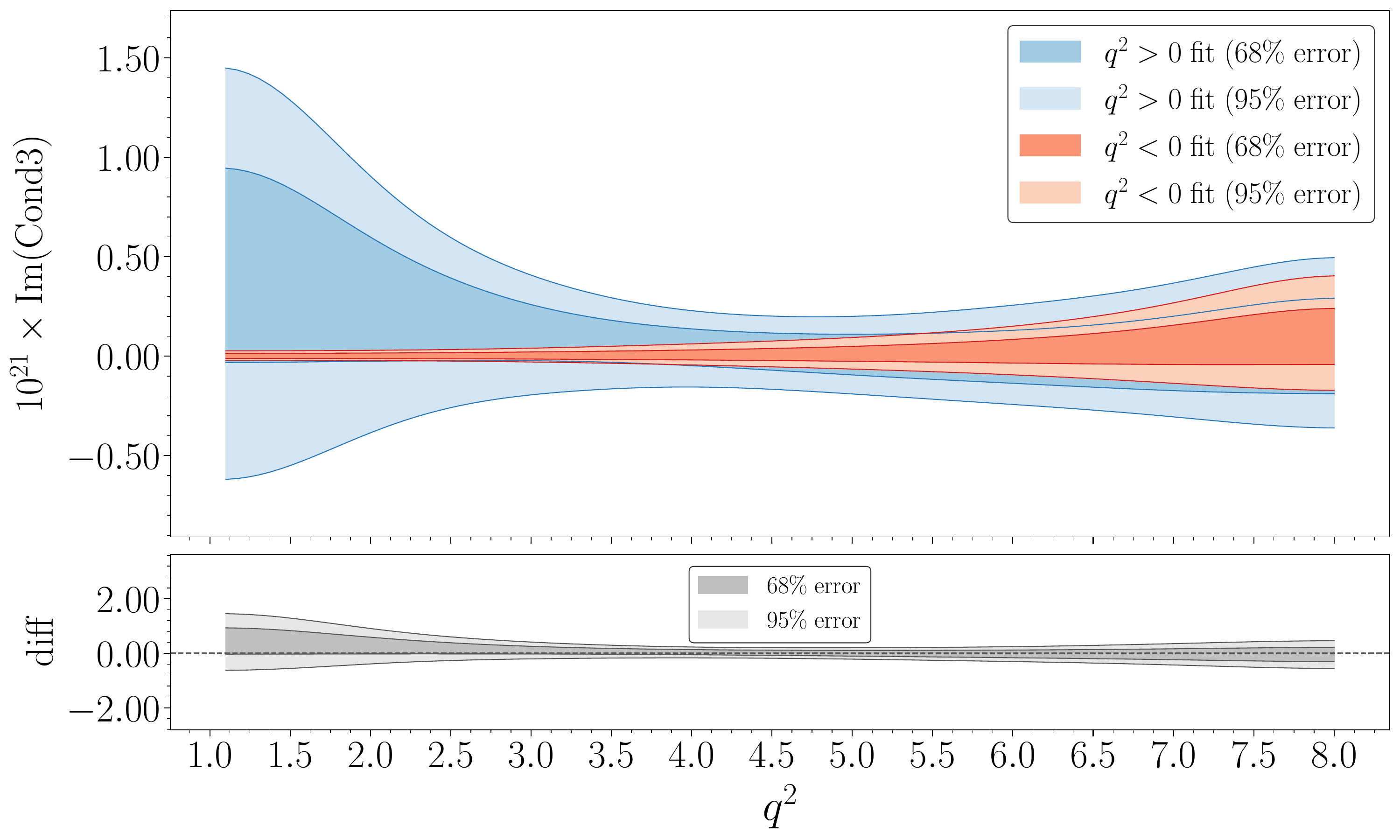}  
%
\includegraphics[height=150pt]{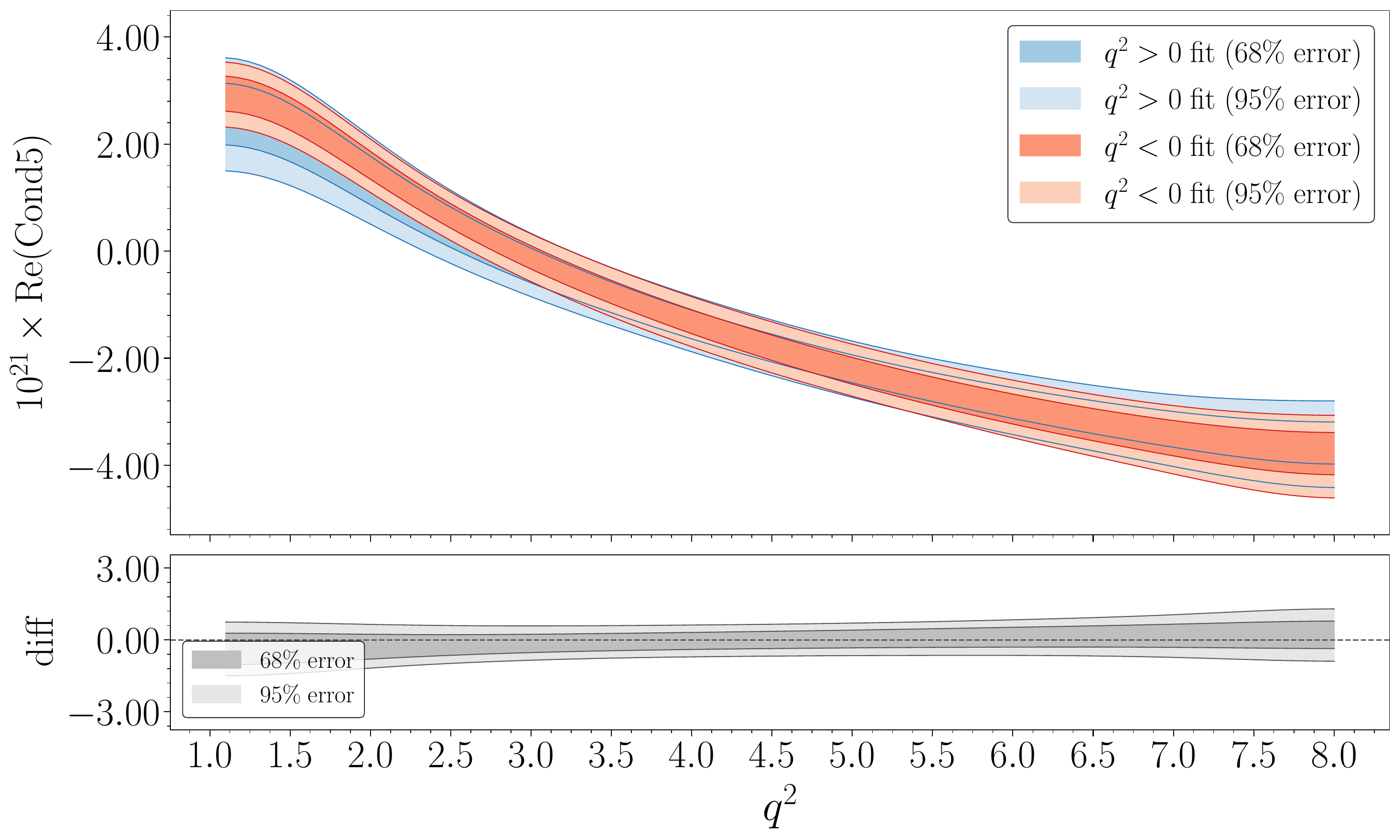}  
\includegraphics[height=150pt]{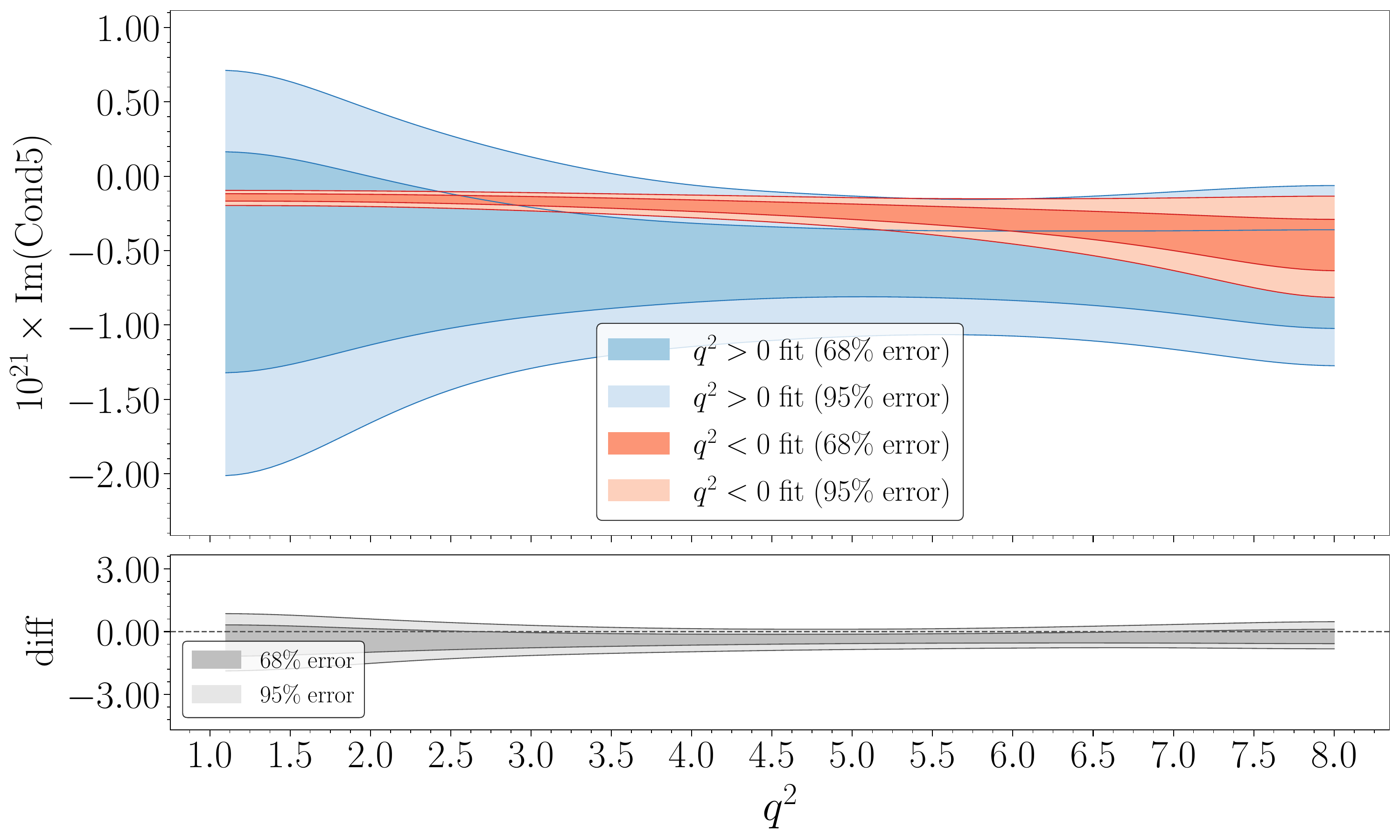} 
\includegraphics[height=150pt]{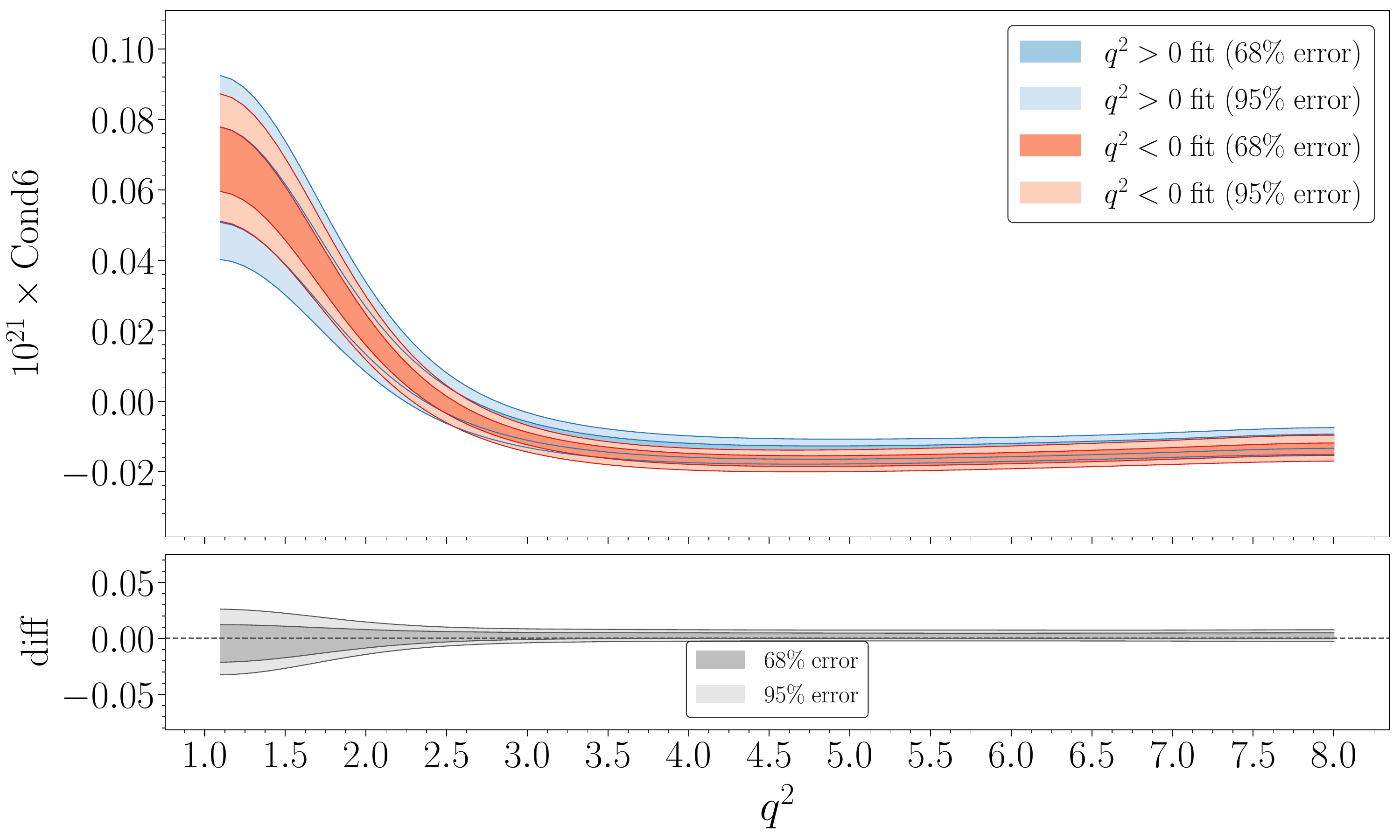}  
\includegraphics[height=150pt]{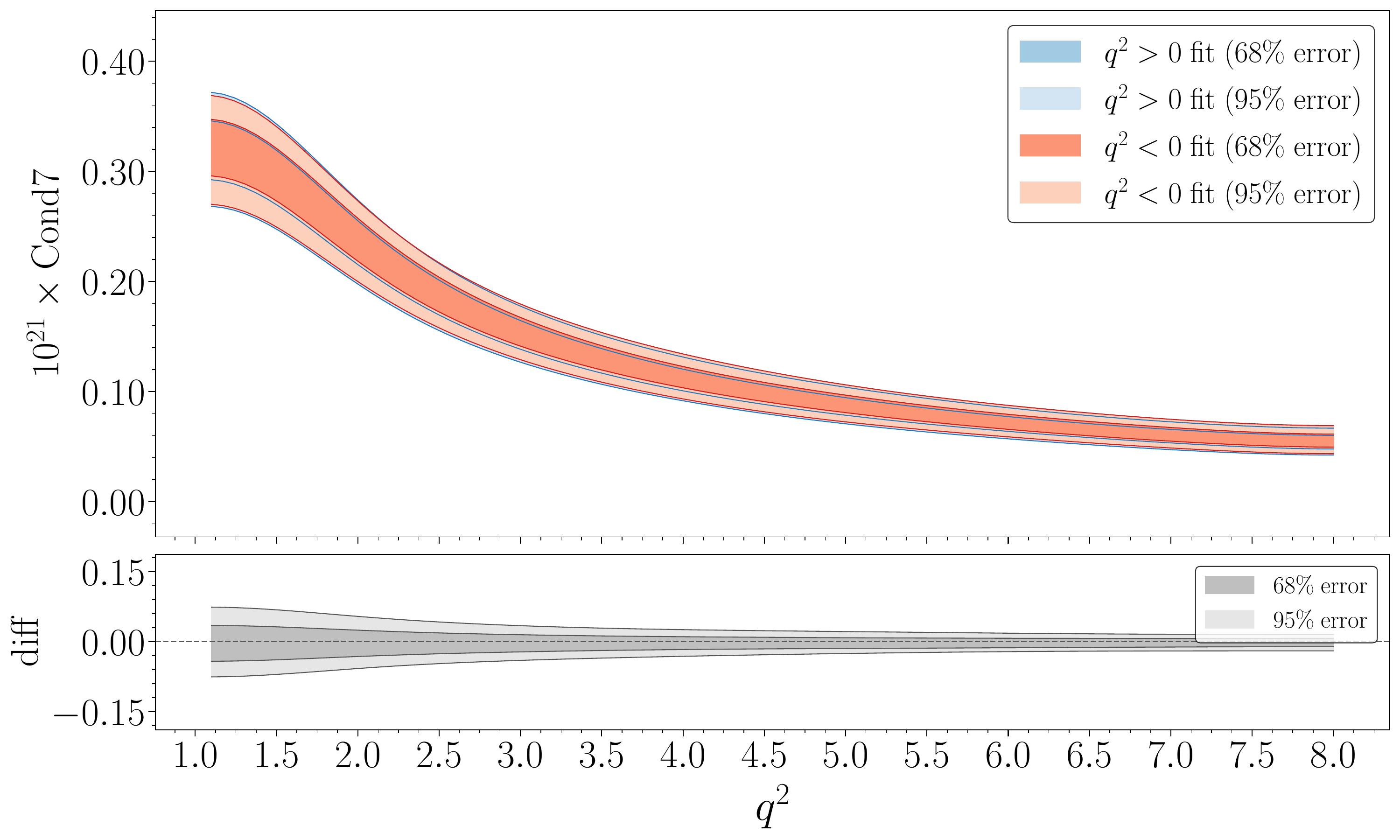} 
\caption{Illustration of conditions 1, 2, 3, 5, 6 and 7. These use the same convention as in Fig.~\ref{condition4}.}
\label{condition5}\end{figure}
\end{widetext}

\clearpage

\section{LEPTON MASS MODIFIED DISTRIBUTION}\label{app:lepmass}

In the theory convention the four-body angular distribution is written:
\begin{widetext}\begin{equation}
\begin{aligned}
    \label{dist}
&\frac{d^4\Gamma}{dq^2\,d\!\cos\theta_K\,d\!\cos\theta_l\,d\phi}=\frac9{32\pi} \bigg[
J_{1s} \sin^2\theta_K + J_{1c} \cos^2\theta_K + (J_{2s} \sin^2\theta_K + J_{2c} \cos^2\theta_K) \cos 2\theta_l\\[1.5mm]
&\hspace{2.7cm}+ J_3 \sin^2\theta_K \sin^2\theta_l \cos 2\phi + J_4 \sin 2\theta_K \sin 2\theta_l \cos\phi  + J_5 \sin 2\theta_K \sin\theta_l \cos\phi \\[1.5mm]
&\hspace{2.7cm}+ (J_{6s} \sin^2\theta_K +  {J_{6c} \cos^2\theta_K})  \cos\theta_l    
+ J_7 \sin 2\theta_K \sin\theta_l \sin\phi  + J_8 \sin 2\theta_K \sin 2\theta_l \sin\phi \\[1.5mm]
&\hspace{2.7cm}+ J_9 \sin^2\theta_K \sin^2\theta_l \sin 2\phi \bigg]\,X+S
\end{aligned}
\end{equation}
\end{widetext}
where S stands for the S-wave part of the amplitude (See Refs.\cite{Descotes-Genon:2015uva,Descotes-Genon:2013vna, Matias:2012xw} for definitions). 

There is a way to define a simpler distribution that contains only the $J$'s required to define the $P_i$ observables and $F_L$ and that 
avoids the conflicting terms $J_{1s}$ and $J_{1c}$ that are proportional to  $F_T$ and $F_L$, respectively, in the massless limit, but that they contain corrections in the massive case. 

The following difference of two distributions is a function of  only the terms that are needed to define the relevant observables:

\begin{equation}
d\hat\Gamma=d\Gamma(\theta_K,\theta_\ell,\phi)-d\Gamma(\theta_K,\theta_\ell+\frac{\pi}{2},\phi)
\end{equation}
where
\begin{widetext}
\begin{equation}
\begin{aligned}
d\hat\Gamma=&\frac9{32\pi} \bigg[
2 (J_{2s} \sin^2\theta_K + J_{2c} \cos^2\theta_K) \cos 2\theta_l\\[1.5mm]
&- J_3 \sin^2\theta_K \cos2\theta_l \cos 2\phi + 2 J_4 \sin 2\theta_K \sin 2\theta_l \cos\phi  + J_5 \sin 2\theta_K (\sin\theta_l-\cos\theta_l)\cos\phi \\[1.5mm]
&+ (J_{6s} \sin^2\theta_K +  {J_{6c} \cos^2\theta_K})  (\sin\theta_l+\cos\theta_l)    
+ J_7 \sin 2\theta_K  \sin\phi (\sin\theta_l-\cos\theta_l)+ 2 J_8 \sin 2\theta_K \sin 2\theta_l \sin\phi \\[1.5mm]
&- J_9 \sin^2\theta_K \cos2\theta_l \sin 2\phi \bigg]\,X+\hat S
\end{aligned}
\end{equation}
\end{widetext}
where $\hat{S}$ stands for the transformed S-wave under the same condition. This new distribution could be useful to avoid problems of precision associated to the extraction of observables when the lepton mass is taken into account.

This difference of distributions does not help with the extraction of the observables from the method of moments when the lepton mass is not neglected, as the degeneracy amongst $S_{2}^{s}$, $S_{2}^{c}$ and $\tilde{S}_{2a}^{c}$ remains. However, for a fit of the data that includes the \mkpi dependence, as in the current generation of analyses, there should be an advantage from this strategy. This is due to the reduction of the degrees of freedom in such a fit.

\section{DISENTANGLING S-WAVE CONTRIBUTIONS AND MASSIVE LEPTONS}
\label{app:swaveandmasses}

To demonstrate the treatment of data with an S-wave component, the pseudo-experiments of the main text have been augmented with a simple S-wave model. The S-wave contribution relative to the P-wave is set to be 0 at the kinematic endpoints ($\qsq=0\gev^{2}$ and $\qsq=19\gev^{2}$) and $\sim10\%$ at $\qsq=9\gev^{2}$. Here, the interference between P- and S-wave has been neglected in the generation of the pseudo-data for the sake of simplicity. The P-wave events are generated with a relativistic Breit Wigner in \mkpi and the S-wave with a LASS parametrisation~\cite{Aston:1987ir,Rui:2017hks}.

In the case that lepton masses are non-negligible one must extract $S_{1}^{c}$, $S_{1}^{s}$, $S_{2}^{c}$ and $S_{2}^{s}$ as they are not trivially related to each other. The respective angular functions of these observables are not orthogonal, such that they can only be obtained from a combination of the moments. In the absence of S-wave contributions, the angular observables are obtained from the moments as per Eq.~\eqref{eq:experiment:masslessmoments}.

Assuming the leptons are massless (and ignoring tensor contributions), one may use the relations~\cite{Altmannshofer_2009}
\begin{align}
    J_{1}^{c} = -J_{2}^{c} ,&& J_{1}^{s} = 3J_{2}^{s} ,&& J_{1a}^{c} = -J_{2a}^{c},
\end{align}

\noindent in which case determining the observables from the moments is trivial, as in Ref.~\cite{Aaij:2115087}. The analyst can pick a middle ground and not use all the relations of massless leptons, to avoid the most egregious biases from lepton masses, as long as there is a unique solution for the observables from the moments.

By taking the leptons as massless one can study the \mkpi dependence of the observables for $\qsq\gtrsim 4\gev^{2}$, which achieves several aims. The \mkpi dependence of the P-wave observables will be assessed for the first time, shedding new light on the narrow-width approximation for the $K^{\ast{}0}$. One can also extract new information about the S-wave observables, in particular if the commonly used functions (the LASS parametrisation or some isobar model) are accurate. It will also become clear if the amplitudes that describe the S-wave should be considered to be the same at low and high values of \mkpi, as is currently assumed by the factorisation of \mkpi from the other kinematic variables in the calculation of the  angular observables. The main text highlights the possibility of achieving a fully two-dimensional map in \mkpi and \qsq of the P-wave observables (without assuming factorisation) in Fig.~\ref{sec:experiment:fig:twodexample}.

Alternatively, the analyst may incorporate information from the kinematic variable \mkpi. Considering \mkpi, the differential decay rate may be written as
\begin{align}
    \frac{\dd\Gamma}{\dd\qsq \dd\Omega \dd\mkpi} =\nonumber\\
    \frac{9}{32\pi}\left[ \sum_{i = 1s\to 9} J_{i}(\qsq) f_{i}(\Omega)|f_{P}(\mkpi|\qsq)|^{2}\right] \nonumber\\
    + \frac{1}{4\pi}\left[ \sum_{i=1ac, 2ac} J_{i}(\qsq)f_{i}(\Omega) |f_{S}(\mkpi|\qsq)|^{2} \right]\nonumber \\
    + \frac{1}{4\pi}\left[\sum_{i = 1bc\to \tilde{S}_{8}} J_{i}(\qsq) f_{i}(\Omega) f_{s}(\mkpi|\qsq)f_{P}(\mkpi|\qsq)\right],
\end{align}
where $\Omega$ represents the three decay angles. The first line describes the P-wave decay, the second is the S-wave and the third the interference between P- and S-wave. The terms $f_{P}(\mkpi|\qsq)$ and $f_{S}(\mkpi|\qsq)$ are complex analytic functions describing the P- and S-wave distributions in \mkpi for a given point in \qsq. There is an implicit assumption that \mkpi factorises from $\Omega$. This should be checked at high \qsq using the massless lepton assumption to determine if the angular observables are functions of \mkpi as well as \qsq.

\begin{figure}[!h]
    \centering
    \includegraphics[width=0.98\linewidth, page = 1]{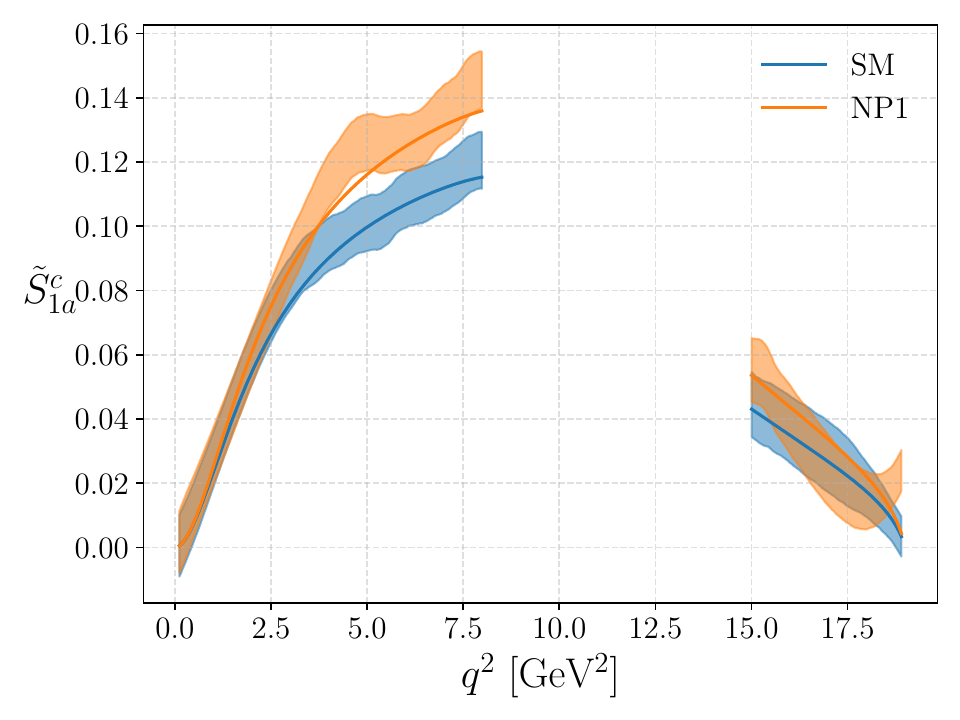}
    \includegraphics[width=0.98\linewidth, page = 2]{Plots/experiment/si_observables_with_swave_trim.pdf}
    \caption{The determination of (top) $\tilde{S}_{1a}^{c}$ and (bottom) $\tilde{S}_{2a}^{c}$ when including the \mkpi description of the calculation of the moments. Shown here is the result of a single pseudo-experiment with the same S-wave amplitudes. The difference between SM and NP are due to the different predicted branching fractions of the P-wave components of the decay.}
    \label{app:swave:fig:swave}
\end{figure}

\begin{figure}[!h]
    \centering
    \includegraphics[width=0.98\linewidth, page = 3]{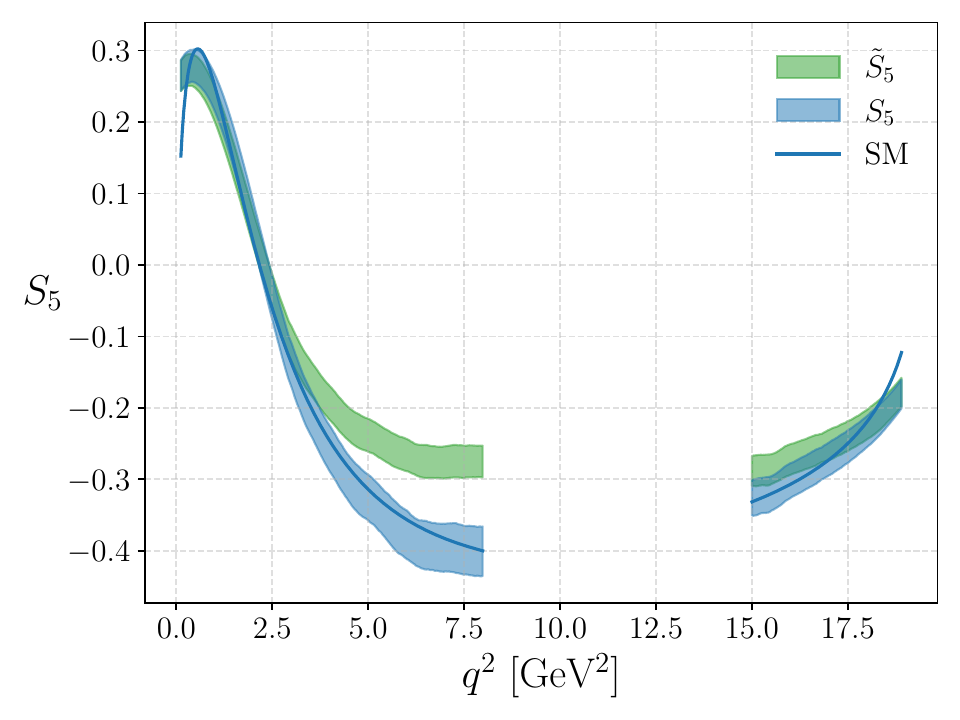}
    \caption{The determination of (green band) $\tilde{S}_{5}$ and (blue band) $S_{5}$ with the moments, compared with (blue line) the true predicted value of $S_{5}$ as from Table~\ref{table:WilsonCoefficientsSMandNP}. The need to renormalise the observables that have been extracted from the moments to account for S-wave contributions is clear.}
    \label{app:swave:fig:reals5}
\end{figure}

If the angular observables are not functions of \mkpi, then $f_{S}(\mkpi|\qsq)$ and $f_{P}(\mkpi|\qsq)$ may be incorporated into the method of moments, for example as
\begin{align}
    M_{1}^{s} = \int_{\rm min}^{\max}\int \frac{{\rm d}\Gamma}{{\rm d}\Omega{\rm d}\mkpi} \sin^{2}\theta_{K} |f_{P}(\mkpi|\qsq)|^{2} {\rm d}\Omega {\rm d}\mkpi,
\end{align}
and analogously for the other moments. The angular functions are by and large orthogonal, so this is only pertinent to the coefficients $S_{1}^{s}$, $S_{1}^{c}$, $S_{s}^{s}$, $S_{s}^{c}$, $\tilde{S}_{1a}^{c}$ and $\tilde{S}_{2a}^{c}$. The precise form of $f_{P}(\mkpi|\qsq)$ and $f_{S}(\mkpi|\qsq)$ are up to the analyst and inevitably contribute as a source of systematic uncertainty in the experimental analysis. One must choose that the functions be normalised over the analysed \mkpi range
\begin{align}
    \int_{\rm min}^{\rm max} |f_{P}(\mkpi|\qsq)|^{2} {\rm d}\mkpi = 1 ,\nonumber\\ \int_{\rm min}^{\rm max} |f_{S}(\mkpi|\qsq)|^{2} {\rm d}\mkpi = 1 .
\end{align}

Defining the integrals
\begin{align}
    A = \int_{\rm min}^{\rm max} |f_{P}(\mkpi|\qsq)|^{4} {\rm d}\mkpi ,\nonumber\\
    B = \int_{\rm min}^{\rm max} |f_{S}(\mkpi|\qsq)|^{4} {\rm d}\mkpi ,\nonumber\\
    C = \int_{\rm min}^{\rm max} |f_{P}(\mkpi|\qsq)|^{2} |f_{S}(\mkpi|\qsq)|^{2} {\rm d}\mkpi ,
\end{align}

one can write down the moments
\begin{widetext}
\vspace*{-0.5cm}
\begin{align}
    M_{1}^{s} = \frac{4}{9} C (3 \soneac-\stwoac)+\frac{1}{10} A(3 \sonect+12 \sonest-\stwoct-4 \stwost), \nonumber\\
    M_{1}^{c} = \frac{2}{9} C (3 \soneac-\stwoac)+\frac{1}{20} A(9 \sonect+6 \sonest-3 \stwoct-2 \stwost), \nonumber\\
    M_{2}^{s} = \frac{1}{50} A(-5 \sonect-20 \sonest+7 \stwoct+28 \stwost)-\frac{4}{45} C (5 \soneac-7 \stwoac), \nonumber\\
    M_{2}^{c} = \frac{1}{100} A(-15 \sonect-10 \sonest+21 \stwoct+14 \stwost)-\frac{2}{45} C (5 \soneac-7 \stwoac), \nonumber\\
    M_{1a}^{c} = \frac{1}{60} (3 C (-5 \sonect-10 \sonest+7 \sonect+14 \stwost)-8 B(5 \soneac-7 \stwoac)), \nonumber\\
    M_{2a}^{c} = \frac{1}{12} (3 C (3\sonect+6 \sonest-\stwoct-2 \stwost)+24 B\soneac-8 B\stwoac).
\end{align}
\end{widetext}

One then finds
\begin{widetext}
\begin{align}
    \tilde{S}_{1}^{s} = \frac{1}{48A(C^{2} - AB)}\left[ AC(20M_{2a}^{c} + 28M_{1a}^{c}) - (\frac{9}{5}AB-C^{2})(35M_{1}^{s} + 25M_{2}^{s}) + (\frac{3}{5}AB - C^{2})(70M_{1}^{c} + 50M_{2}^{c})  \right] , \nonumber\\
    \tilde{S}_{2}^{s} = \frac{5}{48A(C^{2}-AB)}\left[AC(12M_{2a}^{c} + 4M_{1a}^{c}) - (\frac{9}{5}AB-C^{2})(5M_{1}^{s} - 15M_{2}^{s}) + (\frac{3}{5}AB - C^{2})(10M_{1}^{c} + 30M_{2}^{c}) \right] , \nonumber\\
    \tilde{S}_{1}^{c} = \frac{1}{24A(C^{2}-AB)}\left[AC(10M_{2a}^{c} + 14M_{1a}^{c}) - (\frac{6}{5}AB-C^{2})(70M_{1}^{c} + 50M_{2}^{c}) + (\frac{3}{5}AB - C^{2})(35M_{1}^{s} + 25M_{2}^{s}) \right] , \nonumber\\
    \tilde{S}_{2}^{c} = \frac{5}{24A(C^{2}-AB)}\left[AC(6M_{2a}^{c} + 2M_{1a}^{c}) - (\frac{6}{5}AB - C^{2})(10M_{1}^{c} + 30M_{2}^{c}) + (\frac{3}{5}AB - C^{2})(5M_{1}^{s} + 15M_{2}^{s})  \right] , \nonumber\\
    \tilde{S}_{1a}^{c} = \frac{1}{32(C^{2}-AB)}\left[ -15AM_{2a}^{c} - 21AM_{1a}^{c} + C(21M_{1}^{s} + 21M_{1}^{c} + 15M_{2}^{s} + 15M_{2}^{c})\right] , \nonumber\\
    \tilde{S}_{2a}^{c} = \frac{15}{32(C^{2}-AB)}\left[ -3AM_{2a}^{c} - AM_{1a}^{c} + C(M_{1}^{s} + M_{1}^{c} + 3M_{2}^{s} + 3M_{2}^{c})\right] .
    \label{app:swave:eq:proper}
\end{align}
\end{widetext}
Setting $C = 0$, one recovers the P-wave only relations of Eq.~\eqref{eq:experiment:masslessmoments}. This discussion is moot for the observables describing the interference between the P- and S-wave contributions, $\tilde{S}_{1b}^{c}$, $\tilde{S}_{2b}^{c}$, $\tilde{S}_{4}$, $\tilde{S}_{5}$, $\tilde{S}_{7}$ and $\tilde{S}_{8}$. These are already normalised by $\Gamma_{P} + \Gamma_{S}$ and are orthogonal to the other angular functions and so may be extracted directly from their respective moments.

Using the approach of including \mkpi in the moments calculations, as in Eq.~\eqref{app:swave:eq:proper} one has the correct extraction of $\tilde{S}_{1a}^{c}$ and $\tilde{S}_{2a}^{c}$, as shown in Fig.~\ref{app:swave:fig:swave}. Consequently one corrects the normalisation of the P-wave observables with a factor of $1/(1 - 2\tilde{S}_{1a}^{c} + \frac{2}{3}\tilde{S}_{2a}^{c})$. The effect of this renormalisation, and consequently its importance, is evident in Fig.~\ref{app:swave:fig:reals5}.

It should be noted that concerns due to S-wave contributions only pertains to decays of the type $B\to V\ell^+\ell^-$, where there may be several partial waves contributing. When applying this method to $B\to P\ell^+\ell^-$ decays, such as $B^{+}\to \Kp\mu^{+}\mu^{-}$ or $B^{+}\to\pip\mu^{+}\mu^{-}$, there are no other contributing partial waves and so the unbinned extraction of the angular coefficients is trivial.

\bigskip

\bibliography{bibliography}
\end{document}